\documentclass[a4paper,11pt]{article}
\pdfoutput=1 
\usepackage{jheppubMod} 
\usepackage{graphicx}
\usepackage{color}
\usepackage{array,arydshln}
\usepackage{amssymb,amsmath}
\usepackage{tabularray}
\usepackage{mathtools}
\usepackage[T1]{fontenc} 
\newcommand{\red}[1]{{\color[rgb]{0,0,0} #1}}

\newcommand{\green}[1]{{\color[rgb]{0, 0, 0} #1}}

\newcommand{\GeV}{{\text{GeV}}}
\newcommand{\ellp}{\ell^\prime}
\newcommand{\tb}[1]{t_{\beta_{#1}} }
\newcommand{\cotb}[1]{\frac{1}{ t_{\beta_{#1}} }}

\definecolor{navy}{rgb}{0.9,1.,1}

\title{
Anomaly-free axion dark matter in three Higgs doublet model and its phenomenological implications
}

\preprint{TU-1145}

\author{
{\large Kodai Sakurai, Fuminobu Takahashi}
\\*[20pt]
{\it \normalsize 
Department of Physics, Tohoku University, 
Sendai, Miyagi 980-8578, Japan} \\*[5pt]
}

\emailAdd{kodai.sakurai.e3@tohoku.ac.jp}
\emailAdd{fumi@tohoku.ac.jp}

\vspace{20pt}

\abstract{
We study phenomenological implications of an axion that arises as a pseudo Nambu-Goldstone boson due to the spontaneous breaking of anomaly-free global flavor symmetry. 
One interesting possibility for such anomaly-free axion to explain dark matter (DM) is when it has a mass of order keV and an intermediate scale decay constant, \red{since} it can be explored through direct search experiments, X-ray observations, various stellar cooling processes, and the misalignment mechanism naturally explains the DM abundance.
 As a concrete renormalizable model of such axion, we consider an extended Higgs sector with global flavor symmetry, which consists of three Higgs doublet fields and three singlet Higgs fields with $U(1)_{\rm B-L}$ charges.
  We identify viable parameter regions that satisfy theoretical bounds on the Higgs potential and various experimental limits on this model, and evaluate the mass spectra of the axion and extra Higgs bosons.
  \red{
  We find that even an anomaly-free axion can generally couple to photons through mixing with CP-odd Higgs, and that its strength depends on the vacuum expectation values of the Higgs doublets as well as the axion mass.
  As a result,
  }
 \red{the ratios} of the vacuum expectation values of the Higgs doublets are tightly constrained \red{to satisfy the X-ray constraints}. 
\red{We show the favored parameter region where axion DM explains the XENON1T excess.}
We also demonstrate that the axion-electron coupling is correlated with the extra Higgs boson masses and mixing angles for CP-even Higgs bosons. Thus, if the axion is detected in future observations, the extra Higgs boson masses and the coupling of the standard model-like Higgs boson with the weak gauge bosons are restricted.
This is a good example of the synergy between searches for the axion DM and the BSM around the electroweak scale.
  }

\begin{document} 
\maketitle
\flushbottom

\section{Introduction}
The Standard Model (SM) has been scrutinized to a very high degree of accuracy by various experiments.  However, it does not include a candidate for dark matter (DM), which accounts for about a quarter of the current energy density of the universe.  For this reason, DM is considered to be solid evidence of physics beyond the SM.

The identity of DM is still unknown, and one plausible explanation is that it is composed of unknown particles. Various candidates for particle DM have been proposed, and one of the most promising ones is the axion~\cite{Peccei:1977hh,Peccei:1977ur,Weinberg:1977ma,Wilczek:1977pj}. The axion is a pseudo Nambu-Goldstone (NG) boson that appears due to the spontaneous breaking of the global U(1) symmetry.
As such it can easily satisfy the stability requirement for DM because it has naturally a light mass and very weak interactions with the SM particles, if the symmetry breaking scale is sufficiently high. Because of its light mass, the axion is expected to be generally away from its potential minimum in the early universe, so it starts oscillating around the potential minimum when the Hubble parameter becomes roughly equal to the curvature of the potential. The oscillation energy is a natural explanation for the cold axion DM. This is known as the misalignment mechanism~\cite{Preskill:1982cy,Abbott:1982af,Dine:1982ah}. The advantage of axion DM is that it can naturally explain the stability and generation of DM. See Refs.~\cite{Jaeckel:2010ni,Ringwald:2012hr,Arias:2012az,Graham:2015ouw,Marsh:2015xka,Irastorza:2018dyq, DiLuzio:2020wdo} for reviews.

A number of experiments are being conducted to explore the axion DM, and various limits have been placed on its interactions with the SM particles. In particular, axions coupled to photons are often referred to as axion-like particles (ALPs). Light axions are easily produced in hot stellar interiors, which affects stellar evolution by carrying energy outside the star. For instance, the interaction between the axion and photons is restricted by horizontal branch stars~\cite{Ayala:2014pea}, and the interaction between the axion and electrons is restricted by observations of the tip of red giant branch~\cite{Capozzi:2020cbu,Straniero:2020iyi} and white dwarfs~\cite{Battich:2016htm,Corsico:2016okh,MillerBertolami:2014rka}. It is also known that when the axion mass exceeds ${\cal O}(10)$ eV, the observational limit on the {UV and X-ray} photon flux produced by axion decay becomes very tight.  Assuming the anomalous coupling with photons, the decay constant of the axion of keV-scale mass must be above the GUT scale. Such axion DM with the GUT-scale decay constant is far beyond the sensitivity expected in current and near-future direct search experiments, and its contribution to the stellar cooling process is negligibly small.

The axion couplings to gauge bosons are model-dependent, and
various theoretical possibilities  have been discussed.\footnote{For instance, the axion coupling to ordinary or hidden photons can be enhanced~\cite{Higaki:2016yqk,Farina:2016tgd} in a context of the clockwork QCD axion~\cite{Higaki:2015jag}.}
Among them, there is an interesting axion model that escapes the above tight limits from X-ray observations, which was proposed in Ref.~\cite{Nakayama:2014cza} as the anomaly-free axion model. 
For simplicity let us consider a case where the axion is only coupled to leptons.
Then, the axion coupling to photons is caused by one-loop diagrams in which the leptons are running in the loop. 
The anomalous coupling is obviously absent if the charge assignment on the leptons is such that the electromagnetic anomaly is canceled, but the effect of threshold corrections allows the axion to have a suppressed, but finite, coupling to 
photons. On the other hand, the interaction between the axion and the leptons is not particularly suppressed. This lack of anomalous coupling to photons significantly relaxes the severe limits from the  X-ray observations, and allows the axion with a decay constant of intermediate scales to become the dominant component of DM, and to be explored in direct experiments and by cooling processes in stellar objects. Especially in the case of keV-scale mass axions, the misalignment mechanism can naturally explain the observed DM abundance if their decay constant is on the intermediate scale.

Recently, the hint of an excess in the electron recoil events observed in the XENON1T experiment that cannot be explained by the previously known background has attracted much attention~\cite{XENON:2020rca}. The possibility that the excess event is due to a tritium contribution cannot be ruled out, but various DM candidates have been considered to explain it.  Among them, the anomaly-free axion with keV mass is very interesting because it can explain the excess in the XENON1T experiment and at the same time it can explain stellar cooling anomalies~\cite{Takahashi:2020bpq}.

With the above motivation, the purpose of this paper is to investigate a possible UV completion of the anomaly-free axion with keV-scale mass and the decay constant at an intermediate scale. In particular, we consider the three Higgs doublet model (3HDM) as such UV completion. The set-up based on the 3HDM was proposed in Refs.~\cite{Nakayama:2014cza,Takahashi:2020bpq}, but the detailed analysis of the Higgs sector and various phenomenological bounds
have not been investigated so far.
The 3HDM was first proposed in the context of CP violation in the Higgs sector in the pioneering  works~\cite{Weinberg:1976hu,Branco:1980sz} where discrete symmetries were imposed for natural flavor conservation~\cite{Glashow:1976nt}. 
In the 3HDM,  possible interaction terms and the number of parameters crucially depend on the imposed symmetry.
Possible symmetry groups implemented in 3HDMs are surveyed in \green{Refs.~\cite{Ivanov:2011ae,Ivanov:2012fp,Ivanov:2012ry,Keus:2013hya,Ivanov:2015mwl,deMedeirosVarzielas:2019rrp,Darvishi:2019dbh,deMedeirosVarzielas:2021zqs,Varzielas:2022lye}}. 
Furthermore, phenomenological studies are performed in the 3HDM  with  CP4 symmetry~\cite{Ferreira:2017tvy,Ivanov:2021pnr}, 
 $Z_3$ symmetry~\cite{Yagyu:2016whx,Das:2019yad,Boto:2021qgu,Chakraborti:2021bpy}, 
 $Z_2$ symmetries~\cite{Cree:2011uy,Akeroyd:2016ssd, Akeroyd:2020nfj,Logan:2020mdz,Akeroyd:2021fpf,Davoudiasl:2019lcg,Davoudiasl:2021syn}, 
 Sp(6) symmetry~\cite{Darvishi:2021txa}, 
 U(1)$\times Z_2$ flavor symmetry~\cite{Das:2021oik}, 
 and $S_3$ symmetry~\cite{Das:2014fea}. 
Also, eight extra Higgs bosons are predicted in addition to the SM-like Higgs boson with the mass of 125~GeV. 
For this reason, the phenomenology of 3HDM is very rich, and we will focus on the symmetry and matter content that would fit with the anomaly-free axion.
{The anomaly-free axion has also been studied from other phenomenological aspects such as lepton-flavor violation~\cite{Calibbi:2020jvd,Han:2020dwo,Han:2022iig} as well as  inflation~\cite{Takahashi:2020uio}.}

This paper is organized as follows. In Sec.~\ref{sec:model} we give the set-up of the 3HDM, and we study the mass spectrum of the axion and CP-odd Higgs bosons in Sec.~\ref{sec:mass}. We take into account various theoretical and experimental bounds on the model parameters and identify the viable parameter region in Sec.~\ref{sec:con}. The implications for the direct DM search experiments as well as future X-ray observations are studied in Sec.~\ref{sec:XENON1T} and Sec.~\ref{sec:Xray}, respectively. \red{
In particular, we will show that the anomaly-free axion generally acquires a coupling to photons via the mixing with CP-odd Higgs bosons, and we discuss its implications for the X-ray constraints on the model parameters. 
} The last section is devoted to conclusions.

\section{Three Higgs doublet model with $B-L$ Higgs bosons}\label{sec:model}
 We first provide the set-up of the 3HDM with $B-L$ Higgs fields, which is essentially same as the one proposed in Ref.~\cite{Nakayama:2014cza}, but the matter content of the $B-L$ Higgs sector is slightly simplified. We impose a global $U(1)_F$ flavor symmetry on leptons in such a way that \red{its electromagnetic anomaly vanishes, and} the anomalous coupling of the axion to photons is canceled \red{if the mixing effect is negligible}. In this case, the axion-photon coupling arises from the threshold corrections in the lepton loop diagrams. \red{The effect of the mixing between the axion and the CP-odd Higgs will be discussed later in this paper.}

We introduce three Higgs doublet fields $\phi_{i}$ ($i=1,2,3$), and three Higgs singlet fields with $U(1)_{\rm B-L}$ charge +2,  $S_{0}$, $S_{1}$, $S_{\bar{2}}$, where the subscript of $S$ (not $\phi$) denotes the flavor charge.
The assignment of the $U(1)_{F}$ flavor charge $q$ for these Higgs fields and left-handed (right-handed) lepton fields $L_\ell$ ($\ell_R$) $(\ell=e,\mu, \tau)$ are presented in Table~\ref{tab:Q}.
Supposing $L_e$ and $e_R$ are charged under the $U(1)_{F}$ symmetry, there are two possible combinations for the charge assignment of the right-handed leptons, and we call them Type-A and Type-B, respectively. 
Quarks are assumed to be neutral under the $U(1)_{F}$ symmetry.

\begin{table}[t]
  \begin{center}
  \begin{tabular}{c||c|c|c||c|c|c||c|c|c||c|c|c}
   $U(1)_F$ charge $q$ &$\phi_{1}$&$\phi_{2}$&$\phi_{3}$&$S_{0}$&$S_{1}$&$S_{\bar{2}}$&$L_e$&$L_\mu$&$L_\tau$&$e_R$&$\mu_{R}$&$\tau_{R}$\\ \hline
   Type-A& $-3$&$3$&$0$&$0$&$1$&$-2$&$1$&$0$&$-1$&$-2$&0&2\\
   Type-B& $-3$&$3$&$0$&$0$&$1$&$-2$&$1$&{$-1$}&{$0$}&$-2$&2&0\\
  \end{tabular}
  \end{center}
  \caption{The charge assignment for the global U(1)$_F$  flavor symmetry.}
  \label{tab:Q}
  \end{table}%

While soft breaking terms for the $U(1)_{F}$ symmetry are introduced in the Higgs potential, $U(1)_{\rm B-L}$ symmetry spontaneously breaks after the $B-L$ Higgs singlet fields acquire vacuum expectation values (VEVs).
In the following, we describe the Higgs potential and the Yukawa Lagrangian. 
Hereafter, we use shorthanded notation for trigonometric functions, $s_\theta\equiv\sin\theta $, $c_\theta\equiv\cos\theta $, and $t_\theta\equiv\tan\theta$ as needed. 

 As we will see later, the charged lepton Yukawa is already diagonalized in the flavor basis. As pointed out in Ref.~\cite{Nakayama:2014cza}, if we further introduce three right-handed neutrinos with flavor charges $(1, -1, 0)$, and if the VEVs of the $B-L$ Higgs are of similar magnitude, then light neutrino masses and the large mixing angles can be realized by the seesaw mechanism~\cite{Minkowski:1977sc,Yanagida:1979as,Ramond:1979py,Glashow:1979nm}. In the following, we will focus on the correlation of the axion and heavy Higgs bosons.

\subsection{Higgs potential}
The Higgs potential we consider can be divided into three parts as
\begin{align}\label{eq:gpotential}
V_{}=V_{\rm 3HDM}+V_{\rm B-L}+V_{I}\;.
\end{align}
where $V_{\rm 3HDM}$ ($V_{\rm B-L}$) denotes the potential for the three Higgs doublets $\phi_i$ ($B-L$ Higgs $S_j$), and $V_{I}$ corresponds to interaction terms between $\phi_i$ and $S_j$. 

The potential for the three Higgs doublets, $V_{\rm 3HDM}$, is given by
\begin{eqnarray}
  V_{\rm 3HDM} &=& m_{11}^2(\phi_1^\dagger \phi_1) +  m_{22}^2(\phi_2^\dagger
  \phi_2) +  m_{33}^2 (\phi_3^\dagger \phi_3)    \nonumber \\
   &&+\lambda_1(\phi_1^\dagger
  \phi_1)^2 + \lambda_2(\phi_2^\dagger \phi_2)^2 +
  \lambda_3(\phi_3^\dagger \phi_3)^2  \nonumber \\
  && + \lambda_4 (\phi_1^\dagger \phi_1)( \phi_2^\dagger \phi_2) + \lambda_5 (\phi_1^\dagger \phi_1)( \phi_3^\dagger \phi_3) +
  \lambda_6 (\phi_2^\dagger \phi_2)( \phi_3^\dagger \phi_3) \nonumber \\
  && + \lambda_7 (\phi_1^\dagger \phi_2)( \phi_2^\dagger \phi_1) +
  \lambda_8 (\phi_1^\dagger \phi_3)( \phi_3^\dagger \phi_1)+ \lambda_9
  (\phi_2^\dagger \phi_3)( \phi_3^\dagger \phi_2) \nonumber \\
  %
  %
  && + \left[\lambda_{10} (\phi_3^\dagger \phi_1)( \phi_3^\dagger \phi_2) + {\rm h.c.} \right]+V_{\rm soft} \,,
  \label{e:potential}
\end{eqnarray}
where the soft breaking terms are given by
\begin{align}
V_{\rm soft}=
-\left[m_{12}^2(\phi_1^\dagger \phi_2) 
  + m_{23}^2(\phi_2^\dagger\phi_3) 
  +  m_{13}^2 (\phi_1^\dagger \phi_3) 
  + {\rm h.c.} \right] \; .
\end{align}
The first term $m_{12}^{2} (\phi_1^\dagger \phi_2)$ breaks $U(1)_{F}$ down to its $Z_{6}$ subgroup, while the other terms $m_{13}^{2}(\phi_1^\dagger\phi_3)$ and $m_{23}^{2}(\phi_2^\dagger\phi_3)$ break $U(1)_{F}$ to $Z_{3}$. 
In this paper, we focus on the pattern of the symmetry breaking of $U(1)_F \rightarrow Z_6$, taking $m_{12}^2\neq0$, $m_{13}^2=0$ and $m_{23}^2=0$. 
While we only introduce $m^2_{12}$, the terms similar to $m_{13}^{2}$ and $m_{23}^{2}$ are dynamically generated from the interaction terms $V_{ I}$ as we will see shortly. 
Without loss of generality, one can take $m_{12}$ and $\lambda_{10}$ to be real by using the phase degrees of freedom of  $\phi_2$ and $\phi_3$. 
We also note that the above potential corresponds to the one in the $Z_{3}$ invariant 3HDM with the $U(1)_F$ symmetry breaking terms other than $m_{12}^2$ set to $0$ 
in Ref.~\cite{Chakraborti:2021bpy}.

Under the $U(1)_{F}$ and $U(1)_{\rm B-L}$ symmetries, the Higgs potential in the $B-L$ Higgs sector and the interaction terms between the Higgs doublet fields and $B-L$ Higgs fields are given by\footnote{
  Our purpose is to induce the axion mass by the soft breaking terms in $V_{\rm 3HDM}$, namely the Higgs sector related to the electroweak symmetry breaking.
  Hence, we do not consider the soft breaking term $\mu_{1\bar{2}}S_{1}^{\dagger}S_{\bar{2}}$. in $V_{\rm B-L}$. In fact this is ensured by considering only the explicit breaking that 
  preserves the $Z_6$ subgroup. }
\begin{align}
V_{B-L}&=
\sum_{i=0,1,\bar{2}}(\mu_{i}|S_{i}|^{2}+\kappa_{i}|S_{i}|^{4})+\kappa_{01}|S_{0}|^{2}|S_{1}|^{2}+\kappa_{0\bar{2}}|S_{0}|^{2}|S_{\bar{2}}|^{2}+\kappa_{1\bar{2}}|S_{1}|^{2}|S_{\bar{2}}|^{2} \notag \\
&+\green{\kappa_{0110}|S_{0}^\dagger S_{1}|^2+\kappa_{0\bar{2}\bar{2}0}|S_{0}^\dagger S_{\bar{2}}|^2+\kappa_{1\bar{2}\bar{2}1}|S_{1}^\dagger S_{\bar{2}}|^2} \;,\\
\label{eq:V_I}
V_{I}&=
 \sum_{m=0,1,\bar{2}}\sum_{n=1,2,3}\kappa_{m\phi n}|S_{m}|^{2}(\phi_{n}^{\dagger}\phi_{n}) \notag \\
&+\left[ \kappa_{1\bar{2}\phi1\phi3}S_{1}^{\dagger}S_{\bar{2}}(\phi_{1}^{\dagger}\phi_{3})
+\kappa_{\bar{2}1\phi2\phi3}S_{\bar{2}}^{\dagger}S_{1}(\phi_{2}^{\dagger}\phi_{3})+{\rm h.c.}\right]\; .
\end{align}
The dimensionless parameters $\kappa_{1\bar{2}\phi1\phi3}$ and $\kappa_{\bar{2}1\phi2\phi3}$ are taken to be real while one of them is generally complex. 

In the phase after the electroweak symmetry breaking (EWSB), the component fields of the Higgs doublets can be given by
\begin{eqnarray}
  \phi_k = \frac{1}{\sqrt{2}}
  \begin{pmatrix} \sqrt{2}\, w_k^+ \\  v_k + h_k
    + i z_k  \end{pmatrix} \;, \quad k = 1,2,3,
  \label{e:scalvev}
\end{eqnarray}
where $v_k$ represents the vacuum expectation value (VEV) of the CP-even components.
We parameterize the VEVs as~\cite{Das:2019yad}
\begin{eqnarray}\label{eq:vevs}
  v_1 = v \cos\beta_1 \cos\beta_2 \;, \quad v_2 = v \sin{\beta_1} \cos\beta_2 \;, \quad
  v_3 = v \sin{\beta_2}\;,
\end{eqnarray}
The electroweak VEV is then obtained by $v = \sqrt{v_1^2+v_2^2+v_3^2}$.
On the other hand, assuming that $U(1)_{\rm B-L}$ is already spontaneously broken, we parameterize the $B-L$ Higgs fields as
\begin{align}\label{eq:BLscalar}
S_{j}=\frac{1}{\sqrt{2}}(v_{S_{j}}+\rho_{j})e^{\frac{iQ_{j}}{f_{a}}\tilde{a}}\ \ \ (j=0,\ 1,\ \bar{2})\;,
\end{align}
where $\tilde{a}$ denotes the pseudo NG boson associated with the breaking of the $U(1)_F$ symmetry, $Q_{j}$ denotes the $U(1)_{F}$ charge  of $S_{j}$ (see Table.~\ref{tab:Q}), and
 $f_{a}$ denotes the decay constant for the $\tilde{a}$.
 Note that $f_a$ is generally expressed by a linear combination of the VEVs of the $B-L$ Higgs and $\phi_{1,2}$, but we drop the contribution of $\phi_i$ by assuming the $B-L$ breaking scale is much larger than the electroweak scale. Specifically,
we assume later that  $v_{S_{j}}$ and $f_{a}$ are of $\mathcal{O}(10^{10}\mathchar`-10^{12})~\GeV$. 
\green{We also note that the definition of the pseudo scalar $\tilde{a}$ is the one for the specific case of $v_{S_1}=v_{S_{\bar{2}}}$, but we use it for simplicity. }
Although there are the other NG boson for the $U(1)_{\rm{B-L}}$ symmetry and a heavy CP-odd Higgs boson in the $B-L$ sector, they are integrated out in our study.
We also integrated out the three CP-even Higgs bosons $\rho_{j}$ after we set their tadpole conditions.

Setting the $B-L$ Higgs fields equal to their VEVs in Eq.~\eqref{eq:V_I}, one obtains terms similar to $m^2_{13}$ and $m^2_{23}$ 
\begin{align}
  V_I \ni -m^{\prime 2}_{13}e^{-3i \frac{\tilde{a}}{f_a}} (\phi_{1}^{\dagger}\phi_{3})
  -m^{\prime 2}_{23} e^{3i \frac{\tilde{a}}{f_a}}(\phi_{2}^{\dagger}\phi_{3})
  +{\rm h.c.}
  \end{align}
with
\begin{align}
  -m^{\prime 2}_{13}=\frac{1}{2}\kappa_{1\bar{2}\phi1\phi3}v_{S_1}v_{S_{\bar{2}}}\; , \quad
  \red{-m^{\prime 2}_{23}=\frac{1}{2}\kappa_{\bar{2}1\phi2\phi3}v_{S_1}v_{S_{\bar{2}}}\; .}
\end{align}
Note that, although these interaction terms do not explicitly violate the $U(1)_F$ flavor symmetry, at least one of them must be nonzero for the axion to obtain nonzero mass. 
This is because the $B-L$ sector where the axion resides and the EW sector where the $U(1)_F$ symmetry is explicitly broken are completely separated, otherwise. Hence we call $m^{\prime 2}_{13}$ and $m^{\prime 2}_{23}$  the soft breaking parameters in the following.
Hereafter, we introduce the following rescaled soft breaking parameters,
\begin{align}\label{eq:bmsq}
  M^2_{12}=\frac{m_{12}^2}{c_{\beta_1}s_{\beta_1}c_{\beta_2}^2}\;,\quad
  M^{\prime 2}_{13}=\frac{m_{13}^{\prime 2}}{c_{\beta_1}c_{\beta_2}s_{\beta_2}}\;,\quad
  M^{\prime 2}_{23}=\frac{m_{23}^{\prime 2}}{s_{\beta_1}c_{\beta_2}s_{\beta_2}}\;,\quad
\end{align}
 and choose them as input parameters. 
 The replacement of Eq.\eqref{eq:bmsq} makes the expressions for the quartic couplings $\lambda_{1-9}$ simple as shown in Appendix~\ref{ap:lambda}. 
 
\red{
Although the rescaled mass parameters are introduced for convenience, this parameterization may obscure the scale of the portal couplings.   
Before closing this section, we discuss typical size of the portal couplings. 
It can be estimated from the minimization conditions 
for Higgs doublet fields $\phi_i$ ($i=1,2,3$), which are given in Eq.\eqref{eq:m11}-\eqref{eq:m33}.
The mass parameters $m^2_{ii}$ ($ii=11,22,33$) should be typically at the EW scale. 
This follows that the order of all the portal couplings should be 
\begin{align}
   \kappa_X\sim \mathcal{O}(v^2/f_a^2)\sim 10^{-16}\ \ \mbox{at}\ \ f_a=10^{10} {\rm GeV}\;,
\end{align}
where the subscript $X$ can be $ 1 \bar{2}  \phi 1\phi 3$, $ \bar{2} 1 \phi 2\phi 3$, or $m\phi n$ $(m,n=0,1,\bar{2})$.
As diseased in Sec.\ref{sec:mass}, the corresponding mass scale of the axion is $1~{\rm keV}$ if $f_a\sim 10^{10}$ GeV and $m_{12}\sim 100~{\rm GeV}$. 
Since all the portal couplings are small, our model is natural in the sense of 't Hooft~\cite{tHooft:1979rat} (also see the discussion of naturalness in the axion model, e.g. Ref.~\cite{Volkas:1988cm}).
}

\subsection{Physical states after EWSB}
\subsubsection{Definition of physical states}
In the potential obtained by integrating out heavy degrees of freedom, we have the six charged Higgs bosons $w_{k}^{\pm}$, the four CP-odd Higgs bosons $z_{k}$, $\tilde{a}$ and the three CP-even Higgs bosons $h_{k}$.
They are related to the mass eigenstates by orthogonal transformations.

For the charged Higgs bosons, the physical states are given by 
\begin{equation}
  \begin{pmatrix} G^\pm\\ H_1^\pm \\ H_2^\pm\end{pmatrix} =R_{+}
    \begin{pmatrix} w_1^\pm\\ w_2^\pm \\ w_3^\pm\end{pmatrix}\; , \quad
      \quad
\end{equation}
where $R_+$ is a mixing matrix,
$G^\pm$ are NG bosons eaten by the $W^\pm$ bosons, and $H^\pm_1$ and  $H^\pm_2$ denote the charged Higgs bosons. 
The mixing matrix $R_{+}$ is given by a product of two rotation matrices,
\begin{align}
R_{+}=\mathcal{O}_{\gamma_+}\mathcal{O}_\beta \;,
\end{align}
where they can be parameterized as
\begin{eqnarray}\label{eq:gammaRots}
  {\cal O}_{\gamma_+} =
  \begin{pmatrix}
    1 & 0 & 0 \\
    0 & \cos\gamma_+ & -\sin{\gamma_+} \\
    0 & \sin{\gamma_+} & \cos\gamma_+ \end{pmatrix}  \label{e:Ogamma1} \,,
\end{eqnarray}
and
\begin{eqnarray}\label{eq:betaRot}
  {\cal O}_{\beta} =
  \begin{pmatrix} \cos\beta_2 \cos\beta_1 & \cos\beta_2 \sin{\beta_1}
    &  \sin{\beta_2} \\
    -\sin{\beta_1} & \cos\beta_1  &  0  \\
    -\cos\beta_1 \sin{\beta_2} & -\sin{\beta_1}\sin{\beta_2} & \cos\beta_2
  \end{pmatrix}.
\end{eqnarray}
Here we have followed the parametrization of $\cal{O}_\beta$ given in Ref.~\cite{Das:2019yad}. 
The rotation matrix ${\cal O}_{\beta}$ corresponds to the transformation from the original basis $(\phi_1,\phi_2,\phi_3)$ into the Higgs basis~\cite{Georgi:1978ri,Lavoura:1994fv,Lavoura:1994yu,Davidson:2005cw}. In other words,
by the rotation matrix ${\cal O}_{\beta}$, the NG bosons $G^\pm$ are identified, and the remaining two states are transformed into $(H_1^\pm,H_2^\pm)$ by ${\cal O}_{\gamma_+}$. 

For the CP-even Higgs bosons, the physical states ($H_1,H_2,H_3$) are obtained by 
\begin{eqnarray}
  \label{e:ab3hdm}
\begin{pmatrix}
    H_1 \\
    H_2 \\
    H_3
  \end{pmatrix}
  &=&R_{S} 
  \begin{pmatrix}
    h_1 \\
    h_2 \\
    h_3
  \end{pmatrix}\;,
  \label{CP-even}
\end{eqnarray}
where the mixing matrix  $R_{S}$ is given by
\begin{subequations}
  \label{e:Oa}
  \begin{eqnarray}
    R_{S} &=& {\cal O}_{\alpha_3}   {\cal O}_{\alpha_2} {\cal O}_{\alpha_1} \,,
  \end{eqnarray}
with
  \begin{equation}
    \label{e:R}
          {\cal O}_{\alpha_1} = \begin{pmatrix}
                     \cos \alpha_1 & \sin {\alpha_1} & 0 \\
            -\sin {\alpha_1} & \cos \alpha_1 & 0 \\
            0 & 0 & 1 \end{pmatrix}
          \,, \quad {\cal O}_{\alpha_2} = \begin{pmatrix}
            \cos \alpha_2 & 0 & \sin {\alpha_2}  \\
            0 & 1 & 0 \\
            -\sin {\alpha_2} & 0 & \cos \alpha_2
          \end{pmatrix}\,,  \quad
          {\cal O}_{\alpha_3} = \begin{pmatrix}
            1 & 0 & 0 \\
            0 & \cos \alpha_3 &  \sin {\alpha_3}  \\
            0 & -\sin {\alpha_3} & \cos \alpha_3
          \end{pmatrix}.
  \end{equation}
\end{subequations}
We identify $H_1$ as the SM-like Higgs boson with the mass of 125~GeV and $H_{2,3}$ as the additional CP-even Higgs bosons.

For the CP-odd Higgs bosons, the physical sates are given by
 \begin{equation}
      \begin{pmatrix} G^{0}\\ A_1 \\ A_2\\ a\end{pmatrix} =R_{P}
        \begin{pmatrix} z_1\\ z_2 \\ z_3\\ \tilde{a}\end{pmatrix} ,
        \label{charged-and-CP-odd}
\end{equation}
where $R_P$ is the mixing matrix, $G^0$ is  the neutral NG boson eaten by the $Z$ boson, $a$ is the axion corresponding to the
spontaneous breaking of $U(1)_F$, and $A_1$ and $A_2$ are additional CP-odd Higgs bosons. 
The mixing matrix is expressed by
\begin{subequations}
\begin{equation}
R_{P}=
{\cal R}_{A}{\cal R}_{\beta}\;,
\end{equation}
with $4\times 4$ matrices
\begin{align}
{\cal R}_{A}=
\left(
\begin{array}{c:ccc}
1&0&0&0\\ \hdashline
0&&&\\
0&&\mbox{\smash{\huge  ${ R}_{A}$ }}&\\
0&&&\\
\end{array}
\right)\;,\quad
{\cal R}_{\beta}=
\left(
\begin{array}{ccc:c}
&&&0\\
&\mbox{\smash{\huge  ${\cal O}_{\beta}$ }}&&0\\
&&&0\\ \hdashline
0&0&0&1
\end{array}
\right)\; .
\end{align}
Here we introduce the 3 $\times$ 3 orthogonal matrix $R_A$.
While we numerically  derive $R_A$ in the following numerical calculations, the matrix $R_A$ can also be parameterized by introducing rotation matrices,
\begin{align}\label{eq:gamma_i}
R_A={\cal O}_{\gamma_3}{\cal O}_{\gamma_2}{\cal O}_{\gamma_1}
\end{align}
\end{subequations}
The $3\times 3 $ matrices ${\cal O}_{\gamma_{i}}$ are defined by the replacement of $\alpha_i\to\gamma_i$ ($i=1,2,3$) in  Eq.~\eqref{e:R}. 
For later convenience, we define fields in the Higgs basis as 
 \begin{equation}
      \begin{pmatrix} G^{0}\\ z^\prime_2 \\ z^\prime_3\\ a^\prime \end{pmatrix} ={\cal R}_{\beta}
        \begin{pmatrix} z_1\\ z_2 \\ z_3\\ \tilde{a}\end{pmatrix}\;.
\end{equation}

  \subsubsection{Mass matrices for the Higgs sector}
  The mass matrix of the charged Higgs bosons $H_{1}^{\pm}$ and $H_{2}^{\pm}$ in the Higgs basis is given by
  \begin{subequations}\label{e:BC2}
    \begin{eqnarray}
    {\cal B}_{C}^2 \equiv {\cal O}_{\beta}  {\cal M}_{C}^2  {\cal O}_{\beta}^{T} &=& \begin{pmatrix}
    0 & 0 & 0 \\
    0 & {({\cal B}_C^2)}_{22}  & {({\cal B}_C^2)}_{23} \\
    0 & {({\cal B}_C^2)}_{23} & {({\cal B}_C^2)}_{33} \\
    \end{pmatrix} \;,
    \end{eqnarray}
    where,
    \begin{eqnarray}
 {({\cal B}_C^2)}_{22} &=& \frac{1}{2}\Big[
2 M^{2}_{12} c_{\beta_2}^2+2 M^{\prime 2}_{13} s_{\beta_1}^2 s_{\beta_2}^2+2 M^{\prime 2}_{23} c_{\beta_1}^2 s_{\beta_2}^2 \notag \\
&-&\frac{1}{4} \lambda_{10} v^2 (c_{4 \beta_1}+3) \frac{1}{s_{\beta_1}} \frac{1}{c_{\beta_1}} s_{\beta_2}^2-\lambda_8 v^2 s_{\beta_1}^2 s_{\beta_2}^2-\lambda_9 v^2 c_{\beta_1}^2 s_{\beta_2}^2-\lambda_7 v^2 c_{\beta_2}^2 \Big]\;, \\
{({\cal B}_C^2)}_{23} &=&\frac{1}{2}\Big[
M^{\prime 2}_{13} s_{2 \beta_1} s_{\beta_2}-M^{\prime 2}_{23} s_{2 \beta_1} s_{\beta_2}+\lambda_{10} v^2 c_{2 \beta_1} s_{\beta_2} \notag \\
&-&\lambda_8 v^2 s_{\beta_1} c_{\beta_1} s_{\beta_2}+\lambda_9 v^2 s_{\beta_1} c_{\beta_1} s_{\beta_2} \Big]\;,  \\
{({\cal B}_C^2)}_{33} &=&  \frac{1}{2}\Big[
2 M^{\prime 2}_{13} c_{\beta_1}^2+2 M^{\prime 2}_{23} s_{\beta_1}^2-\lambda_{10} v^2 s_{2 \beta_1}-\lambda_8 v^2 c_{\beta_1}^2-\lambda_9 v^2 s_{\beta_1}^2
\Big]\,.
    \end{eqnarray}
  \end{subequations}
  Performing the further rotation for $({\cal B}_{C})^2$ with ${\cal O}_{\gamma_+}$ yields 
  the masses of $H^\pm_1$ and $H^\pm_2$ and the mixing angle {$\gamma_+$} as
\begin{align}\label{eq:mch1}
  m^2_{H^\pm_1}&=
  \cos^2\gamma_+({\cal B}^2_{C})_{11}-\sin(2\gamma_+)({\cal B}^2_{C})_{12}+\sin^2\gamma_+({\cal B}^2_{C})_{22}\;,  \\
  m^2_{H^\pm_2}&=
  \sin^2\gamma_+({\cal B}^2_{C})_{11}+\sin(2\gamma_+)({\cal B}^2_{C})_{12}+\cos^2\gamma_+({\cal B}^2_{C})_{22}\;,\\ \label{eq:gammap}
\tan(2\gamma_+)&=\frac{({\cal B}^2_{C})_{12}}{({\cal B}^2_{C})_{22}-({\cal B}^2_{C})_{11}}\;.
\end{align}
  The quartic couplings $\lambda_{7}$, $\lambda_{8}$ and $\lambda_{9}$ are expressed in terms of these three physical parameters, $\gamma_{+}$, $m_{H^{\pm}_{1}}$ and $m_{H^{\pm}_{2}}$.

  Similarly, the masses of the CP-even Higgs bosons are represented by
  \begin{eqnarray}\label{e:msdiag}
    {\rm diag}(m^2_{H_1},m^2_{H_2},m^2_{H_3})= R_S {\cal M}_{S}^2  R_S^T\;, 
    \end{eqnarray}
    where the mass matrix for the CP-even Higgs bosons in the basis $(h_{1}, h_{2} , h_{3})^{T}$, $\mathcal{M}^{2}_{S}$, is given by
	\begin{subequations}
	\begin{eqnarray}
{({\cal M}_S^2)}_{11} &=&
\frac{1}{2} c_{\beta_2}^2 \Big[2 M^{2}_{12} s_{\beta_1}^2+t_{\beta_2}^2 (2 M^{\prime 2}_{13}-\lambda_{10} v^2 t_{\beta_1})+4 \lambda_1 v^2 c_{\beta_1}^2\Big]\;, \\ 
{({\cal M}_S^2)}_{12} &=&
\frac{1}{2} \lambda_{10} v^2 s_{\beta_2}^2-s_{\beta_1} c_{\beta_1} c_{\beta_2}^2 \Big(M^{2}_{12}-v^2 (\lambda_4+\lambda_7)\Big)\;,  \\
{({\cal M}_S^2)}_{13} &=&
s_{\beta_2} c_{\beta_2} \Big[v^2 \Big(\lambda_{10} s_{\beta_1}+c_{\beta_1} (\lambda_5+\lambda_8)\Big)-M^{\prime 2}_{13} c_{\beta_1}\Big]\;,  \\
{({\cal M}_S^2)}_{22} &=&
\frac{1}{2} c_{\beta_2}^2 \Big[2 M^{2}_{12} c_{\beta_1}^2+t_{\beta_2}^2 (2 M^{\prime 2}_{23}-\lambda_{10} v^2 \cot{\beta_1})+4 \lambda_{2} v^2 s_{\beta_1}^2\Big]\;,  \\
{({\cal M}_S^2)}_{23} &=&
s_{\beta_2} c_{\beta_2} \Big[v^2 (\lambda_{10} c_{\beta_1}+s_{\beta_1} (\lambda_{6}+\lambda_9))-M^{\prime 2}_{23} s_{\beta_1}\Big]\;, \\
{({\cal M}_S^2)}_{33} &=&
M^{\prime 2}_{13} c_{\beta_1}^2 c_{\beta_2}^2+M^{\prime 2}_{23} s_{\beta_1}^2 c_{\beta_2}^2+2 \lambda_3 v^2 s_{\beta_2}^2
	\end{eqnarray}\label{e:mselement}
\end{subequations}
With Eq.~\eqref{e:msdiag}, the six potential parameters, $\lambda_{i}$ ($i=1-6$) can be expressed by the physical parameters $\alpha_{j}$ and $m_{H_{j}}$ ($j=1,2,3$).

Finally we express the mass matrices for the CP-odd Higgs bosons $A_{1}$, $A_{2}$ and the axion by the mass matrix in the Higgs basis as
	\begin{subequations}
		\begin{eqnarray}\label{e:mp2x2}
		{\cal B}_{P}^2 \equiv {\cal R}_{\beta}  {\cal M}_{P}^2  {\cal R}_{\beta}^{T} &=& \begin{pmatrix}
		0 & 0 & 0 & 0\\
		0 & {({\cal B}_P^2)}_{22} & {({\cal B}_P^2)}_{23} & {({\cal B}_P^2)}_{24} \\
		0 & {({\cal B}_P^2)}_{23} & {({\cal B}_P^2)}_{33} & {({\cal B}_P^2)}_{34} \\
		0 & {({\cal B}_P^2)}_{24} & {({\cal B}_P^2)}_{34} & {({\cal B}_P^2)}_{44} \\
		\end{pmatrix} \,,
		\end{eqnarray}
where  $\mathcal{M}_{P}$ denotes the mass matrix in the basis of $(G^0,z_{2}',z_{3}',a')^{T}$.
The elements of ${{\cal B}_P^2}$ are given by,

\begin{eqnarray}
 		{({\cal B}_P^2)}_{22} &=& 
M^{2}_{12} c_{\beta_2}^2+\frac{s_{\beta_2}^2}{2}  \Big\{c_{2 \beta_1} (-M^{\prime 2}_{13}+M^{\prime 2}_{23}-2 \lambda_{10} v^2 \cot{2 \beta_1})+M^{\prime 2}_{13}+M^{\prime 2}_{23}\Big\}
  ,\\
 {({\cal B}_P^2)}_{23}  &=&
 \frac{1}{2} s_{\beta_2} \Big\{s_{2 \beta_1} (M^{\prime 2}_{13}-M^{\prime 2}_{23})+2 \lambda_{10} v^2 c_{2 \beta_1}\Big\}
  \,,\\
  {({\cal B}_P^2)}_{24}  &=&
  -\frac{3 v s_{\beta_1} c_{\beta_1} s_{\beta_2}^2 c_{\beta_2} (M^{\prime 2}_{13}+M^{\prime 2}_{23})}{f_a}
\,,\\
 		{({\cal B}_P^2)}_{33} &=&  
\frac{1}{2} \Big\{c_{2 \beta_1} (M^{\prime 2}_{13}-M^{\prime 2}_{23})+M^{\prime 2}_{13}+M^{\prime 2}_{23}-2 \lambda_{10} v^2 s_{2 \beta_1}\Big\}
 		\;,\\
 {({\cal B}_P^2)}_{34}  &=&
 -\frac{3 v s_{\beta_2} {c_{\beta_2}} (c_{2 \beta_1} (M^{\prime 2}_{13}+M^{\prime 2}_{23})+M^{\prime 2}_{13}-M^{\prime 2}_{23})}{2 f_a}\;, \\
 {({\cal B}_P^2)}_{44}  &=&
 \frac{9 v^2 s_{\beta_2}^2 c_{\beta_2}^2 (M^{\prime 2}_{13} c_{\beta_1}^2+M^{\prime 2}_{23} s_{\beta_1}^2)}{f_a^2}
 	\;.
 		\end{eqnarray}\label{e:BP2}
	\end{subequations}
The matrix ${{\cal B}_P^2}$ are fully diagonalized by the orthogonal transformation
	\begin{eqnarray}
	{R}_{A}({\cal B}^2_{P}) {R}_{A}^T &=&
  {\rm diag}  ( m^2_{A_1},~ m^2_{A_2},~m^{2}_{a})
	\,,\label{e:PSrot}
	\end{eqnarray}
where we have assumed the mass ordering, $m_a<m_{A_2}<m_{A_1}$. 
For the CP-odd Higgs sector, the masses are chosen as the output parameters~\footnote{If one expresses $R_A$ by Eq.~\eqref{eq:gamma_i}, one has the six physical parameters, $\gamma_i$ ($i=1,2,3$), $m_{A_1}$,$m_{A_2}$ and $m_a$. Since the mass matrix ${{\cal B}_P^2}$ are determined by the five potential parameters $(\lambda_{10},m_{12},m^\prime_{13},m^\prime_{23},f_a)$ for the fixed EW VEVs, one cannot choose all physical parameter as inputs.}
\footnote{\red{While we integrate out heavy Higgs bosons in the B-L sector, 
we mention radiative corrections to the masses for $a$ and $A_i$ from the B-L Higgs bosons $\rho_{m}$ $(m=0,1,\bar{2})$.
Focusing on the portal interaction with $\kappa_{1\bar{2}\phi1\phi3}$,
we can estimate the one-loop corrections to $m_a^2$ and $m_{A_1}^2$ as $\Delta m_a^2\sim-({f_a^2}/{16\pi^2})\kappa_{1\bar{2}\phi1\phi3}({v_1v_3}/{f_a^2})(R_P)_{44}^2\sim (0.1{\rm keV})^2$, $\Delta m_{A_1}^2
   \sim ({f_a^2}/{16\pi^2})\kappa_{1\bar{2}\phi1\phi3}(R_P)_{11}(R_P)_{13} \sim (0.1{\rm  GeV})^2$, where we assume $\kappa_{1\bar{2}\phi1\phi3}=v^2/f_a^2$ and take $f_a\sim 10^{10}{\rm GeV}$, $v_{1,3}\sim v$, $(R_P)_{11,13}\sim 0.1$ and $(R_P)_{44}\sim 1$. Thus, the radiative corrections to the masses from the $B-L$ sector would not be significant. }}. 

Using the tadpole conditions and the mass formulae for the Higgs bosons, the original \green{17} parameters are replaced by the following physical parameters.
\begin{align}
v\;,\ t_{\beta_{1}}\;,\ t_{\beta_{2}}\;,\ \alpha_{i}\;,\ m_{H_{i}}\;,\ \gamma_{+}\;,\  m_{H^{\pm}_{1}}\;,\ m_{H^{\pm}_{2}} \;,\  M_{12}\;,\  M^{\prime}_{13}\;,  M^{\prime}_{23}\;,\  \lambda_{10}
\;,\  \green{f_{a}}
\end{align}
with $i=1, 2, 3$, where the electroweak VEV $v$ and the mass of $H_{1}$ are fixed as $v=246~{\rm GeV}$ and $m_{H_{1}}=125.09~{\rm GeV}$, respectively.
Without loss of generality the domain of the mixing angles are taken to be
 \begin{align}
 \alpha_{1,2,3},\gamma_{+}=\left[-\frac{1}{2}\pi,\frac{1}{2}\pi  \right]\;,\quad
 \beta_{1,2}=\left[0,\frac{\pi}{2}\right]\;.
 \end{align}
The quartic coupling constants $\lambda_i$ ($i=1-9$) in the original Higgs potential can be given in terms of above the 16 parameters \green{except for $f_a$}. 
The analytical formulae are given in Appendix~\ref{ap:lambda}.
\subsection{Yukawa Lagrangian and kinetic terms} 

\begin{table}[t]\centering
\begin{tblr}{ |c||c|c|c|c|
}\hline
 $\xi_{H^{+}_{i}}^{f}$    & $q$  & $e$   & $\ell$ & $\ellp$ \\ \hline
 $H_{1}^{+}$ & $-\frac{1}{t_{\beta_{2}}}s_{\gamma_{+}}$  & $\frac{1}{t_{\beta_{1}}}\frac{c_{\gamma_{+}}}{c_{\beta_{2}}}+t_{\beta_{2}}s_{\gamma_{+}} $     & $-\cotb{2}s_{\gamma_{+}}$ & $-\tb{1}\frac{c_{\gamma_{+}}}{c_{\beta_{2}}}+\tb{2}s_{\gamma_{+}}$\\
 $H_{2}^{+}$    &$\frac{1}{t_{\beta_{2}}}c_{\gamma_{+}}$ &  $-\tb{1}c_{\gamma_{+}} +\cotb{2}\frac{s_{\gamma_{+}}}{c_{\beta_{2}}}$  & $\cotb{2}c_{\gamma_{+}}$ &$-\tb{2}c_{\gamma_{+}}-\tb{1}\frac{s_{\gamma_{+}}}{c_{\beta_{2}}}$  \\ \hline
\end{tblr}
\caption{Coefficients of  the Yukawa coupling for the charged Higgs bosons. \label{tab:ch}}
\vspace{1cm}

\begin{tblr}{|c||c|c|c|c|
  } \hline
   $\xi_{H_{i}}^{f}$    & $q$  & $e$   & $\ell$ & $\ellp$ \\ \hline
   $H_{1}$ & $\frac{s_{\alpha_{2}}}{s_{\beta_{2}}}$  & $\frac{s_{\alpha_{1}} c_{\alpha_{2}}}{s_{\beta_{1}} c_{\beta_{2}}}$     & $\frac{s_{\alpha_{2}}}{s_{\beta_{2}}}$  & $\frac{c_{\alpha_{1}} c_{\alpha_{2}}}{c_{\beta_{1}} c_{\beta_{2}}}$\\
   $H_{2}$    &$\frac{c_{\alpha_{2}} s_{\alpha_{3}}}{s_{\beta_{2}}}$ & $\frac{1}{s_{\beta_{1}} c_{\beta_{2}}}(c_{\alpha_{1}}c_{\alpha_{3}}-s_{\alpha_{1}}s_{\alpha_{2}}s_{\alpha_{3}}) $  & $\frac{c_{\alpha_{2}} s_{\alpha_{3}}}{s_{\beta_{2}}}$ &$\frac{1}{c_{\beta_{1}} c_{\beta_{2}}}(-s_{\alpha_{1}}c_{\alpha_{3}}-c_{\alpha_{1}}s_{\alpha_{2}}s_{\alpha_{3}})$  \\
    $H_{3}$    &$\frac{c_{\alpha_{2}} c_{\alpha_{3}}}{s_{\beta_{2}}}$&  $\frac{1}{s_{\beta_{1}} c_{\beta_{2}}}(-s_{\alpha_{1}}s_{\alpha_{2}}c_{\alpha_{3}}-c_{\alpha_{1}}s_{\alpha_{3}}) $  & $\frac{c_{\alpha_{2}} c_{\alpha_{3}}}{s_{\beta_{2}}}$ & $\frac{1}{c_{\beta_{1}} c_{\beta_{2}}}(-c_{\alpha_{1}}s_{\alpha_{2}}c_{\alpha_{3}}+s_{\alpha_{1}}s_{\alpha_{3}}) $  \\ \hline
  \end{tblr}
  \caption{Coefficients of  the Yukawa coupling for the CP-even Higgs bosons. \label{tab:even}}
  \vspace{1cm}
\begin{tblr}{|c||c|c|c|c|
}\hline
 $\xi_{A_{i},a}^{f}$    & $q$  & $e$   & $\ell$ & $\ellp$ \\ \hline
 $A_{1}$ & $\frac{1}{s_{\beta_2}}(R_{P})_{23}$  & $\frac{1}{s_{\beta_1}c_{\beta_2}}(R_{P})_{22}$  & $\frac{1}{s_{\beta_2}}(R_{P})_{23}$  & $\frac{1}{c_{\beta_1}c_{\beta_2}}(R_{P})_{21}$\\
  $A_{2}$ & $\frac{1}{s_{\beta_2}}(R_{P})_{33}$  & $\frac{1}{s_{\beta_1}c_{\beta_2}}(R_{P})_{32}$  & $\frac{1}{s_{\beta_2}}(R_{P})_{33}$  & $\frac{1}{c_{\beta_1}c_{\beta_2}}(R_{P})_{31}$\\
 $a$ & $\frac{1}{s_{\beta_2}}(R_{P})_{43}$  & $\frac{1}{s_{\beta_1}c_{\beta_2}}(R_{P})_{42}$  & $\frac{1}{s_{\beta_2}}(R_{P})_{43}$  & $\frac{1}{c_{\beta_1}c_{\beta_2}}(R_{P})_{41}$\\ \hline
\end{tblr}
\caption{Coefficients of  the Yukawa coupling for the CP-odd Higgs bosons. \label{tab:oddgene}}
\vspace{1cm}
\end{table}
Under the charge assignment in Table.\ref{tab:Q}, the Yukawa Lagrangian for Type-A and Type-B is  commonly written by
\begin{align}
\mathcal{L}^{}_{Y}&=-(Y_{u})_{ij}\bar{Q}_{i}\tilde{\phi_{3}}(u_{R})_{j}-(Y_{d})_{ij}\bar{Q}_{i}{\phi_{3}}(d_{R})_{j} \notag \\
&-y_{e}\bar{L}_{e}{\phi_{2}}e_{R}-y_{\ell}\bar{L}_{\ell}{\phi_{3}}\ell_{R}-y_{\ellp}\bar{L}_{\ellp}{\phi_{1}}\ellp_{R}+{\rm h.c.}
\end{align}
where $Y_{u}$ and $Y_{d}$ correspond to $3\times 3$ Yukawa coupling matrices for quarks, $y_{e},\ y_{\ell}$, and $\ y_{\ellp}\ (\ell,\ellp=\mu\ {\rm or}\ \tau)$ are the Yukawa coupling constants for the leptons.
Due to the charge assignment of the $U(1)_{F}$ charge, there are no off-diagonal components of the lepton Yukawa matrix. Depending on the charge assignment, 
$\ell$ and $\ellp$ are given by
\begin{align}\label{eq:Yukawatype}
(\ell,\ellp)=
 \begin{dcases}
&(\mu,\tau)\quad \mbox{for Type-A}\\
&(\tau,\mu)\quad \mbox{for Type-B}\;.
 \end{dcases}
\end{align}
In either case,  the Yukawa Lagrangian above can be expressed in terms of the mass eigenstates of the Higgs bosons as
\begin{align}
\mathcal{L}^{M}_{Y}&\ni
\frac{\sqrt{2} }{v}  V_{\rm CKM}\sum_{i=1}^{2} \xi_{H_{i}^{\pm}}^{q}H_{i}^{\pm}\Big\{\bar{u}(m_{u}P_{L}-m_{d}P_{R})d +{\rm h.c.}\Big\} \notag \\
&-\frac{m_{q}}{v}\sum_{i=1}^{3}\xi_{H_{i}}^{q}H_{i}\bar{q}{q}+i{2I_q}\frac{m_{q}}{v}\sum_{i=1}^{2}\xi_{A_{i}}^{q}A_{i}\bar{q}{\gamma_5}{q}  \notag\\
&-\sqrt{2}\frac{m_{l}}{v} \sum_{i=1}^{2} \xi_{H_{i}^{\pm}}^{l}H_{i}^{\pm}\Big\{\bar{\nu}_{L}P_{R}l +{\rm h.c.}\Big\}
-\frac{m_{l}}{v}\sum_{i=1}^{3}\xi_{H_{i}}^{l}H_{i}\bar{l}{l}  \notag\\
&{-i\sum_l}\frac{m_{l}}{v}\sum_{i=1}^{2}\xi_{A_{i}}^{l}A_{i}\bar{l}{\gamma_5}{l} 
{-i{\sum_l}g_{a\ell}a\bar{l}\gamma_5{l}}\;, 
\end{align}
where $V_{\rm CKM}$ denotes the Cabibbo-Kobayashi-Maskawa (CKM) matrix, { and $I_q$ is isospin for the quarks}. 
The last term corresponds to the axion couplings with leptons. The coupling constant $g_{a l}$ can be written by 
\begin{align}\label{eq:gaeex1}
g_{a l}&=\xi_a^l\frac{m_l}{v}  \\ 
&\simeq q_l\frac{m_l}{f_a}\quad (l=e,\mu,\tau) \;,
\label{gae}
\end{align}
where $q_l$ denotes the \red{effective} $U(1)_{F}$ charge for the lepton $l$, {and $m_l$ is the lepton mass. }
\red{For instance, $q_e = 1 - (-2) = 3.$}
 We use  Eq.~\eqref{eq:gaeex1} in the numerical calculations in Sec.~\ref{sec:XENON1T} and~\ref{sec:Xray}. 
The analytical expressions for the axion coupling with charged leptons are given by Appendix~\ref{ap:axioncoup}.
We note that the difference between Eq.~\eqref{eq:gaeex1} and Eq.~\eqref{gae} comes from the breaking of $U(1)_F$ symmetry (i.e., $m_{12}^2$) 
\red{as well as the mixing with the CP-odd Higgs.}
The coefficients, $\xi^{q/l}_{H_{i}^{\pm}}$, $\xi^{q/l}_{H_{i}}$ and {$\xi^{q/l}_{A_{i},a}$} are presented in Tables~\ref{tab:ch}, \ref{tab:even} and \ref{tab:oddgene}, respectively.
Hereafter, we focus on the Type-B Yukawa lagrangian.~\footnote{\green{In the numerical calculations presented in sec.~\ref{sec:XENON1T} and \ref{sec:Xray}, the difference between Type-A and Type-B only appears in the evaluation of the bounds for perturbativity for running coupling constants.}}

From the kinetic terms of the Higgs doublet fields, one can derive the gauge-gauge-Higgs couplings as
\begin{align}
  \mathcal{L}^M_{\rm Kin.}\ni 2\frac{m_W^2}{v}\sum_{i=1}^3\kappa_V^{H_i}H_iW^{+\mu}W^-_{\mu}+\frac{m_Z^2}{v}\sum_{i=1}^3\kappa_V^{H_i}H_iZ^{\mu}Z_{\mu},
  \end{align}
where the scaling factors $\kappa_V^{H_i}$ are given by
\begin{align}
  \kappa^{H_{1}}_{V}&=c_{\alpha_{2}}c_{\alpha_{1}-\beta_{1}}c_{\beta_{2}}+s_{\alpha_{2}}s_{\beta_{2}}\;, \\
  \kappa^{H_{2}}_{V}&=-(s_{\alpha_{1}-\beta_{1}}c_{\alpha_{3}} +c_{\alpha_{1}-\beta_{1}}s_{\alpha_{2}}s_{\alpha_{3}} )c_{\beta_{2}}+c_{\alpha_{2}}s_{\alpha_{3}}s_{\beta_{2}}
  \;, \\
  \kappa^{H_{3}}_{V}&=s_{\alpha_{1}-\beta_{1}}s_{\alpha_{3}}c_{\beta_{2}}+(-c_{\alpha_{1}-\beta_{1}}s_{\alpha_{2}}c_{\beta_{2}}+c_{\alpha_{2}}s_{\beta_{2}}  )c_{\alpha_{3}}
  \;.
  \end{align}

\subsection{Alignment limit} 
Current measurements of the coupling constants of the 125~GeV Higgs boson at the LHC Run~2~\cite{ATLAS:2021vrm,CMS:2020gsy} show that the properties of the discovered Higgs boson are similar to those predicted in the SM. 
Theoretically, this situation can be realized in the so-called alignment limit, 
where the CP-even Higgs boson with the mass of 125 GeV has the same tree-level couplings as the SM.
The alignment limit in two Higgs doublet models (2HDMs) was discussed in e.g., Refs.~\cite{Gunion:2002zf,Carena:2013ooa}, and, in the context of 3HDM, the analytical condition for the limit was systematically derived ~in Ref.~\cite{Das:2019yad}. 
Symmetries for the Higgs potential that naturally lead to the alignment limit were discussed in \green{Refs.~\cite{Darvishi:2021txa,Darvishi:2020teg,Darvishi:2022zag}}.

Since we identify $H_1$ with the SM-like Higgs boson, the alignment limit requires $\kappa_{V}^{H_1}=1$. 
This can be reduced as
\begin{align}\label{eq:kv}
\sin^2\left(\frac{\alpha_1-\beta_1}{2}\right)\cos^2\left(\frac{\alpha_2+\beta_2}{2}\right)
+\cos^2\left(\frac{\alpha_1-\beta_1}{2}\right)\sin^2\left(\frac{\alpha_2-\beta_2}{2}\right)
=0,
\end{align} 
and this equation yields the condition for the alignment limit:
\begin{align}\label{eq:alignment}
  \alpha_1=\beta_1,\ \alpha_2=\beta_2.
\end{align}
As can be easily seen in Table.~\ref{tab:even}, $\xi_{H_1}^f$ becomes unity in this limit. 
One can also see that the mixing matrix for the CP-even Higgs bosons is expressed as
\begin{align}
  R_S={\cal O}_{\alpha_3}{\cal O}_{\beta}
\end{align}
when Eq.~\eqref{eq:alignment} is satisfied. 
This means that, similar to the charged Higgs bosons, the SM-like Higgs boson $H_1$ is diagonalized by the rotation ${\cal O}_{\beta}$ and the remaining two CP-even Higgs states are transformed into the mass eigenstates $(H_2,H_3)$ by the rotation ${\cal O}_{\alpha_3}$.  In this connection, the Yukawa couplings for $(H_2,H_3)$ have the same structure as the  $(H_1^\pm,H_2^\pm)$ in the alignment limit, i.e.,
\begin{align}
  \xi^f_{H_2}=\left.\xi^f_{H^\pm_1}\right|_{\gamma_+\to -\alpha_3}\;,\quad \xi^f_{H_3}=\left.\xi^f_{H^\pm_2}\right|_{\gamma_+\to -\alpha_3}\;,
\end{align}
where the minus sign comes from the different convention in the rotation matrices (see Eqs. \eqref{eq:gammaRots} and ~\eqref{e:R}).

\red{\subsection{Axion-photon coupling}}
\red{
We here give the analytical expression for the axion photon coupling, $g_{a \gamma}$.
It is defined by the following effective Lagrangian,
\begin{align}
{\cal L}_{\rm eff}=-\frac{g_{a\gamma}}{4}aF_{\mu\nu}\tilde{F}^{\mu\nu},
\end{align}
where $F_{\mu\nu}=\partial_\mu A_\nu -\partial_\nu A_\mu$ and $\tilde{F}^{\mu\nu}=\epsilon^{\mu\nu \rho \sigma} F_{\rho \sigma}/2$.
The axion photon-coupling $g_{a \gamma}$ can be derived from the amplitude for the axion decay into two photons. 
 Using the Yukawa interactions of the axion, i.e., Eq.~\eqref{eq:gaeex1}, it can be calculated as 
\begin{align}
   \mathcal{M}&=-g_{a\gamma}\epsilon^{\alpha\beta \mu\nu}p_{1,\alpha}p_{2,\beta}\epsilon^\ast_\mu(p_1)\epsilon^\ast_\nu(p_2)\;, \notag \\
   g_{a\gamma}&=-\frac{2\alpha_{em}}{\pi v}\sum_{\ell} m_\ell^2\xi_{\ell}^aC_0(0,0,m_a^2,m_\ell,m_\ell,m_\ell)
\end{align}
where $p_{1,2}$ denote the external momenta of the photons, and  $\epsilon_\mu(p_{1})$, and $\epsilon_\nu(p_{2})$ are the polarization vectors.
The coefficient of the tensor products in ${\cal M}$ corresponds to the axion-photon coupling $g_{a\gamma}$, 
which is given in terms of the Passarino-Veltman function $C_0$~\cite{Passarino:1978jh}. 
The approximate formula in the case of $m_a \ll m_\ell$ is given by
\begin{align}\label{eq:c0ap}
   C_0(0,0,m_a^2,m_\ell,m_\ell,m_\ell)\simeq   -\frac{1}{2m_\ell^2}
   \left(1+\frac{1}{12}\frac{m_a^2}{m_\ell^2}\right).
\end{align}
Using this approximate formula, we arrive at the axion-photon coupling,
\begin{align}\label{eq:gagamfull}
   g_{a\gamma}=\frac{\alpha_{em}}{\pi v}\sum_{\ell}
   \left[\xi_a^\ell+\frac{m_a^2}{12}\frac{\xi_a^\ell}{m_\ell^2} \right].
\end{align}
The first term  corresponds to the anomalous coupling and the second term $ \xi_a^\ell m_a^2/(12m_\ell^2)$ corresponds to the threshold corrections by the lepton loop diagrams~\cite{Nakayama:2014cza} (see also Ref.~\cite{Pospelov:2008jk}). 
The parameter $\xi_a^\ell$ contains the effect of the mixing between axion and CP-odd Higgs bosons and the breaking of $U(1)_F$ symmetry. 
We note that in the absence of these two effects, $\xi_a^\ell$ can be replaced with
the charge $q_\ell$ thorough the Eq.~\eqref{gae}.
If this is the case, the anomaly terms are completely canceled out and one obtains a further reduced expression of $g_{a\gamma}$,
\begin{align}\label{eq:gagamap}
   g_{a\gamma}=\frac{\alpha_{em}}{12\pi f_a}\sum_{\ell}
   q_\ell\frac{m_a^2}{m_\ell^2},
\end{align}
which was used in the analysis of Refs.~\cite{Nakayama:2014cza,Takahashi:2020bpq}.
However, we would like to emphasize that in the 3HDM model for the anomaly-free axion, 
the axion-photon coupilng generically receives contributions from the mixing with the CP-oddd Higgs and the $U(1)_F$ breaking. We will study its implications for the X-ray constraints later in this paper.

For later convenience, we formulate a relation between $g_{a\gamma}$ and $g_{ae}$. 
From Eq.~\eqref{eq:gagamfull}, it can be written by
\begin{align}\label{eq:gagamwD}
   g_{a\gamma}=\frac{\alpha_{em}m_a^2}{12\pi m_e^3}g_{ae}+\frac{\alpha_{em}}{\pi f_a}|q_e|\Delta,
\end{align}
where $\Delta$ is defined by
\begin{align}
   \Delta&=\frac{1}{|q_e|}\frac{f_a}{v}
   \left[\sum_{\ell=e,\mu,\tau}\xi_a^\ell+\frac{m_a^2}{12}\sum_{\ell=\mu,\tau}\frac{\xi^\ell_a}{m_\ell^2}\right]\;, \\ 
     &=\frac{1}{|q_e|}{f_a}
   \left[\sum_{\ell=e,\mu,\tau}\frac{g_{a\ell}}{m_\ell}+\frac{m_a^2}{12}\sum_{\ell=\mu,\tau}\frac{g_{a \ell}}{m_\ell^3}\right] \;.
\end{align}
Analytical expressions of the axion-lepton couplings $g_{a \ell}$ are presented in the limit of $m_{12}=0$ in Appendix~\ref{ap:axioncoup}. 
The parameter $\Delta$ quantifies the deviation from the limiting case (\ref{eq:gagamap}) normalized by the typical anomalous coupling.
In fact, in the absence of the mixing with the CP-odd Higgs and the $U(1)_F$ symmetry breaking, i.e.,$t_{\beta_1}=1$ and $m_{12}=0$, the anomalous coupling (the first term in $\Delta$) vanishes, and we are left with
tiny threshold corrections from heavier charged leptons, $\Delta=m_a^2\left(q_\mu/m_\mu^2+q_\tau/m_\tau^2\right)/(12|q_e|)\simeq 2.5\times 10^{-12}$ at $m_a=$1\,keV. 
 } \red{In practice, the second term in $\Delta$ is always smaller than the threshold correction due to the electron shown by the first term in Eq.~(\ref{eq:gagamwD}), and so, the axion-photon coupling is mainly determined by
 the two contributions, the threshold correction due to the electron and the (residual) anomalous coupling.
 One can see by using the results in Appendix~\ref{ap:axioncoup} that the two 
 contributions tend to have the opposite sign, and the axion-photon coupling can be extremely small when they are nearly canceled with each other.
 }

\section{Masses for axion and CP-odd Higgs bosons }\label{sec:mass}

In this section, we focus on the axion and CP-odd Higgs bosons and evaluate their masses and dependence on the model parameters.
As discussed in the previous section, the mass eigenvalues can be calculated by diagonalizing the mass matrix for the CP-odd Higgs bosons, Eq.~\eqref{e:PSrot}. However, the axion mass can be roughly estimated from the (4,4) element of the mass matrix ${\cal B}_P^2$, i.e., $m_a^2\sim (m^{2}_{12}v^2)/f^2_a$, if $m^2_{12}\sim m^{\prime 2}_{13} \sim m^{\prime 2}_{23}$, or equivalently, if $\kappa_{1\bar{2}\phi 1 \phi 2}\sim \kappa_{\bar{2} 1\phi 1 \phi 3}\sim m_{12}^2/f_a^2$. 
 This assumption is natural since it implies that the interaction $V_I$ between the EW and $B-L$ sectors is suppressed and does not require any extra fine tuning to obtain the EW scale.
Then, the scale of the decay constant $f_a$ and the soft breaking parameter $m_{12}$ determine the mass scale of the axion. For example, if $m_{12}$ is of order the EW scale and $f_a$ is about $10^{10}$ GeV, the axion mass is of order keV.
For the anomaly-free axion, the keV-scale mass and intermediate-scale decay constant are particularly attractive from a phenomenological point of view. One reason is that, \red{especially} \red{in the absence of the} anomalous coupling to photons \red{as Eq.~\eqref{eq:gagamap}},  the axion is stable on cosmological time scales, making it a good candidate for DM.
\red{Through this limiting expression of $g_{a\gamma}$},  the axion mainly decays into two photons with the lifetime, 
\begin{align}
   \tau_{a\to \gamma\gamma}\simeq 2\times10^{32} \,{\rm sec.} \left(\frac{m_a}{2\,{\rm keV}} \right)^{-7} 
   \left(\frac{f_a/q_e}{10^{10}{\rm GeV}}\right)^2,
\end{align}
which is so long that it can easily satisfy the current limit from the X-ray observations. 
\red{While this is the special case where the effects of breaking of $U(1)_{F}$ and mixing among the axion and the CP-odd Higgs bosons vanish, we will discuss in Sec.~\ref{sec:Xray}} how severe the X-ray bound becomes when the general expression of $g_{a\gamma}$, Eq.~\eqref{eq:gagamfull}, is applied. 
Importantly, the misalignment mechanism can naturally produce the right amount of axion to explain DM \red{in the case where the current limit of X-ray observations can be evaded}.
Also, the axion can explain the excess of electron recoil events observed in the XENON1T experiment through the axion-electron coupling $g_{ae} \sim m_e/f_a \sim 10^{-14}$. 
On the other hand, if the axion accounts for about $10\%$ of DM, it can explain not only the XENON1T excess, but also various stellar cooling anomalies simultaneously~\cite{Takahashi:2020bpq}. Such interesting axion scenarios can be explored by future X-ray observations such as Theseus~\cite{THESEUS:2017qvx,THESEUS:2017wvz}, Athena~\cite{Barret:2018qft}, eROSITA~\cite{eROSITA:2012lfj}, and XRISM~\cite{XRISMScienceTeam:2020rvx} \red{and the direct search experiments such as LZ~\cite{LZ:2021xov} and DARWIN~\cite{DARWIN:2016hyl}}. 

On the other hand, the mass of the CP-odd Higgs bosons must be somewhat heavy in order to satisfy the limits of the direct search at collider experiments. Whether such a mass spectrum can be achieved depends on the model parameters such as the soft breaking mass $m_{12}$ in the Higgs potential. The purpose of this section is to see how the masses of the CP-odd Higgs bosons and axion depend on the model parameters and to understand their behavior intuitively. To this end, here we do not impose any theoretical and experimental constraints that we will discuss in the next section.

\begin{figure}[t]
  \centering
  \includegraphics[scale=0.6]{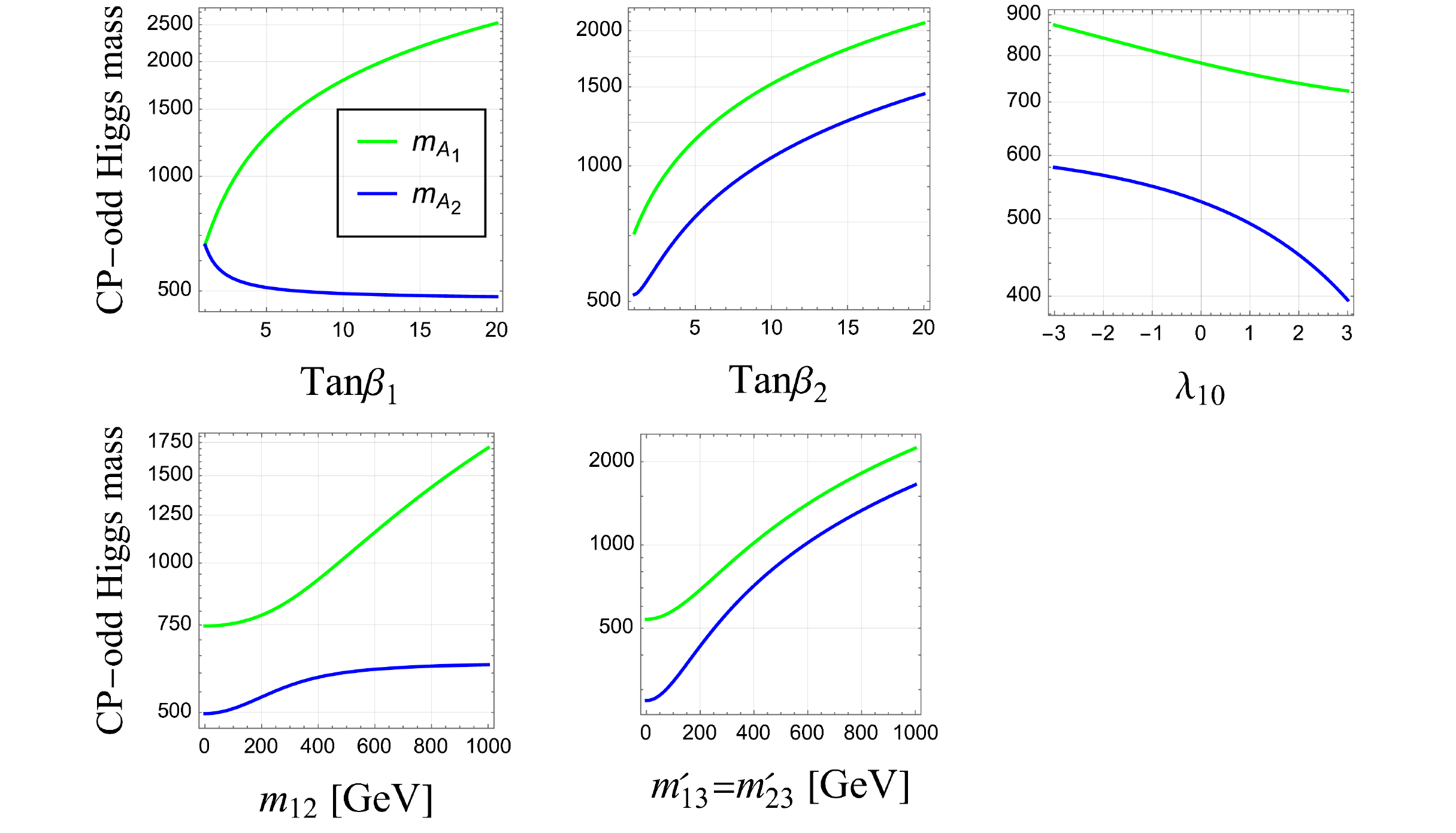}
  \caption{Masses of the CP-odd Higgs bosons $m_{A_{1}}$, $m_{A_{2}}$ as a function of the model parameters. Except for the parameter shown in the horizontal axis, we use the input parameters given in Eq.~\eqref{eq:bpmass}. 
  }
  \label{FIG:MCPodd}
 \end{figure}

 In Fig.~\ref{FIG:MCPodd}, the masses of the CP-odd Higgs bosons are shown as a function of $t_{\beta_{1,2}}$, $\lambda_{10}$  and the soft breaking parameters, $m_{12}, m^\prime_{13}, m^\prime_{23}$. 
The set of the parameters,
\begin{align}\label{eq:bpmass}
    &t_{\beta_{1}}=t_{\beta_{2}}=2\;,\quad
    \lambda_{10}=-2\;,\quad
    f_a=10^{10}~{\rm GeV}\;,\quad
    \notag\\
    &M_{12}^2(c_{\beta_1}s_{\beta_1}c_{\beta_2}^2)=M_{13}^{\prime 2}(c_{\beta_1}c_{\beta_2}s_{\beta_2})=M_{23}^{\prime 2}(s_{\beta_1}c_{\beta_2}s_{\beta_2})=(300\ {\rm GeV})^2 
\end{align}
is chosen 
as a bench mark point, and each model parameter is individually varied in the plot, where we identify $m_{12}$, $m^\prime_{13}$ and $m^\prime_{23}$ as inputs not the re-scaled ones.
As seen in the left top panel,
while $m_{A_1}$ increases as $t_{\beta_1}$, $m_{A_2}$ becomes constant in the large $t_{\beta_2}$ region. 
This behavior can be understood in terms of the diagonal elements of the mass matrix ${\cal B}_P$. 
The (3,3) element dominantly contributes to $m_{A_2}$, and it can be approximately expressed as the constant term $({\cal B}^2_P)_{33}\simeq M^{\prime 2}_{23}$ when $t_{\beta_1}\gg1 $. 
On the other hand, large $t_{\beta_2}$ makes  both $A_1$ and $A_2$ heavy, since both $M^{\prime 2}_{13}$  and  $M^{\prime 2}_{23}$ are enhanced in the (2,2) and (3,3) elements of  ${\cal B}_P$.   
As to the dependence on $\lambda_{10}$, the terms with $\lambda_{10}$ are destructive for the other terms in $({\cal B}^2_P)_{33} $ and $({\cal B}^2_P)_{22}$.

The dependence on the soft breaking parameters is nontrivial, especially for the mass of $A_2$. 
While $m_{A_1}$ increases with $m_{12}^2$,
$m_{A_2}$ becomes almost constant for $m_{12}\gtrsim 400 $ GeV. 
To understand this behavior, let us consider the mixing matrix $R_A$ parameterized by the mixing angles $\gamma_i$ ($i=1,2,3$), see \eqref{eq:gamma_i}. First, note that, due to the hierarchy between $f_a$ and the EW scale, the mixing angles between the axion and $A_{1,2}$ are extremely suppressed.
The remaining angle $\gamma_1$, which is responsible for the mixing of the two CP-odd Higgs states in the basis $R_{\beta} (z_1,\;z_2,\;z_3)$, can be expressed by
\begin{align}\label{eq:gamma1}
\tan{2\gamma_1}\simeq \frac{2({\cal B}^2_P)_{23}}{({\cal B}^2_P)_{22}-({\cal B}^2_P)_{33}}\;.
\end{align}
In terms of the mixing angle $\gamma_1$, $m_{A_2}$ can be written as
\begin{align}\label{eq:mA2}
m_{A_2}^2\simeq\frac{1}{2}\big\{1-\cos(2\gamma_1)\big\}({\cal B}^2_P)_{22}+\frac{1}{2}\big\{1+\cos(2\gamma_1)\big\}({\cal B}^2_P)_{33}
-2\sin(2\gamma_1)({\cal B}^2_P)_{23}\;. 
\end{align}
As $m_{12}$ increases, only $({\cal B}^2_P)_{22}$ increases but $({\cal B}^2_P)_{23}$ and $({\cal B}^2_P)_{33}$ remain the same, so the mixing angle $\gamma_1$ becomes much smaller than unity.
In this case, $m_{A_2}$ is approximately given by the linear combination of $({\cal B}^2_P)_{33}$ and $({\cal B}^2_P)_{23}$
(i.e. the last two terms in Eq.~(\ref{eq:mA2})) which is independent of $m_{12}$.
Conversely, from the middle bottom panel in the lower part of Fig.~\ref{FIG:MCPodd}, one can see that the $m_{A_2}$ increases with $m_{13}^\prime$. 
Note however that this is due to the result of setting $m_{13}^\prime \sim m_{23}^\prime$. 
When either  $m^\prime_{13}\gg m^\prime_{23}\sim v$ or $m^\prime_{23}\gg m^\prime_{13}\sim v$, $m_{A_2}$ is bounded above, which is approximately determined by $ m^\prime_{23}$ for the former case and $ m^\prime_{13}$ for the later case.  
 For the former case, one can check that terms including $m^\prime_{13}$ in the right-handed side of Eq.~\eqref{eq:mA2} are indeed canceled out when $m_{13}^{\prime }$ is enough large to be able to neglect other terms in the mass matrix  ${\cal B}_P$. 

\begin{figure}[t]
  \centering
  \includegraphics[scale=0.39]{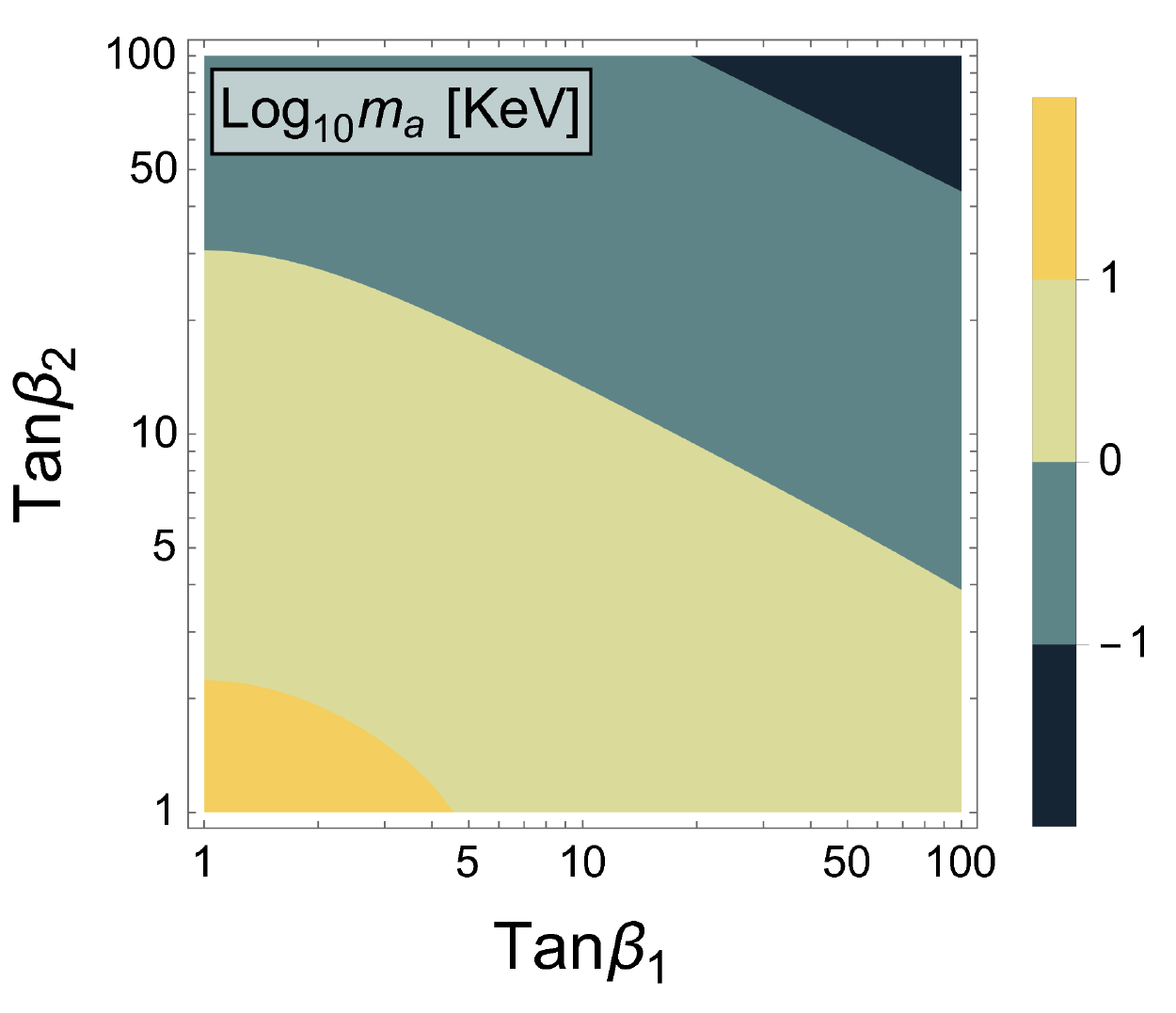}
  \includegraphics[scale=0.39]{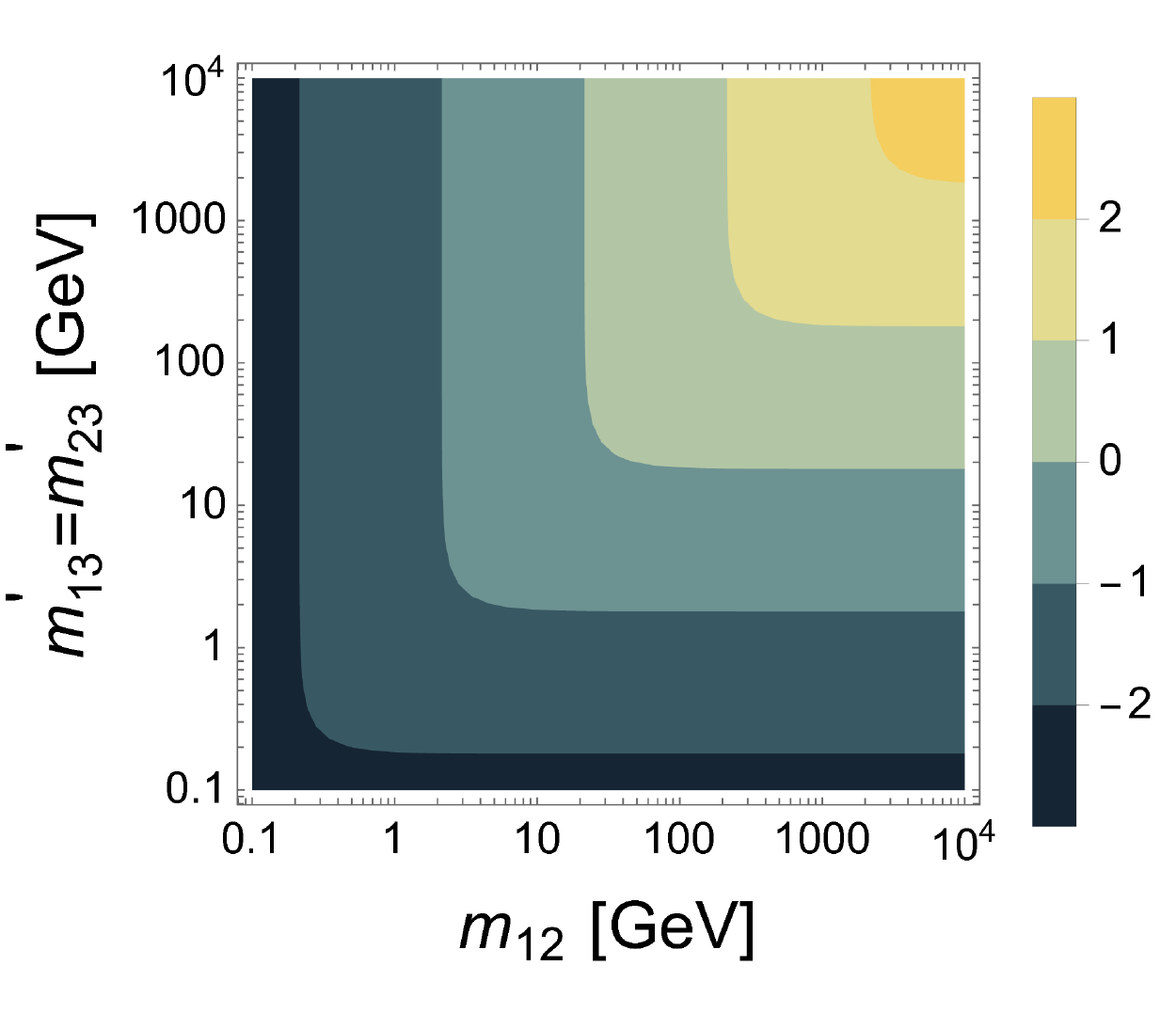}  
  \includegraphics[scale=0.39]{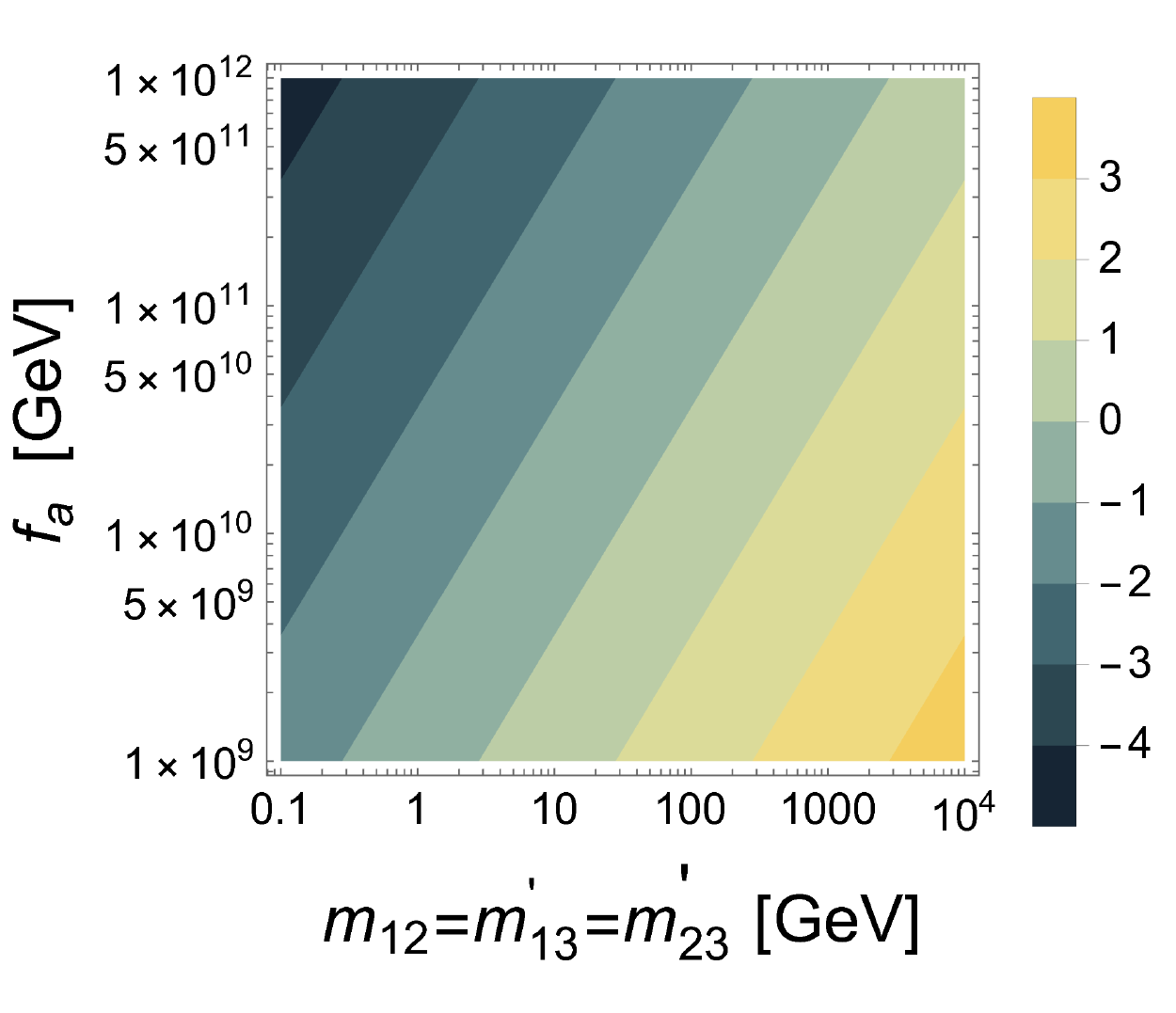}  
  \caption{ Dependence of the axion mass on the model parameters. The left panel shows $\log_{10}(m_a)$ in the parameter space of ($\tb{1}$, $\tb{2}$), where other inputs are fixed as {given in Eq.~\eqref{eq:bpmass}. 
  The middle and right panel shows $\log_{10}(m_a)$ in the plane $(m_{12},~m'_{13})$ and $(m_{12},f_a)$, respectively. 
While $m'_{13}=m'_{23}$ is taken in the middle panel, all soft breaking parameters are degenerate in the right panel. }
 }
  \label{FIG:ALPcont}
 \end{figure}

In Fig.~\ref{FIG:ALPcont}, the axion mass is shown in the plane of ($\tb{1}$,$\tb{2}$), ($m_{12}$,$m^\prime_{13}$) and ($m_{12}$,$f_a$) from left to right. 
We set $m^{\prime 2}_{13}=m^{\prime 2}_{23}$ and $m_{12}^2=m^{\prime 2}_{13}=m^{\prime 2}_{23}$ in the middle  and right panels, respectively. In each panel we vary only the parameters corresponding to the horizontal and vertical axis, while the other parameters are fixed as Eq.~\eqref{eq:bpmass}. 

In the left panel, one see that negative correlation between $m_a$ and $t_{\beta_{1,2}}$. 
{This behavior can be understood from the fact that $m_{12}\phi_1^\dagger \phi_2$ is only the soft breaking term introduced in this model and thus gives the axion mass. 
When we take $t_{\beta_2}\gg 1$, the effect of $U(1)_F$ breaking is suppressed by $c_{\beta_2}^2$. 
On the other hand, it shrinks by $c_{\beta_1}$ in case of $t_{\beta_1}\gg 1$. 
Thereby, compared with the dependence on $\tb{2}$, $m_a$ slightly decreases with $t_{\beta_1}$. }

From the middle panel, one can see the characteristic soft mass dependence of $m_a$. 
Namely, the axion mass increases only when all of the soft breaking masses increase simultaneously. This is because for the axion to have mass, in addition to the explicit breaking of the U(1) flavor symmetry $m_{12}$, either $m_{13}^{\prime}$ or $m_{23}^{\prime}$  must be nonzero to transmit the breaking in the EW sector to the $B-L$ sector where the axion lives.
 This behaviors can be understood  by supposing  the case of
 $t_{\beta_1}=1$ and $m_{13}^{\prime 2}=m_{23}^{\prime2}$. 
 In this specific parameter choice, the CP-odd Higgs field in the Higgs basis $z_3^\prime$ corresponds to the mass eigenstate and mixing happens only in the remaining fields $z^\prime_2$ and $a^\prime$. 
 The approximate formula for the mass of axion can be obtained as 
 \begin{align}\label{eq:malpap}
     m_a^2\simeq
\begin{dcases}
&18c_{\beta_2}^2m^2_{12} \frac{v^2 }{f_a^2} \left[ \left(1-\frac{\sqrt{2}}{t_{\beta_2}}  \frac{m^2_{12}}{m^{\prime 2}_{13}}\right)-9 c_{\beta_2}^2\frac{v^2}{f_a^2}  \left(1-\frac{3 \sqrt{2}}{t_{\beta_2}}  \frac{m^2_{12}}{m^{\prime 2}_{13}}\right) \right]
\quad (m_{12}^2\sim v^2\ll m_{13}^{\prime 2}) \\
&9s_{\beta_2}m^{\prime 2}_{13}\frac{  v^2 }{f_a^2} \left[\sqrt{2} c_{\beta_2}-\frac{m^{\prime 2}_{13}}{m^2_{12}} s_{\beta_2} \right]
\quad\quad\quad\quad\quad\quad\quad\quad\quad\quad\quad\quad\quad\  (m_{13}^{\prime 2}\sim v^2 \ll  m_{12}^2) \\
&\frac{9 m^2_{12} v^2 t_{\beta_2} \left[4 f_a^2 (4 s_{2 \beta_2}+\sqrt{2} c_{2 \beta_2}+3 \sqrt{2})+9 \sqrt{2} v^2 (c_{4{\beta_2}}-1) \right]}{2 f_a^4 (\sqrt{2} t_{\beta_2}+2)^3} \quad (v^2\ll m_{13}^{\prime 2}= m_{12}^2)
 \end{dcases}     
 \end{align}
 where we have assumed that $X^2 \ll f^2_a$ ($X=m_{12},\; m_{13}^{\prime}$ or $v$) and neglected terms of the order ${\cal O}(X^6/f_a^6)$. 
 { The soft breaking parameters $m_{13}^{\prime}$ and $m_{23}^{\prime}$ have little effect on the axion mass in the case of $m^2_{12}\sim v^2\ll m^{\prime 2}_{13}=m^{\prime 2}_{23}$. 
 The case of $m^{\prime 2}_{13}=m^{\prime 2}_{23}\sim v^2\ll m^{ 2}_{12}$
 is similar, and  $m_{12}^2$ has little effect on the axion mass.
  }
 One can see that the axion mass increases when all the soft breaking masses increase and become larger than $v$.
 In addition, the behavior of $m_a$ in the right panel obeys the last line of \eqref{eq:malpap}.

To summarize this section, we have shown the following properties for the mass scales of $A_{1,2}$ and $a$;
\begin{itemize}
      \item Typical scale of the mass of $A_1$ and $A_2$ is mostly determined by the soft breaking masses. 

Thus, the CP-odd Higgs bosons become heavy and get decoupled by taking
{two of the three breaking parameters are sufficiently larger than EW VEV.}
       \item 
      The axion mass increases when all of the soft breaking masses, $m_{12}^2$, $m_{13}^{\prime 2}$, $m_{23}^{\prime 2}$,   increase simultaneously. On the other hand, the axion mass is suppressed when  $m_{12}^2$ or both $m_{13}^{\prime 2}$ and $m_{23}^{\prime 2}$ are  small.

      \item If the soft breaking masses are of order the EW scale, the axion mass is of the order of keV for $f_a\sim{\cal O}(10^{10}\mathchar`-10^{11})$ GeV. 
\end{itemize}

\section{Theoretical and experimental bounds on 3HDM }\label{sec:con}
Here we discuss theoretical and experimental constraints on the model parameters.
For the former, we take into account conditions for the potential bounded from below (BFB) and perturbative unitarity, and {perturbativity} on the running coupling constants. 
For the latter, we consider constraints from the EW oblique parameters, $B$ meson decays, and $B$ meson mixing. We will impose these constraints on 
the model parameters in the numerical calculations in Sec.~\ref{sec:XENON1T} and \ref{sec:Xray}. 

\subsection{The potential bounded from below and perturbative unitarity}
In order to obtain the stable minimum after the EWSB,
the Higgs potential should be bounded from below in any direction of the Higgs fields.
While the $B-L$ Higgs fields are involved in the original potential \eqref{eq:gpotential},
the radial modes of the $B-L$ Higgs fields are integrated out in our analysis.
Hence, we focus on the conditions for the 3HDM potential \eqref{e:potential} to be bounded from below.
In the pioneering work of Ref.~\cite{Klimenko:1984qx}, the BFB conditions were derived for a potential that involves two Higgs doublets and one Higgs singlet.
More recently,  the BFB conditions for the 3HDM with $Z_{3}$ symmetry were derived in Ref.~\cite{Boto:2021qgu}.
As mentioned in Sec.~\ref{sec:model}, 
the potential \eqref{e:potential} can be obtained by setting the $U(1)_F$ symmetry breaking terms except for $m_{12}^2$ to zero in their $Z_{3}$ invariant potential. Therefore, we can simply read off the BFB conditions
from the results of Ref.~\cite{Boto:2021qgu} as
\begin{align}
&\lambda_1>0,\ \lambda_2>0,\ \lambda_3>0,  \\
&\left\{\beta>-\sqrt{\lambda_1\lambda_3},\  \gamma>-\sqrt{\lambda_2\lambda_3},\  \alpha>-\sqrt{\lambda_1\lambda_2},\  \beta\geq -\gamma \sqrt{\lambda_1/\lambda_2}\right\} \notag \\
&\cup\left\{\sqrt{\lambda_2\lambda_3}>\gamma>-\sqrt{\lambda_2\lambda_3},\ \quad-\gamma\sqrt{\lambda_1/\lambda_2}\geq\beta > -\sqrt{\lambda_1\lambda_3},\ \quad \lambda_3\alpha>\beta\gamma - \sqrt{\Delta_\alpha \Delta_\gamma}\right\} ,
\end{align}
with
\begin{align}
\alpha&= \frac{1}{2}(\lambda_4+\lambda_7) ,\nonumber\\[8pt]
\beta&= \frac{1}{2}(\lambda_5+\lambda_8-2|\lambda_{10}|) ,\nonumber\\[8pt]
\gamma&= \frac{1}{2}(\lambda_6+\lambda_9-2|\lambda_{10}|) .
\end{align}
and
\begin{equation}
\Delta_\alpha=\beta^2-\lambda_1\lambda_3,
\quad \Delta_\gamma=\gamma^2-\lambda_2\lambda_3.
\end{equation}

The partial wave unitarity bound for the elastic $2\to2$ scattering processes in the high energy limit restricts scalar couplings in the Higgs potential. 
In Ref.~\cite{Lee:1977eg}, the unitarity bounds were applied to the SM
to derive an upper limit on the mass of the Higgs boson.
A similar argument can be made for the extended Higgs models.
In Ref.~\cite{Bento:2017eti}, the tree-level unitarity bounds were derived in the framework of 3HDM with $Z_{3}$ symmetry.
Following the results of Ref.~\cite{Bento:2017eti}, 
we can obtain eigenvalues $\Lambda_{i}$ ($i=1-21$) for the partial waves of the $S$-matrix amplitudes for $2\to2$  scattering processes with the replacement
\begin{align}
r_{1-3}=\lambda_{1-3},\
r_{4-9}=\frac{1}{2}\lambda_{4-9},\
c_{11}=\frac{1}{2}\lambda_{10}\; ,
\end{align}
where $r_i$ ($i=1-9$) and $c_{11}$ denote the quartic couplings in 3HDM with the $Z_3$ symmetry defined in Ref.~\cite{Bento:2017eti}. 
The criterion that the partial wave amplitudes satisfy the unitarity is given by
\begin{align}
|\Lambda_{i}|<8\pi.
\end{align} 

\subsection{Perturbativity on running coupling constants}

The degree of freedom of the axion appears from the $B-L$ sector in the model presented in Sec.~\ref{sec:model}. 
Since we consider a new physics scenario where the $U(1)_{\rm B-L}$ symmetry is spontaneously broken at the intermediate scale $\Lambda = {\cal O}(10^{10-12})$ GeV, 
our model should be well behaved up to the breaking scale of $U(1)_{\rm B-L}$. 
To ensure that the model does not lose predictivity, we require that the Landau pole should not appear at the scale $\Lambda$ or below by imposing the following criterion
\begin{align}\label{eq:triviality}
    |\lambda_i(\Lambda)|< 4\pi\quad (i=1\mathchar`-10)\;, \quad
    y_{f}^2(\Lambda)<4\pi
    \quad (f=t,\; b,\;  \tau,\; \mu,\; e )
    \;.
        \end{align}
We identify the breaking scale of $U(1)_{\rm B-L}$ with  $\Lambda= f_a$. 
The running coupling constants $\lambda_i(\Lambda)$ and 
$y_{f}(\Lambda)$ are evaluated by solving the renormalization group equations (RGEs). 
We use the $\beta$ functions at 1-loop level. 
Their analytical expressions are given in Appendix~\ref{ap:RGE}. 

\begin{figure}[t]
 \centering
 \includegraphics[scale=0.5]{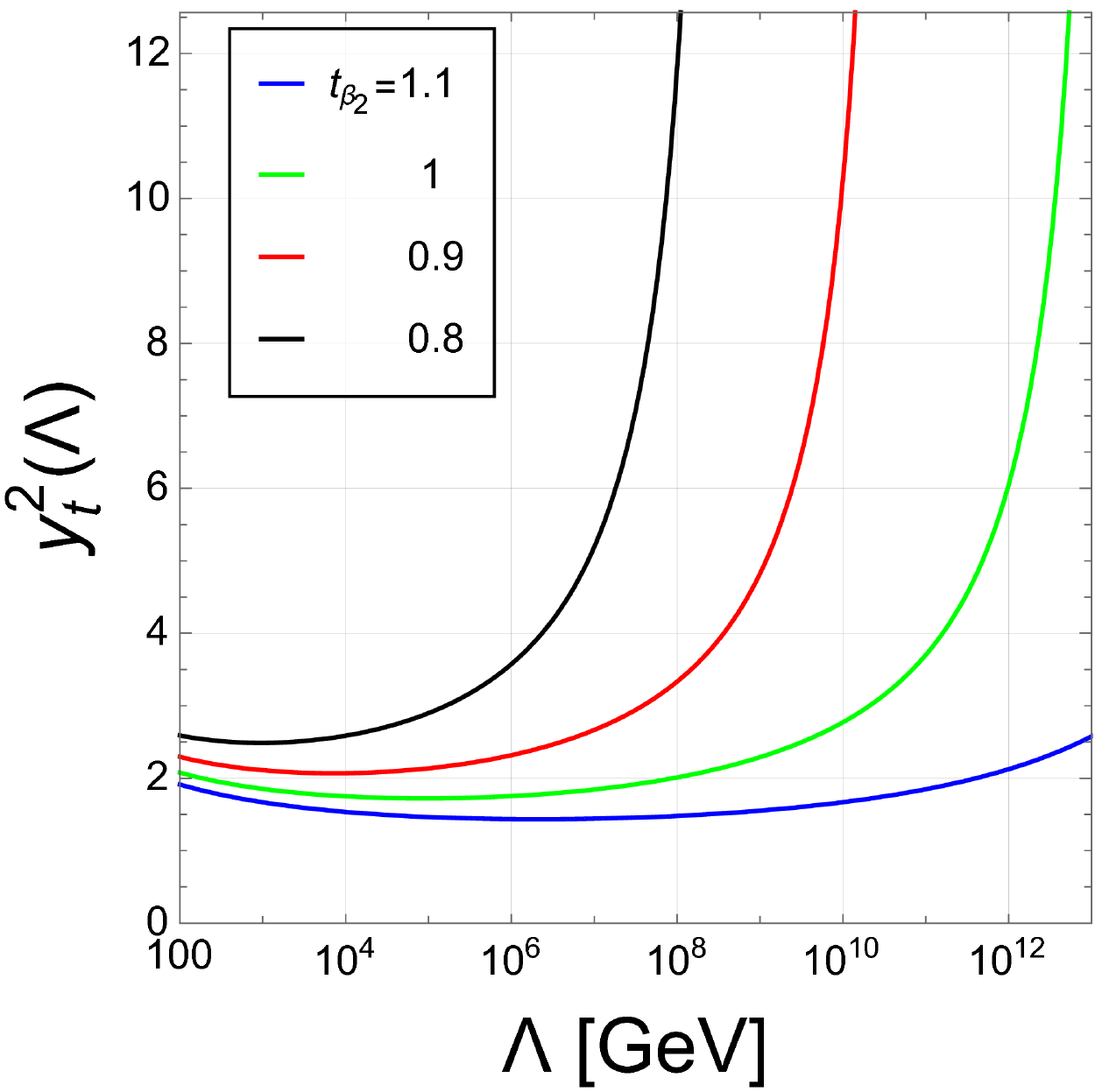}
 \caption{ Running of the top Yukawa coupling as a function of the cutoff scale.  }
 \label{FIG:RGEyt}
\end{figure}

Through the conditions \eqref{eq:triviality} for the quartic scalar couplings, all the input parameters are constrained. 
On the other hand, the conditions for the Yukawa couplings are only relevant for $\beta_1$ and $\beta_2$, since their renormalization group (RG) flows with the 1-loop $\beta$ functions are independent of the quartic scalar couplings. 
In particular, the top Yukawa coupling $y_t=\sqrt{2} m_t/(v s_{\beta_2})$ could blow up immediately when $\tb{2}$ is not large. 
To see this,
we show  in Fig~\ref{FIG:RGEyt} the RG flows of the top coupling $y_{t}(\Lambda)$ with different values of $t_{\beta_{2}}$. 
The Landau pole appears at $\Lambda<10^{10}{\rm GeV}$ when we set $t_{\beta_{2}}=0.9$ or smaller, and there is no Landau pole at $\Lambda < 10^{12}\,$GeV when  $t_{\beta_{2}}= 1$ or larger.
Since we consider the decay constant $f_a$ of order $10^{10}$ - $10^{12}$ GeV in the following numerical analysis, we take the lower bound on $t_{\beta_{2}}$ as 
\begin{align}
t_{\beta_{2}}\gtrsim 1. 
\end{align}

\subsection{Electroweak $S$ and $T$ parameters}

Here we discuss the limits of the electroweak precision measurements for heavy particles.
First, the electroweak $\rho$ parameter does not deviate from unity at the tree level in the multiple Higgs doublet models. 
{However, all the additional Higgs bosons contribute to it at 1-loop level.
The loop corrections to the $\rho$ parameter are described by electroweak oblique parameters.} 

The electroweak oblique parameters, which parameterize new physics effects for the gauge boson self-energies, were first proposed in Ref.~\cite{Peskin:1991sw}.
Their analytical expressions in the multi-Higgs doublet models were calculated in Refs.~\cite{Grimus:2007if,Grimus:2008nb}.
With the definitions of $S$ and $T$ parameters given in Ref.~\cite{Hagiwara:1994pw}, their analytical expressions in the 3HDM with axion can be derived from the new scalar boson loop contributions to the gauge boson two-point functions. 
We give the general formulae for the $S$ and $T$ parameters in Appendix~\ref{ap:ST}, which are used in the numerical calculations in Sec.~\ref{sec:XENON1T} and \ref{sec:Xray}. 
Since the general formulae for the $S$ and $T$ parameters are somewhat lengthy, 
we here describe the analytical expressions in the alignment limit. 
New physics contributions to the $S$ and $T$ parameters,  $\Delta S=S_{\rm 3HDM}-S_{\rm SM}$ and $\Delta T=T_{\rm 3HDM}-T_{\rm SM} $, are written in terms of the Passarino-Veltman functions~\cite{Passarino:1978jh} as
\begin{align}
4\pi\Delta {S}&=
s_{\alpha_3+{\gamma_1}}^2 \left\{\Delta B_5(m_Z^2,m_{H_3},m_{A_1})+\Delta B_5(m_Z^2,m_{H_2},m_{A_2})\right\}
\notag\\
&+c_{\alpha_3+{\gamma_1}}^2 \left\{\Delta B_5(m_Z^2,m_{H_2},m_{A_1})+\Delta B_5(m_Z^2,m_{H_3},m_{A_2})\right\}
\notag\\
&-\Delta B_5(m_Z^2,m_{H^\pm_1},m_{H^\pm_1})-\Delta B_5(m_Z^2,m_{H^\pm_2},m_{H^\pm_2})\;, \\
\Delta {T}&=
-\frac{G_f}{8\sqrt{2}\pi^{2}\alpha_{em} } \Big[c_{{\gamma_1}-{\gamma_+}}^2 B_5(0,m_{A_1},m_{H^\pm_1})-c_{\alpha_3+{\gamma_1}}^2 B_5(0,m_{H_2},m_{A_1})
\notag\\
&+s_{{\gamma_1}-{\gamma_+}}^2 B_5(0,m_{A_1},m_{H^\pm_2})-s_{\alpha_3+{\gamma_1}}^2 B_5(0,m_{H_3},m_{A_1})
\notag\\
&+s_{{\gamma_1}-{\gamma_+}}^2 B_5(0,m_{A_2},m_{H^\pm_1})-s_{\alpha_3+{\gamma_1}}^2 B_5(0,m_{H_2},m_{A_2})
\notag\\
&+c_{{\gamma_1}-{\gamma_+}}^2 B_5(0,m_{A_2},m_{H^\pm_2})-c_{\alpha_3+{\gamma_1}}^2 B_5(0,m_{H_3},m_{A_2})
\notag\\
&+c_{\alpha_3+{\gamma_+}}^2 B_5(0,m_{H_2},m_{H^\pm_1})+s_{\alpha_3+{\gamma_+}}^2 B_5(0,m_{H_3},m_{H^\pm_1})
\notag\\
&+s_{\alpha_3+{\gamma_+}}^2 B_5(0,m_{H_2},m_{H^\pm_2})+c_{\alpha_3+{\gamma_+}}^2 B_5(0,m_{H_3},m_{H^\pm_2})\Big],
\end{align}
where 
\begin{align}
\Delta B_{5}(q^{2},X,Y)=\frac{B_{5}(0;X,Y) - B_{5}(q^{2};X,Y) }{q^{2}}\;.
\end{align}
These are obtained from the general expressions given in Eqs.~\eqref{eq:DeltaS} and~\eqref{eq:DeltaT}, taking $(\alpha_1,\alpha_2)$=$(\beta_1,\beta_2)$, and assuming that the mixing matrix for the CP-odd Higgs bosons $R_A$ is parameterized by \eqref{eq:gamma_i} with $\gamma_2=\gamma_3$=0. 
The $B_5$ function is evaluated by using {\tt LoopTools}~\cite{Hahn:1998yk}. 
We use the experimental values for the $S$ and $T$ parameters given in Ref.~\cite{Haller:2018nnx}. 

\begin{figure}[t]
 \centering
     \includegraphics[scale=0.5]{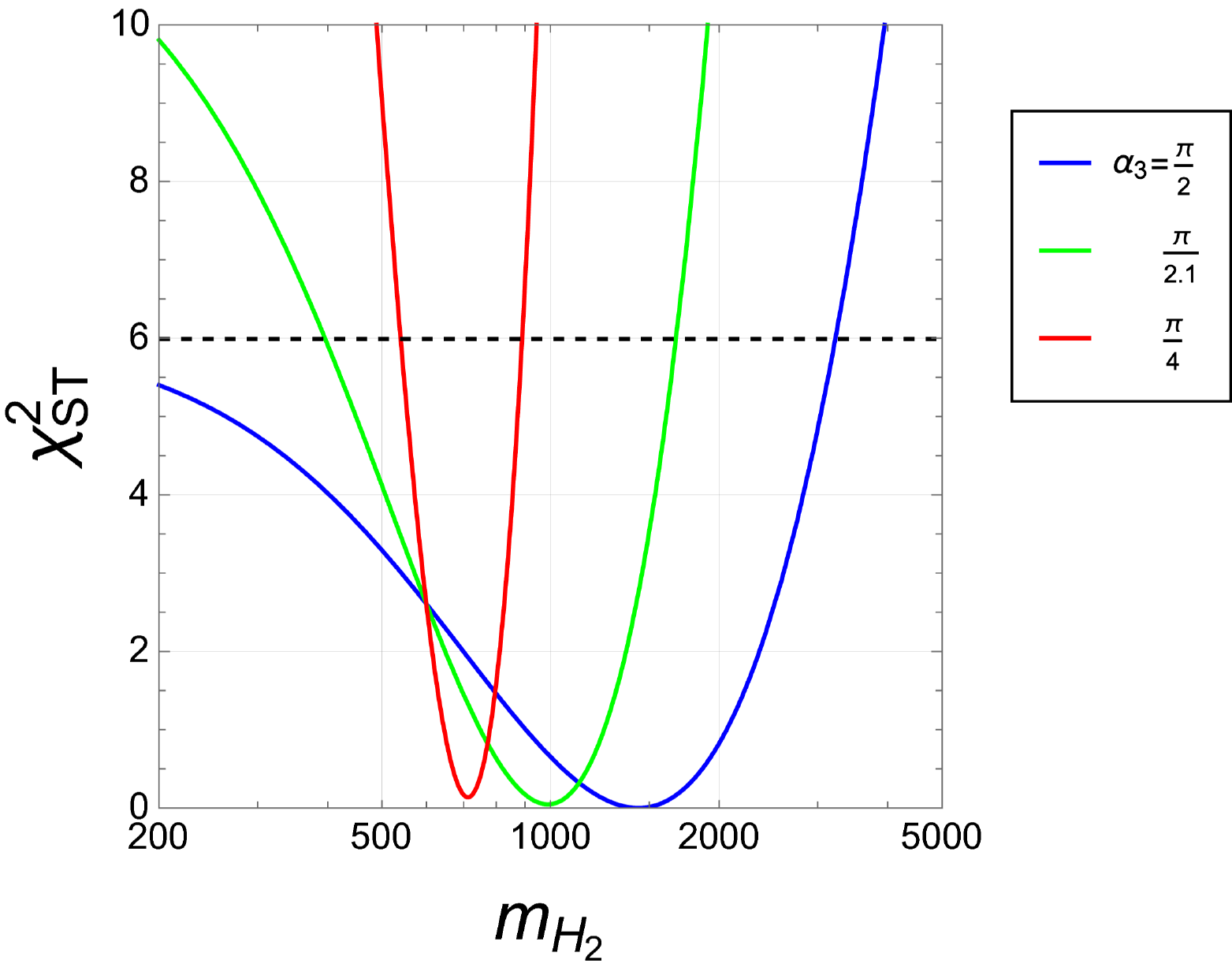}\hspace{0cm}
 \caption{ The $\chi^2$ fit  for the $S$ and $T$ parameters as a function of  $m_{H_{2}}$, where we vary the mixing angle $\alpha_{3}$. 
 }
 \label{FIG:STpara}
\end{figure}

In Fig.~\ref{FIG:STpara}, to illustrate the parameter space favored by the constraints from the $S$ and $T$ parameters, we show the $\chi^2$ values for the $S$ and $T$ parameters with the correlation coefficient $+0.92$~\cite{Haller:2018nnx} as a function of the mass of $H_2$ with $\alpha_3=\pi/4\ ({\rm red\ curve}),\; \pi/2.1,\ ({\rm blue\ curve})\; \pi/2$ (green curve). 
The other parameters are taken as 
\begin{align}\label{eq:BPST}
&\tb{1}=10\;,\quad \tb{2}=2\;,\quad M^2_{12}=M_{13}^{\prime 2}=M_{23}^{\prime 2}=(500\ {\rm GeV})^2\;,  
\lambda_{10}=-0.45\;, \notag \\
&\gamma_+=\frac{\pi}{2}\;, \quad
\alpha_1=\beta_1\;, \quad
\alpha_2=\beta_2\;, 
\notag \\
&m_{H_3}=600\ {\rm GeV}\;,\quad
m_{H_1^\pm}=600\ {\rm GeV}\;,\quad
m_{H_2^\pm}=500\ {\rm GeV}\;. 
\end{align}
Using this set of the parameters, the mass and mixing angles of the CP-odd Higgs boson are obtained as
\begin{align}
    m_{A_1}=601\ {\rm GeV} \;,\quad m_{A_2}=500\ {\rm GeV}, \quad \gamma_1=0.07\pi \;. 
\end{align}
The large mass difference between $m_{H_2}$ and the masses of other heavy Higgs bosons tends to be in conflict with the experimental results for each value of $\alpha_3$.  
Setting $\alpha_3=\pi/4\ (\pi/2.1)$ yields the bounds for the mass of $H_2$,  $550\ (400) \lesssim m_{H_2} \lesssim 900\ {\rm GeV}\ (1700~{\rm GeV})$. 
Although $m_{A_1}=m_{H^\pm_1}$ and $m_{A_2}=m_{H^\pm_2}=m_{H_3}$ are satisfied in Eq.~\eqref{eq:BPST}, the constraint becomes much tighter in case that the mass degeneracy among these additional Higgs states is assumed. 
Note that the contributions from the axion to the $S$ and $T$ parameters are negligible due to the tiny mixing angles; i.e., $\gamma_2,\ \gamma_3~\simeq 0$.

\subsection{Flavor constraints}

The 3HDM model parameter space is limited by measurements of $B$ meson rare decays and $B$ meson mixing. A particularly strong constraint is given by $B \to X_s \gamma $, which is altered from the SM prediction by the additional contributions of the loop diagram of the charged Higgs boson.
The Heavy Flavor Averaging Group (HFAG) gives the the experimental value for ${\rm BR}(B \to X_s \gamma)$ as~\cite{HFLAV:2019otj}
\begin{align}
  {\rm BR}(B \to X_s \gamma)_{\rm exp}=(3.32\pm 0.15)\times 10^{-4}
\end{align}
with the cut off for the photon energy $E_\gamma>1.6$ {\rm GeV}.
The precise evaluation of the SM prediction with QCD corrections have been performed at NLO~\cite{Adel:1993ah,Misiak:1994zw,Ali:1995bi,Pott:1995if,Greub:1996tg,Chetyrkin:1996vx,Greub:1997hf,Buras:1997xf,Buras:2001mq} and at NNLO~\cite{Misiak:2006zs,Misiak:2020vlo}. 
Effects of the charged Higgs boson loop contributions for $B \to X_s \gamma$ with NLO QCD ~\cite{Ciuchini:1997xe,Borzumati:1998tg,Borzumati:1998nx,Ciafaloni:1997un,Bobeth:1999ww} and NNLO QCD~\cite{Hermann:2012fc,Misiak:2015xwa,Misiak:2020vlo} have been investigated in the 2HDMs (Also, see the evaluation in Type II 3HDM, in Ref.~\cite{Chakraborti:2021bpy}). 
We evaluate $ {\rm BR}(B \to X_s \gamma)$ for our model by the linearized formula given in Ref.~\cite{Misiak:2020vlo}:
\begin{align}\label{eq:bsgam}
  {\rm BR}(B \to X_s \gamma)=(3.4\pm 0.17)\times 10^{-4}-8.25\Delta C_7-2.10\Delta C_8, 
\end{align} 
where the first term contains theoretical uncertainties. 
The second and third terms, $\Delta C_{7,8}$, denote additional new physics contributions to the Wilson coefficients $C_7$ and $C_8$~\footnote{The corresponding operators are given by $\mathcal{O}_7=e/(16\pi^2)m_b(\bar{s}_L\sigma^{\mu\nu}b_R)F_{\mu\nu}$ and $\mathcal{O}_8=g_s/(16\pi^2)m_b(\bar{s}_L\sigma^{\mu\nu}T^ab_R)G^a_{\mu\nu}$.} at the scale of the EW theory $\mu_0$, which is taken to be $\mu_0=160$ {\rm GeV}. 
Making use of the explicit formula of $\Delta C_{7,8}$ in Ref.~\cite{Hermann:2012fc}, we include the contributions of $H_1^\pm$ and $H_2^\pm$ to Eq.~\eqref{eq:bsgam}.

We also take into account the constraint from the mass difference $\Delta M_s$ for the mass eigenstates in the $B_s$-$\overline{B}_s$ system. 
 The experimental value is taken from the result of HFAG~\cite{HFLAV:2019otj},
\begin{align}
\left( \Delta M_s \right)_{\rm exp}
  =17.757\pm0.021\ {\rm ps}^{-1}.
\end{align}
The analytical formula is given by~\cite{Barger:1989fj,Enomoto:2015wbn}
\begin{align}\label{eq:delMs}
  \Delta M_s=\frac{G_F^2 m_W^2 m_{B_s}}{6\pi^2}|V_{tb}^\ast V_{ts}|^2f_{B_s}^2B_{B_s}\eta_{B_s}C_{VLL},
\end{align} 
where
$m_{B_s}$ and $f_{B_s}$ denote the mass and decay constant of the $B_s$ meson.
$C_{VLL}$ is the Wilson coefficient at EW scale $\mu_0$ for the operator $O_{VLL}=\bar{s}^\alpha \gamma_\mu (1-\gamma_5)b^\alpha \bar{s}^\beta \gamma^\mu (1-\gamma_5) b^{\beta}$ with $\alpha$, $\beta$ being the color indices. 
It is composed of the contributions from exchanges of the $W$ boson and the charged Higgs bosons in the box diagrams, and the explicit formula is given in Appendix~\ref{ap:DMs}. 
The effect of QCD running from the scale $\mu_0$ to that of the ${B_s}$-$\bar{B}_s$ system is encoded by the evolution factor $\eta_{B_s}$.
Non-perturbative QCD effects are included by the bag parameter $B_{B_q}$. 
In Eq.~\eqref{eq:delMs}, the contributions with the small mass fraction $x_b=m_b^2/m_W^2$ is omitted.  
For the numerical evaluation of $\Delta M_s$, we use the following input values~\cite{CKMfit2021} 
\begin{align}
    f_{B_s}&= (228.8 \pm 0.7 \pm 1.9)\ {\rm MeV}\;,\quad B_{B_s}=1.327 \pm 0.016 \pm 0.030\;,\   \notag \\
    \eta_{B_s}&= 0.5510 \pm 0 \pm 0.0022\;,
\end{align}
where the second (third) values stands for statistical and systematic uncertainties (systematic theoretical uncertainties). 

\begin{figure}[t]
  \centering
   \includegraphics[scale=0.6]{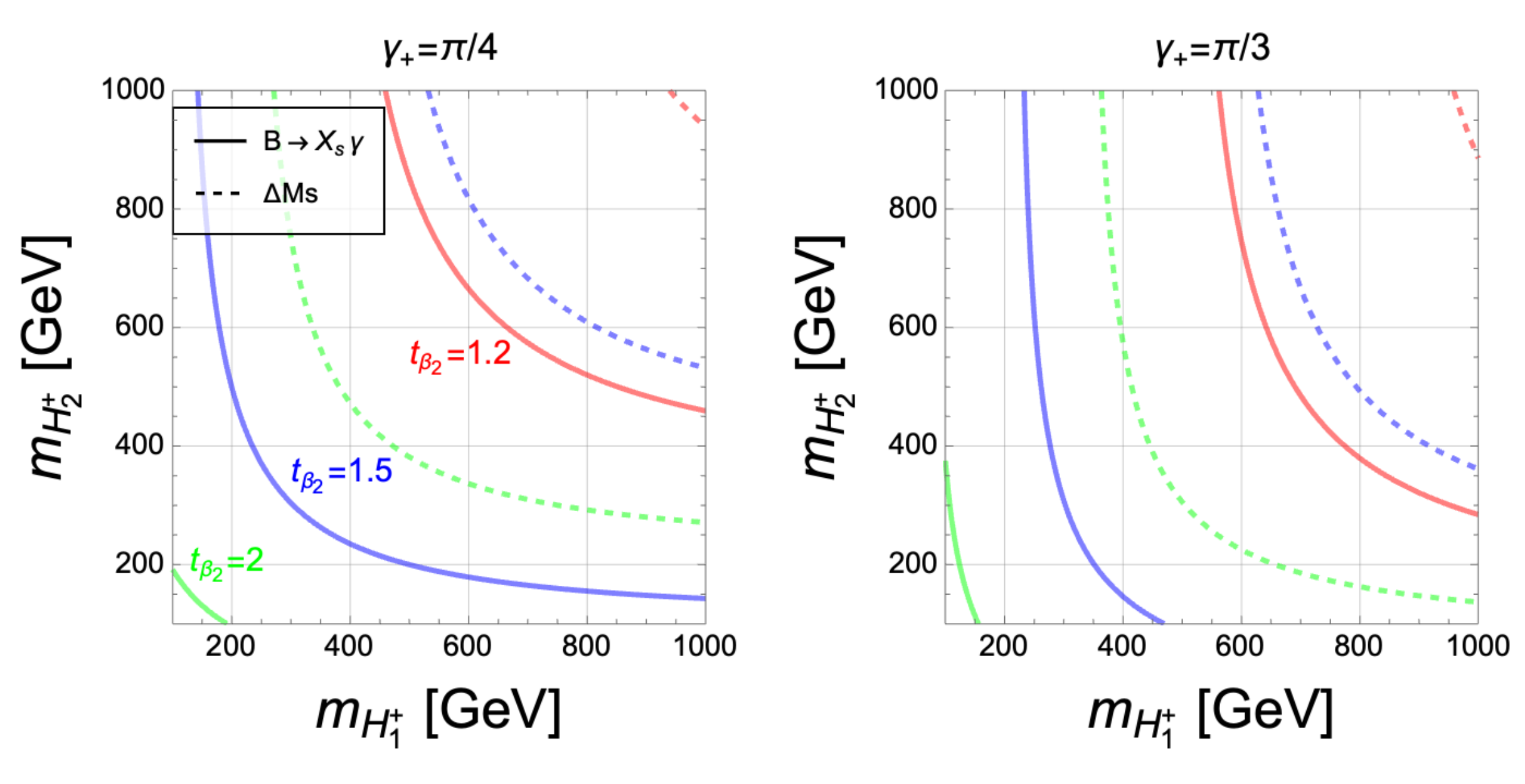}\hspace{0mm}
  \caption{ Lower bounds on the masses of the charged Higgs bosons from the $B\to X_s \gamma$ and the $B_s$-$\bar{B}_s$ oscillation parameter $\Delta M_s$.
  The solid (dotted) line shows the result of  $B\to X_s \gamma$ ($\Delta M_s$).
  We take $t_{\beta_2}$=1.2 (red), 1.5 (blue), 2 (green), and the mixing angles for the charged Higgs bosons are fixed as $\gamma_{+}=\pi/4$ (left) and $\gamma_{+}=\pi/3$ (right). }
  \label{fig:flavor}
 \end{figure}

Contributions from the charged Higgs bosons to $B\to X_s\gamma$ and $\Delta M_s$ are controlled by their masses $m_{H^\pm_{1,2}}$ and the quark Yukawa couplings $\xi^{H^\pm}_{1,2}$. 
Thus, these measurements give the lower bounds on the masses of charged Higgs bosons for fixed $\tb{2}$ and $\gamma_+$ as shown in Fig~\ref{fig:flavor}, where we choose $\tb{2}$ = 1.2\ $\mbox{(red line)}$, 1.5\ \mbox{(blue line)}, 2\ \mbox{(green line)}, and $\gamma_+= \pi/4\ \mbox{(left panel)}$, $\pi/3\ \mbox{(right panel)}$. 
One can clearly see that the constraint from $\Delta M_s$ is more severer than ${\rm BR}(B\to X_s\gamma)$. 
At {$(\tb{2},\gamma_+)=(1.5,\pi/4)$} and $m_{H^\pm_1}\simeq 600~(800)$ {\rm GeV}, e.g., $m_{H^\pm_2}\gtrsim 800\ (600)\ {\rm GeV}$ are required by $\Delta M_s$. 
Since the quark Yukawa coupling $\xi^{H^\pm}_{1,2}$ is inversely proportional to the $\tb{2}$, the lower bounds for the masses are relaxed by taking larger $\tb{2}$. 
On the other hand, if $\gamma_+$ is close to $\gamma_{+}=\pi/2$ (0), loop contributions of  $H^{\pm}_{2}$ ($H^{\pm}_{1}$) almost decouple in both of $B\to X_s\gamma$ and $B_s$-$\overline{B}_s$ mixing. 
We emphasize that the Yukawa structure of the quark sector for our model is the same as Type-I and Type-X 3HDMs, since, at the LO, only the up- and down-type quark Yukawa 
couplings are relevant for $B\to X_s \gamma$. 
Hence, the flavor constraints for the Type-I and Type-X 3HDMs are similar to our results.

\section{Phenomenological implications of the XENON1T excess}\label{sec:XENON1T}
The axion mass is determined by the ratio of the soft breaking masses $m_{ij}^{(\prime)2}$  and the decay constant $f_a$, as described in Sec.~\ref{sec:mass}. The mass spectrum of the extra Higgs bosons also depends on the soft breaking masses $m_{ij}^{(\prime )2}$. This implies that for a given decay constant, the axion mass can be related to the mass spectrum of the heavy Higgs bosons. To see if this picture is correct, we study the implications of the XENON1T excess in the electron recoil event~\cite{XENON:2020rca} for the 3HDM with the $B-L$ Higgs bosons. 
We will examine the parameter region which explains the  XENON1T 
excess and satisfies all the constraints presented in the previous section.

\red{The favored range of the axion mass that can explain the XENON1T excess at 95 \% confidence level
is given by ~\cite{Bloch:2020uzh},\footnote{According to Ref.~\cite{XENON:2020rca} by the
XENON1T group, the suggested axion mass at 68 \% CL is given by $2.3 \pm 0.2$ keV. 
}}
\begin{align}
\label{eq:maX1t}
2.1\ {\rm keV}\lesssim m_a \lesssim 3.1 \ {\rm keV} \;
\end{align}
with the best fit value of the axion-electron coupling $g_{ae} = 4\times 10^{-14} $. 
\red{
To make both $g_{ae}$ and $m_a$ consistent with the XENON1T excess,
in this section we fix the axion decay constant as
\begin{align}
f_a=3m_e/(4.0\times 10^{-14})\simeq 3.8\times 10^{10}~{\rm GeV}
\end{align}}
 and use Eq.~\eqref{eq:maX1t} when necessary.
\subsection{The viable parameter space}

We have performed numerical calculations to find a parameter region that would explain the XENON1T excess while satisfying the experimental limits described in the previous section. To this end we assume degenerate masses for the additional Higgs bosons
\begin{align}\label{eq:defmPhi}
m_{H_{2}}=m_{H_{3}}=m_{H_{1}^{\pm}}=m_{H_{2}^{\pm}}\equiv m_{\Phi}
\end{align}
to satisfy the constraints from the $S$ and $T$ parameters, and we take the alignment limit (\ref{eq:alignment}), $\alpha_{1}=\beta_{1}$ and $\alpha_{2}=\beta_{2}$.
\red{We also take $\tb{1}$=1 in order to avoid the constraint from the current X-ray observations. The detailed discussions are presented in Sec.~\ref{sec:Xray}.}
In the numerical analysis we vary the \red{remaining} input parameters in the following ranges,
\begin{align}\label{eq:scanr}
\tb{2}&=[1,\;10]\;,\quad \ \gamma_{+},\ \alpha_{3}=[-\frac{\pi}{2},\frac{\pi}{2}]\;,\quad 
\lambda_{10}=[-5,\;5],\;\quad \notag \\
{m_\Phi}
&= [200{\rm GeV},\;1000 {\rm GeV}]\;, \notag \\
 M^2_{12}&=[0,\;(1000\ {\rm GeV})^2] \;,\quad  M^{\prime 2}_{13}=M^{\prime 2}_{23}=[0,\;(1000\ {\rm GeV})^2]\;.
\end{align}
where the lower limits of $t_{\beta_2}$ are chosen so as to satisfy the constraints from the RG running of the top Yukawa coupling, and the measurements of the $B$ meson decay $B\to X_s \gamma$ and $B_s$-$\bar{B}_s$ mixing. 
With the scan range Eq.~\eqref{eq:scanr}, we obtain $2.3\times 10^{-14}\lesssim g_{ae}\lesssim 4\times 10^{-14}$, which is within the range indicated in Ref.~\cite{Bloch:2020uzh}. 
\red{Furthermore, the range of axion-photon coupling is $4.5\times 10^{-20}{\rm GeV}^{-1}\lesssim g_{a\gamma}\lesssim 1.5\times 10^{-18}{\rm GeV}^{-1}$. We have checked that there are no parameter points excluded by the current limit of X-ray observations in this case.
}

\begin{figure}[t]
 \centering
  \includegraphics[scale=0.6]{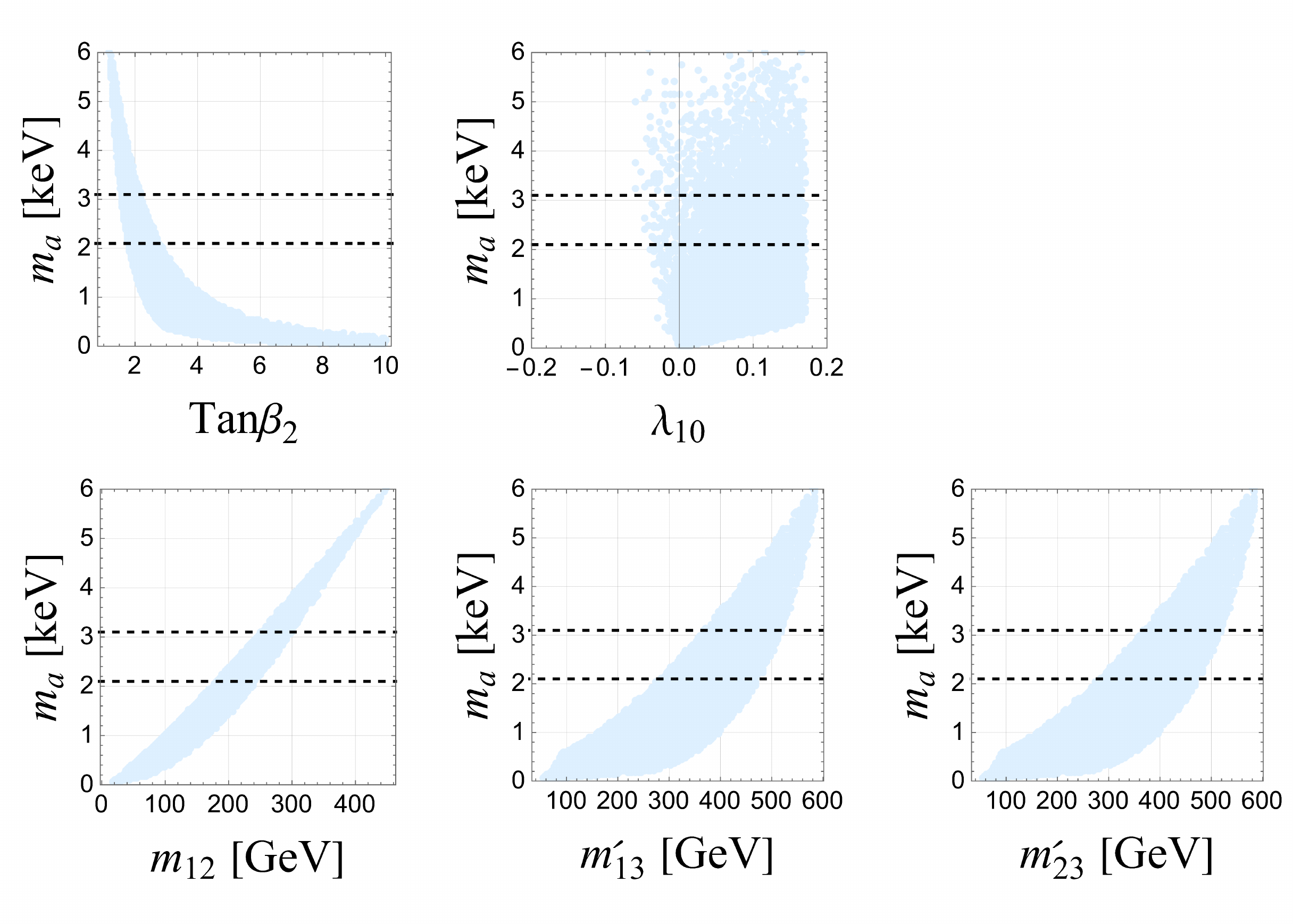}\hspace{3cm}
 \caption{The axion mass as a function of the model parameters in the scanned range of \eqref{eq:scanr}.
  All the theoretical constraints and the experimental limits discussed in the Sec.~\ref{sec:con} are imposed. 
  }
 \label{FIG:alpmasstr}
\end{figure}

In Fig.~\ref{FIG:alpmasstr}, we show the axion mass as a function of the relevant input parameters.
The region between the black dotted lines indicates the mass suggested by the XENON1T excess. 
The negative correlation between $m_a$ and \red{{$t_{\beta_{ 2}}$}} seen in Fig.~\ref{FIG:ALPcont} is also confirmed in Fig.~\ref{FIG:alpmasstr}. 
One of the intriguing observations here is that the XENON1T excess restricts the range of $\tb{2}$, i.e., $\tb{2}\lesssim 3 $. 
\red{The range of the soft breaking mass is} also limited as \green{$180\ {\rm GeV}\lesssim m_{12}\lesssim \red{300}\ {\rm GeV}$}. 
Note that the upper bounds of these parameters {are correlated with} the masses of the additional Higgs bosons as will be seen shortly.

\begin{figure}[t]
 \centering
  \includegraphics[scale=0.5]{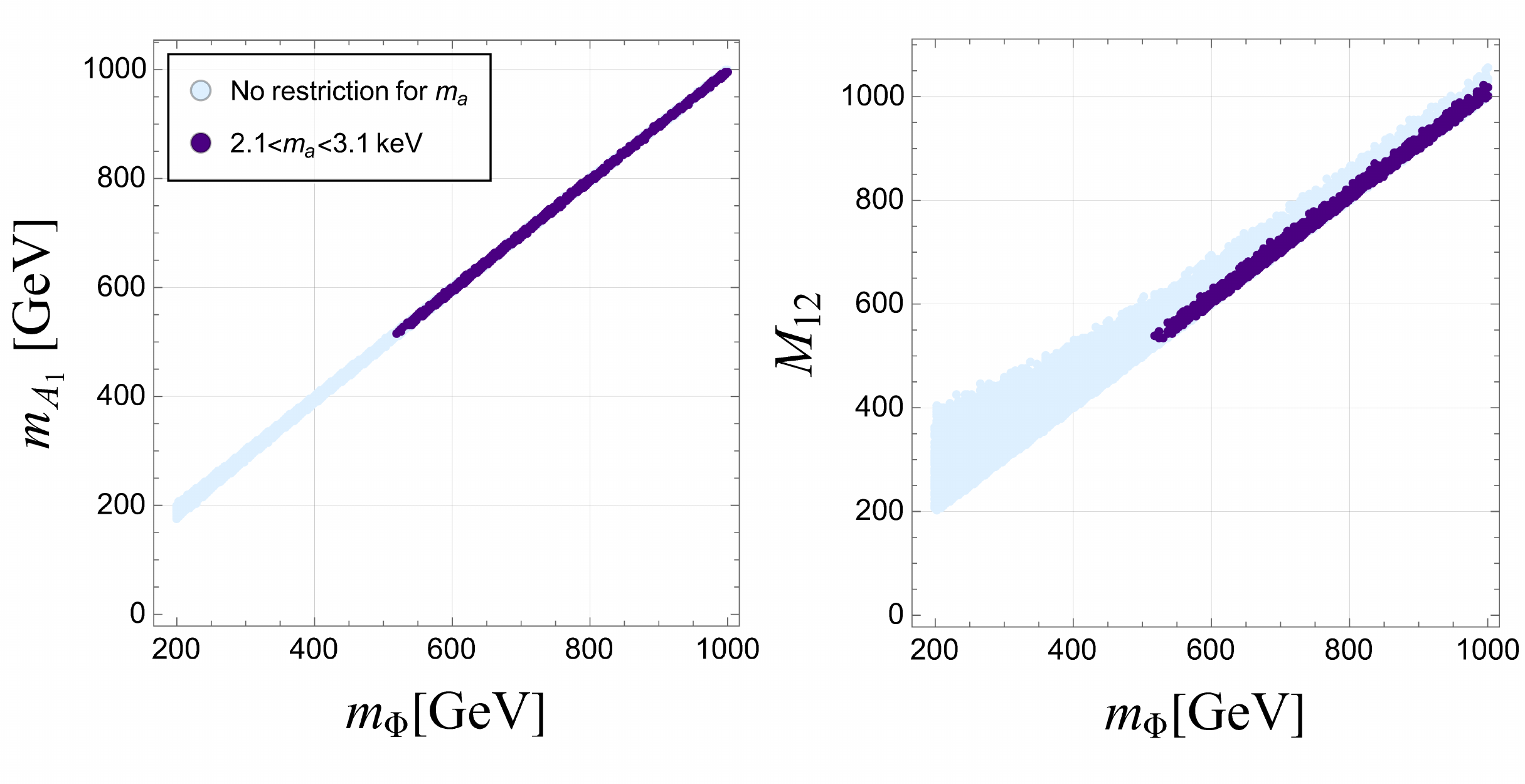}\hspace{0cm}
  \caption{Correlation between $m_{A_2}$ ($M_{12}$) and \green{$m_{\Phi}$} in the left (right) panel. \green{The mass $m_{\Phi}$ is defined in Eq.~\eqref{eq:defmPhi}.}
  {All the theoretical constraints and the experimental limits discussed in the Sec.~\ref{sec:con} are imposed.}   }
 \label{FIG:Heavymass}
\end{figure}

In Fig~\ref{FIG:Heavymass}, we show the correlation between $m_{A_1}$ and $m_{\Phi}$, as well as $m_{A_1}$ and $M_{12}$. 
The light blue points correspond to the region that satisfies all the theoretical and experimental constraints, 
and the dark blue points correspond to the region where the axion mass is further restricted to explain the XENON1T excess. 
While the favored mass range of the axion by the XENON1T excess bounds the soft breaking masses $m_{ij}^{(\prime)}$ around the EW scale, this does not necessarily mean that the mass scale of the additional Higgs bosons is similarly restricted. 
Even if one imposes $m_a$ in the range of Eq.~\eqref{eq:maX1t}, all the additional Higgs bosons can decouple by taking $M_{12}^2\sim M_{13}^{\prime 2}\sim M_{23}^{\prime 2}\gg v$. 
Since these parameters are rescaled by the trigonometric functions of the angles $\beta_1$ and $\beta_2$ (see Eq.~\eqref{eq:bmsq}), the scale of $M_{ij}^{(\prime)}$ does not necessarily correspond to that of $m_{ij}^{(\prime)}$.  
Note also that the lower bound on the mass of additional Higgs bosons, \red{$m_{\Phi} \gtrsim 510~{\rm GeV}$}, is given when the axion mass is in the range of Eq.~\eqref{eq:maX1t}. 
This lower bound comes from the combination of the XENON1T excess and the constraint from the $B_s$ - $\bar{B}_s$ mixing. 
If the mass of additional Higgs boson is \green{less than 1 TeV}, {$\tb{2}$ should not be large to keep $m_a\sim {\cal O}(1)$ keV~(see the top middle panel in Fig.~\ref{FIG:alpmasstr})}.
In contrast, to evade the constraint from the $B_s$ - $\bar{B}_s$ mixing, $\tb{2}\gtrsim 2$ is required for $m_{\Phi}\simeq 450$ {\rm GeV}. 
As a result, if the masses of the additional Higgs bosons were lower, the XENON1T excess would conflict with the measurement of the $B$ meson mixing.

From the {right} panel of Fig.~\ref{FIG:Heavymass}, \green{one can see that the value of $m_\Phi$ is not so different from $M_{12}$}. 
This is mostly caused by the constraints from the RG evolution of the scalar quartic couplings. 
A large hierarchy among the mass of the additional Higgs bosons and the rescaled soft breaking parameters enlarges the scalar quartic couplings.    
If one set $m_{\Phi}=M_{12}=M^\prime_{13}=M^\prime_{23}$ in the alignment limit,
the analytical expressions of the quartic couplings can be reduced as 
\begin{align}
    &\lambda_{1}
    =
    \frac{1}{4}t_{\beta_1}t_{\beta_2}^2 \frac{1}{c_{\beta_1}^2} \lambda_{10} v^2  
    +\frac{1}{2} \frac{m_{H_1}^2}{v^2}\;,\;
    \lambda_{2}
  =\frac{1}{4}\frac{t_{\beta_2}^2}{t_{\beta_1}} \frac{1}{s^2_{\beta_1}}\lambda_{10} v^2   +\frac{1}{2} \frac{m_{H_1}^2}{v^2}\;,\;
  \\ 
  &2\lambda_{3}=\lambda_{5}=\lambda_{6}=\frac{m_{H_1}^2}{v^2}\;,\;
  \lambda_{4}=\frac{m^2_{H_1}}{v^2}-\frac{1}{2}\frac{t_{\beta_2}^2}{s_{\beta_1}c_{\beta_1}}\lambda_{10}\;,\; \\
  &\lambda_7=0\;,\;\quad \frac{1}{t_{\beta_1}}\lambda_8=t_{\beta_1}\lambda_9=-\lambda_{10}\;, 
\end{align}
Namely, in this case, these couplings do not depend on the masses of the additional Higgs bosons and the soft breaking masses, and so, the constraints from the RG evolution can be evaded.  
We also note that a further requirement, $\lambda_{10}=0$ leads to $\lambda_7=\lambda_8=\lambda_9=0$ and $\lambda_1=\lambda_2=\lambda_3=2\lambda_4=2\lambda_5=2\lambda_6$. 
In this limit, the 3HDM potential $V_{\rm 3HDM}$ 
except for the soft breaking terms $V_{\rm soft}$
has an Sp(6) symmetry~\cite{Darvishi:2021txa}, which is explicitly broken by the terms with the coefficients $M_{12}^2$ or $\kappa_{m \phi n}v^2_{S_m}\ (m=1,\bar{2},\ n=1,2,3)$.  In fact,
the allowed region for $\lambda_{10}$ is  $\red{-0.1}\lesssim \lambda_{10} \lesssim 0.17$, which contains this limit.
Hence, the Sp(6) symmetry would be the desired (approximate) symmetry for the 3HDM part of the Higgs potential to satisfy all the theoretical and experimental constraints. 

\begin{figure}[t]
 \centering
  \includegraphics[scale=0.5]{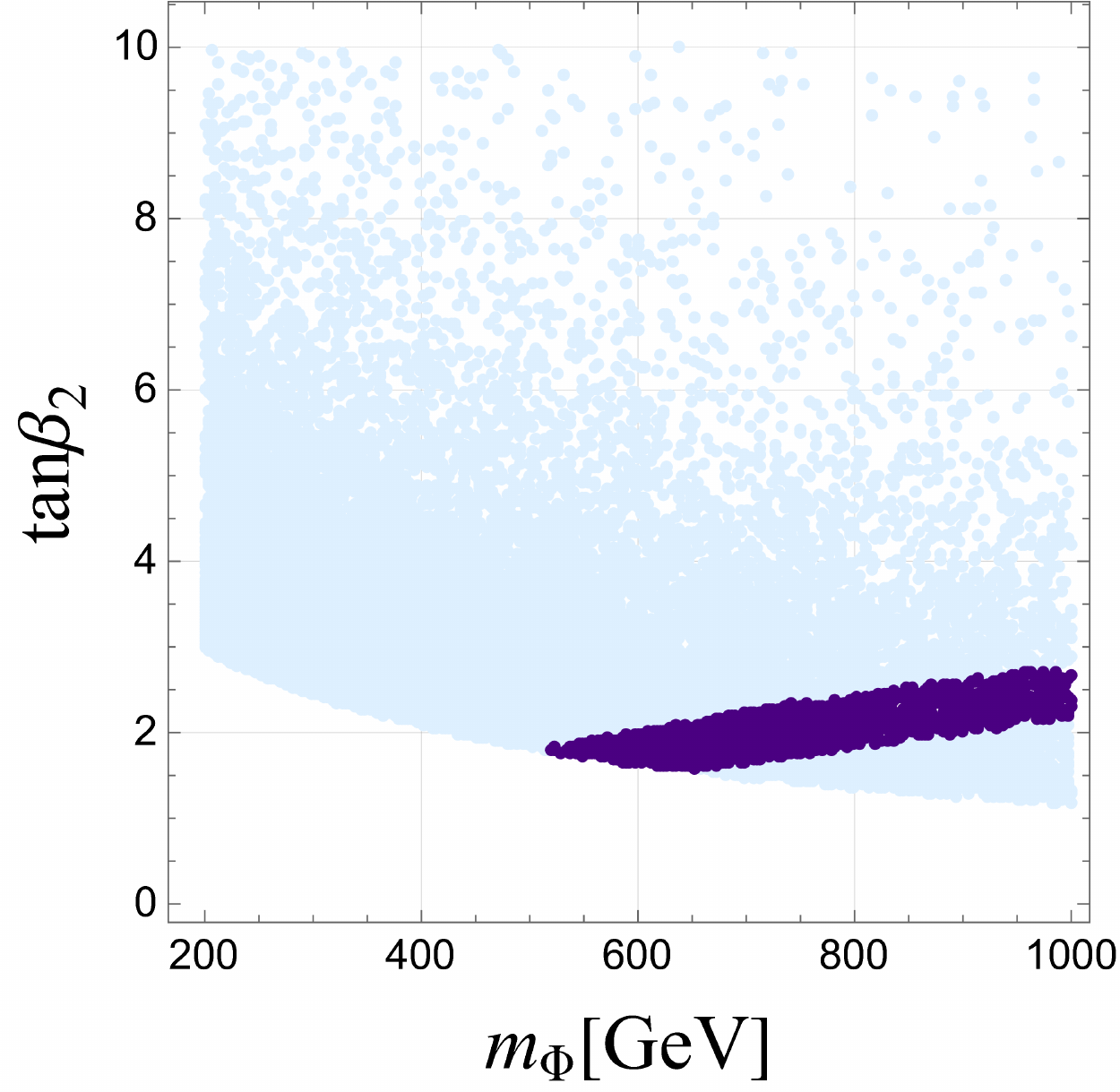}\hspace{0cm}
 \caption{The allowed regions of the masses of the extra Higgs bosons \green{$m_\Phi$} and \red{$\tb{2}$}.
 \green{The mass $m_{\Phi}$ is defined in Eq.~\eqref{eq:defmPhi}.}
 All the theoretical constraints and the experimental limits discussed in the Sec.~\ref{sec:con} are imposed.  }
 \label{FIG:HeavymassTB}
\end{figure}

Another consequence for heavy Higgs bosons from the XENON1T excess is that there is a correlation between the mass $m_{\Phi}$ and the ratio of the VEVs $\tb{2}$ as shown in Fig.~\ref{FIG:HeavymassTB}. 
For $m_{\Phi}=800~{\rm GeV}$, \red{the allowed range of $\tb{2}$ is} $1.3\lesssim \tb{2}\lesssim 2.4$. 
\red{The range is} enlarged for heavier additional Higgs bosons.
 The Yukawa couplings for the additional Higgs bosons are controlled by $\tb{1}$ and $\tb{2}$ for fixed mixing angles.  Thus, if extra Higgs bosons are found in collider experiments, the decay properties of the extra Higgs bosons may allow us to test whether they are consistent with the anomaly-free axion that explains the XENON1T excess.

We also comment that the parameter regions favored by the XENON1T excess may be explored by the future measurements of the $B$ meson mixing. 
In Ref.~\cite{Charles:2020dfl}, projected sensitivity for new physics effect to the  $B$ meson mixing is studied, considering planned LHCb Upgrade II~\cite{LHCb:2018roe} and a possible upgrade of Belle II~\cite{BelleIIVXD} as well as FCC-ee as a tera-$Z$ factory. 
We find that the future 95\% sensitivity for $B_s$ mixing by LHCb 300 ${\rm fb^{-1}}$ and Belle II 250 ${\rm ab^{-1}}$ can probe \green{$m_{\Phi}\lesssim 940~(700)~{\rm GeV}$ for $(t_{\beta_2},\gamma_+)=(1.75,\pi/2)$ ($(2,\pi/2)$)}.

\subsection{Predictions for the SM-like Higgs boson decays}

Here let us illustrate the extent to which the decays of the SM-like Higgs boson $H_1$ are deviated from the SM prediction in the allowed parameter space satisfying all the theoretical and experimental constraints. 
As we consider the alignment limit, the decay of the SM-like Higgs boson into weak gauge bosons and the fermions are the same as the SM predictions at the tree level.
On the other hand, 
the decays of $H_1\to \gamma\gamma$ can deviate from the SM one through the charged Higgs boson loop diagrams.
Let us express the deviation from the SM prediction for $H_1\to \gamma\gamma$ 
in terms of the modifier defined by
\begin{align}\label{eq:defDR}
    \Delta \kappa_{\gamma}=\sqrt{\frac{\Gamma(H_1\to \gamma\gamma)}{\Gamma(H_1\to \gamma\gamma)^{\rm SM}}}-1\;, 
\end{align}
\green{where $\Gamma(H_1\to \gamma\gamma)^{\rm SM}$ denotes the decay rate in the SM.}
The precision measurements of the Higgs boson coupling will be performed in the future collier experiments. 
At the HL-LHC (ILC 250 GeV), the sensitivity of the coupling modifier for $H_1\to \gamma\gamma$ can reach 1.6\% (1.4\%)~\cite{deBlas:2019rxi}. 
Furthermore, in Ref.~\cite{deBlas:2019rxi} the combined sensitivity of FCC-ee, FCC-eh, FCC-hh is estimated as 0.31\%. 
For the evaluation of the decay rate in the 3HDM, we use the following  analytical formula,
\begin{align}
\Gamma(H_1\to\gamma\gamma)&=\frac{\alpha^{2}_{\rm EM} }{256\pi^{3}}\frac{m^{3}_{H_1}}{v^{2}}
\Bigg | \kappa^{H_1}_{V}F_{1}\Bigg(\frac{4m_{W}^{2}}{m^{2}_{H_1}} \Bigg)
+N_{f}^{c}Q_{f}^{2}\xi^{f}_{H_1}\sum_{f}F_{1/2}\Bigg(\frac{4m_{f}^{2}}{m^{2}_{H_1}} \Bigg) \notag \\
&+\sum_{j}\frac{m_{W}^{2} \lambda_{H_1 H^{+}_{j} H^{-}_{j} } } {2c_{W}^{2} m_{H^{+}_{j}}^{2}}F_{0}\Bigg(\frac{4m_{H^{+}_{j}}^{2} }{m^{2}_{H_1}} \Bigg)\Bigg|^{2}\;,
\end{align}
For the explicit forms of the loop functions $F_{1}$, $F_{1/2}$ and $F_{0}$, we refer the reader to Ref.~\cite{Djouadi:2005gi,Djouadi:2005gj}. 
In the alignment limit, the scalar couplings for the charged Higgs boson are given by
\begin{align}
  \lambda_{H_1H_1^\pm H_1^\pm}v&=
-2 M^{2}_{12} c_{\beta_2}^2 c_{\gamma_+}^2-2 M^{\prime 2}_{13} (s_{\beta_1} s_{\beta_2} c_{\gamma_+}-c_{\beta_1} s_{\gamma_+})^2 \notag \\
&-2 M^{\prime 2}_{23} (c_{\beta_1} s_{\beta_2} c_{\gamma_+}+s_{\beta_1} s_{\gamma_+})^2+m_{H_1}^2+2 m_{H^\pm_1}^2,\\
   \lambda_{H_1H_2^\pm H_2^\pm}v&= 
-2 M^{2}_{12} c_{\beta_2}^2 s_{\gamma_+}^2-2 M^{\prime 2}_{13} (s_{\beta_1} s_{\beta_2} s_{\gamma_+}+c_{\beta_1} c_{\gamma_+})^2 \notag \\
&-2 M^{\prime 2}_{23} (s_{\beta_1} c_{\gamma_+}-c_{\beta_1} s_{\beta_2} s_{\gamma_+})^2+m_{H_1}^2+2 m_{H^\pm_2}^2.
\end{align}

\begin{figure}[t]
  \centering
  \includegraphics[scale=0.7]{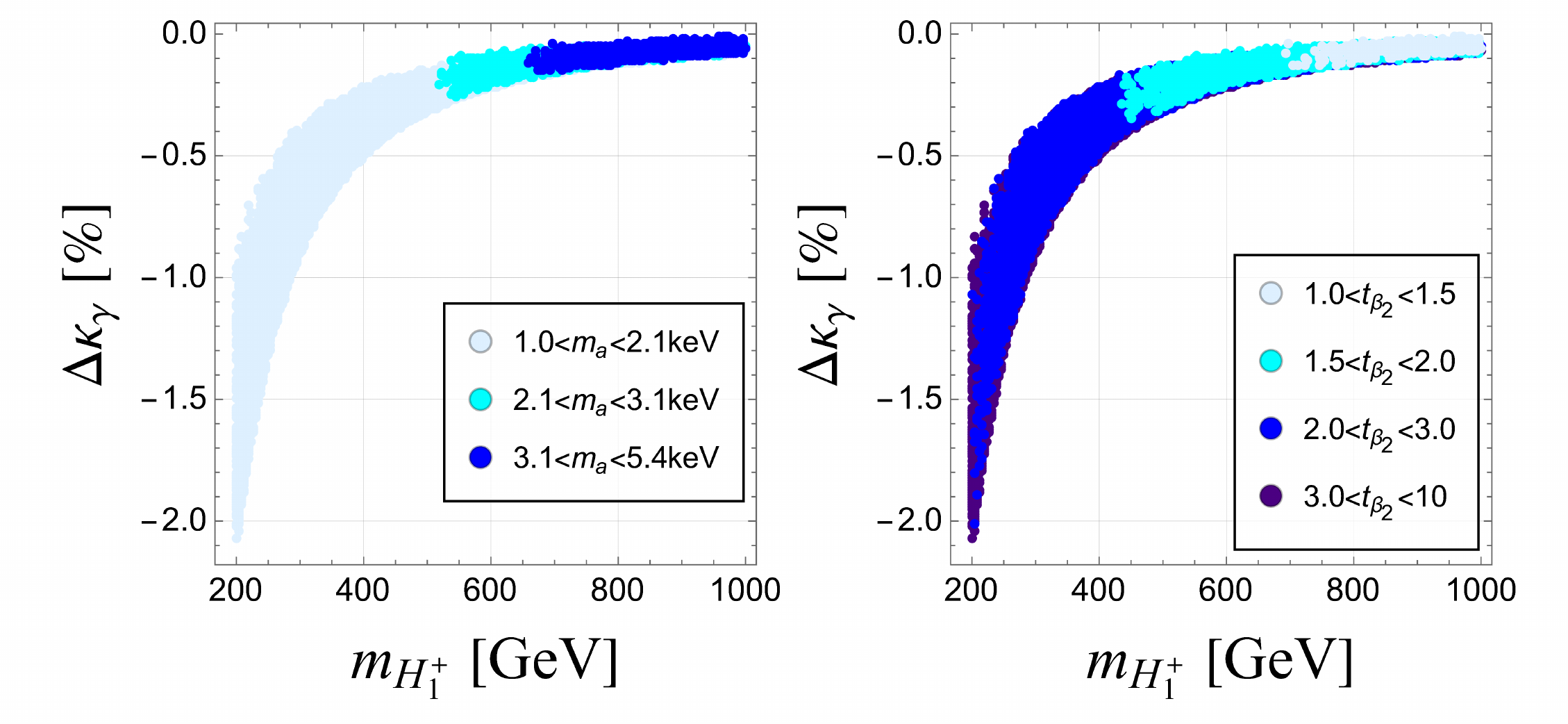}\hspace{3cm}
    \caption{Predictions for deviations from the SM in $H_1\to \gamma\gamma$. 
    Light intensity corresponds to the value of $m_a$ and $t_{\beta_2}$ in the left and right panels, respectively. 
    $\Delta \kappa_{\gamma}$ is defined in Eq.~\eqref{eq:defDR}. The input parameters are scanned in the range of Eq.~\eqref{eq:scanr}, taking into account the constraints discussed in Sec.~\ref{sec:con}. 
    }
  \label{FIG:hgamgma}
 \end{figure}

We show the numerical results for $\Delta \kappa_{\gamma}$ in  Fig.~\ref{FIG:hgamgma}, where the different color of the points corresponds to the values of $m_a$ and $t_{\beta_2}$ in the left and right panels, respectively. 
Since we take the alignment limit, the deviations from the SM predictions purely come from the contributions from the charged Higgs boson loop diagrams. 
Remarkably, contributions of the charged Higgs bosons do not decouple in case of \green{$m_{H^\pm_1}\lesssim 600\ {\rm GeV}$}. 
This is due to the constraint from $\Delta M_s$. 
As seen in the right panel, relatively high $\tb{2}$ is required for lighter $H_{w1}^\pm$ by the constraint. 
We have checked that $\Delta \kappa_\gamma\simeq 0$ can be realized even in the region \green{$m_{H^\pm_1}\lesssim 600\ {\rm GeV}$} if the constraint from $\Delta M_s$ is switched off.

The deviation $\Delta \kappa_\gamma$ is maximized at $m^2_{H_1^\pm}=200$ GeV and exceeds $-2\%$. 
In this case, the corresponding axion mass is less than 2.1 keV. 
If the axion mass lies in the range favored by the XENON1T excess, the deviation shrinks, i.e., \green{$-0.22\%\lesssim\Delta \kappa_\gamma\lesssim0\%$}, which would be difficult to detect at the HL-LHC and ILC 250 GeV. 
Equivalently, if the deviation of $H_1\gamma\gamma$ is found within the range \green{$\Delta \kappa_\gamma\lesssim-0.25\%$}, the axion should be lighter than 2.1 keV. 

We have also calculated the deviation in the self-coupling of the SM-like Higgs boson $\lambda_{H_1H_1H_1}$ using the effective potential method.  
In contrast to the case of $H_1\to \gamma \gamma$, 
all additional Higgs bosons can contribute at the 1-loop level. 
We find that the magnitude of the deviation in $\lambda_{H_1H_1H_1}$ is not comparable with the projected sensitivity in the future collider experiments with the order of 10\%~\cite{Fujii:2017vwa,Cepeda:2019klc,deBlas:2019rxi,Goncalves:2018qas}. 
The main reason for this result is the strong constraint from the RG evolution of the scalar quartic couplings. 
The non-decoupling effects of the additional Higgs bosons are highly suppressed. 

\section{Exploration of extra Higgs bosons from X-ray observations}\label{sec:Xray}

We have discussed in the previous section implications for the extra Higgs bosons and predictions in the SM-like Higgs boson decays in the scenario where the axion has properties suggested by the XENON1T excess.
An interesting aspect of such axion from a cosmological point of view is that it can naturally explain the observed DM abundance, and it can also be probed by the future X-ray observatories such as Theseus~\cite{THESEUS:2017qvx,THESEUS:2017wvz}, Athena~\cite{Barret:2018qft}, eROSITA~\cite{eROSITA:2012lfj}, and XRISM~\cite{XRISMScienceTeam:2020rvx}. 
While in the previous section the axion decay constant $f_a$ is fixed to reproduce the value of $g_{ae}$ suggested by the XENON1T excess, we here vary $f_a$ and investigate the connection between the axion coupling with electron $g_{ae}$ and the mass of the additional Higgs bosons.
\red{We also demonstrate the parameter regions allowed by the current X-ray observations or probed by future X-ray observations.}
We then present the expected mass spectrum of the extra Higgs bosons if the axion is detected in the future X-ray observatories.  
In addition, one may be interested in the effect on the alignment parameters for the CP-even Higgs bosons $\alpha_1$ and $\alpha_2$ when considering the testability of the axion. 
To see how much these parameters can deviate from the alignment limit depending on the axion mass and the axion-electron coupling, 
we numerically evaluate the possible size of deviations in the SM-like Higgs boson with the weak gauge bosons $\kappa_V$.

\subsection{Scale of the masses of extra Higgs bosons}
\red{\subsubsection*{Bounds from X-ray observations }}
\begin{figure}[t!]
  \centering
  \includegraphics[scale=0.8]{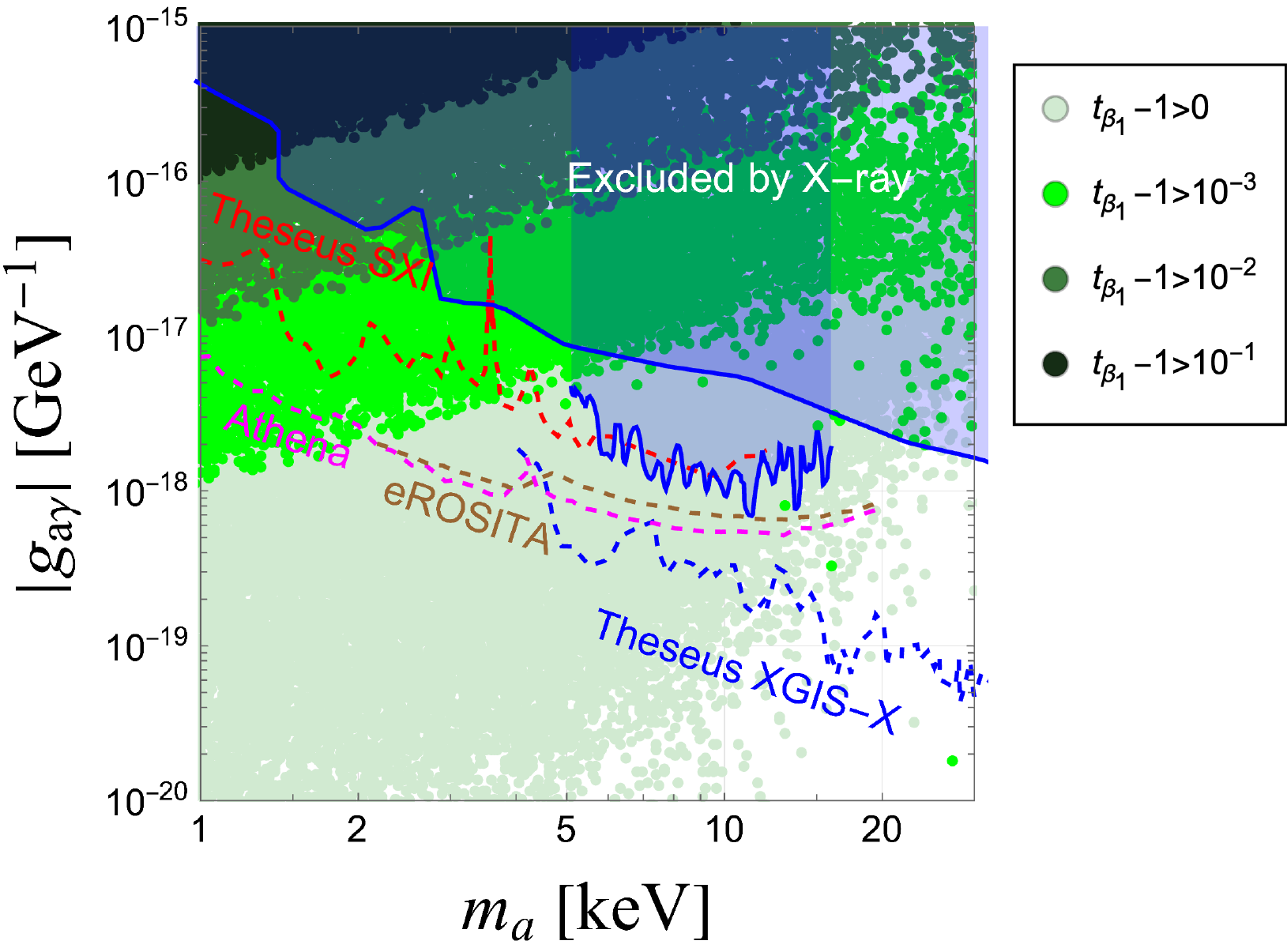}
  \caption{ 
  \red{
  The parameter region excluded by the X-ray observations is indicated by the blue shaded region in the plane of ($g_{a\gamma}, m_a$). 
 Future sensitivities are also shown by the colored dashed lines. 
 The size of $t_{\beta_1}$ is shown by the color difference of the points.
}
   }
  \label{FIG:gagamXray}
 \end{figure}
We first show the allowed parameter regions to satisfy the constraint from X-ray observations by scanning the model parameters. 
We take all the dimentionful parameters degenerate, i.e., 
\begin{align} \label{eq:degenemass}
m_{\Phi}=M_{12}=M_{13}'=M_{23}',
\end{align}
and the alignment limit for the simplicity. 
The remaining parameters are scanned in the following ranges
\begin{align}\label{eq:scangae}
    &m_{\Phi}=[200~{\rm GeV},10~{\rm TeV}]\;,\quad \red{t_{\beta_1}=[1,5]}\;,\quad t_{\beta_2}=[1,10] \;,\quad
    \lambda_{10}=[-5,5]\;,\quad  \notag \\
    &\alpha_3,\gamma_+=\left[-\frac{\pi}{2},\frac{\pi}{2}\right]\;, f_a=[1\times 10^{10}~{\rm GeV},5\times 10^{12}~{\rm GeV}]\;.
\end{align}
taking into account the constraints presented in Sec.~\ref{sec:con}. 

\red{In Fig.~\ref{FIG:gagamXray}, we show the current X-ray constraints on the axion-photon coupling by
 the blue shaded regions
~\cite{Pospelov:2008jk,Boyarsky:2018tvu,Tremaine:1979we,Boyarsky:2008ju,Gorbunov:2008ka,Savchenko:2019qnn,Shi:1998km,Serpico:2005bc,Shaposhnikov:2008pf,Laine:2008pg,Canetti:2012kh,Graham:2013gfa,Redondo:2013lna,Fabbrichesi:2020wbt,Boyarsky:2007ge}. 
In particular, the bound from XMM-Newton~\cite{Foster:2021ngm} gives the strongest constraint in $5~{\rm keV}<m_a<16~{\rm keV}$.
Also shown are the future sensitivities of the Theseus XGIS-X (blue dashed line), Theseus SXI ( red dashed line), Athena (pink dashed line), and eROSITA (brown dashed line), which are taken from the projection limits in Refs~\cite{Morgan:2020rwc,Dekker:2021bos}.
As can be seen, most of the parameter points with $t_{\beta_1}>1.1$ are excluded by the current X-ray bounds. 
Furthermore, the region with $t_{\beta_1}>1.001$ can be probed by the future observations. 
This result can be understood from the analytical expression for $g_{a\gamma}$ of Eq.~\eqref{eq:gagamwD}. While the first term is suppressed by $m_a^2/m_e^2$, the second term is controlled by the quantity $\Delta$, which involves the effect of the mixing among the axion and the CP-odd Higgs bosons. 
In the case of $t_{\beta_1}=1$, $\Delta$ is highly suppressed as mentioned in Sec.~\ref{sec:model}, and thus the constraint from the X-ray can be evaded. On the other hand, if $\Delta$ is $\mathcal{O}(1)$, the corresponding  $g_{a\gamma}$ becomes too large to satisfy the current X-ray constraints. 
\red{
One can see that $g_{a \gamma}$ can be extremely small when $t_{\beta _1}$ is close to unity. This is because, as mentioned before, $g_{a \gamma}$  mainly receives two contributions with an opposite sign, and a cancellation could take place.
This should be contrasted to the axion-electron coupling, whose magnitude is always
determined by the ratio of the electron mass and decay constant.
Note that the apparent tight constraint on $t_{\beta _1}$ can be ameliorated by considering larger values of $t_{\beta_2}$ (see Appendix \ref{ap:axioncoup}).
}
In the following numerical calculations, we impose the current X-ray bounds in addition to the theoretical and experimental constraints discussed in Sec.~\ref{sec:con}. 
}

\red{\subsubsection*{Cosmological abundance of axion}}
We next discuss the production mechanism of the axion and its abundance. 
The axion can be produced through the misalignment mechanism and/or thermal production in the early Universe. 
Thermally produced axion with $\mathcal{O}$(1)~keV is regarded as warm DM. Hence, it suffers from bounds on galactic-scale structure formation, i.e., Lyman-$\alpha$ forest
observations~\cite{Viel:2005qj,Irsic:2017ixq,Kamada:2019kpe}, if it saturates all components of DM. 
To safely evade the constraint, we assume that the axion is primarily produced by the misalignment mechanism\cite{Preskill:1982cy,Abbott:1982af,Dine:1982ah}. 

When the Compton length of axion is larger than the Hubble scale, by the Hubble friction the axion is fixed at a certain field value $\theta_i(=a_i/f_a)$, which is called the initial misalignment angle. 
As the Universe cools down, the Hubble scale becomes comparable with the axion mass at a certain point.
Then, at  
\begin{equation}\label{eq:ALPosci}
  m_a\simeq 3H(T_{\rm osc})  
\end{equation}
the axion starts to oscillate around the potential minimum. 
Shortly thereafter the axion abundance gets fixed.
The temperature at the onset of the oscillation, $T_{\rm osc}$ can be estimated from Eq.~\eqref{eq:ALPosci} and the Hubble-temperature relation $H(T)=1.66 g_{\ast}^{1/2}T^2/m_{pl}$ as
\begin{align}
  T_{\rm osc}\sim \left(\frac{m_{pl} m_a}{4.98 g_{\ast}^{1/2}}\right)^{1/2}.
\end{align}
Then the density of the axion can be estimated as
\begin{align}
  \Omega_a=\frac{\rho_a(T)}{\rho_0}=\frac{m_a n_a(T_{\rm osc})}{\rho_a}\frac{g_s(T)}{g_s(T_{\rm osc})}\left(\frac{T}{T_{\rm osc}}\right)^3\;,
\end{align}
with the critical energy density $\rho_0=3m_{pl}^2H_0^2/(8\pi)$. 
The number density at $T_{\rm osc}$ reads
\begin{align}
  n_a(T_{\rm osc})=\frac{1}{2}m_af_a^2\theta_i^2 F(\theta_i)\;,
\end{align}
where $F(\theta_i)$ is the anharmonicity factor $F(\theta_i)=\ln(e/(1-\theta_i^2/\pi^2))^{7/6}$~\cite{Lyth:1991ub,Visinelli:2009zm}, with $e$ being Napier's constant, by which the effect of quartic couplings in the axion potential is included. 
The factor $F(\theta_i)$ affects the number density when the $\theta_i$ is not small. 
The cosmological abundance of the axion substantially depends on the initial condition for the axion fields, i.e., $\theta_i$, the mass $m_a$ and the decay constant $f_a$. 
Thus, the correct DM abundance $\Omega_{\rm DM}h^2\simeq 0.12$ can be explained by taking an appropriate value of $\theta_i$ for fixed $m_a$ and $f_a$. In particular, the initial angle is of order unity for $m_a = {\cal O}(1)$\,keV and $f_a = {\cal O}(10^{10})$\,GeV.

\begin{figure}[t!]
  \centering
  \includegraphics[scale=0.8]{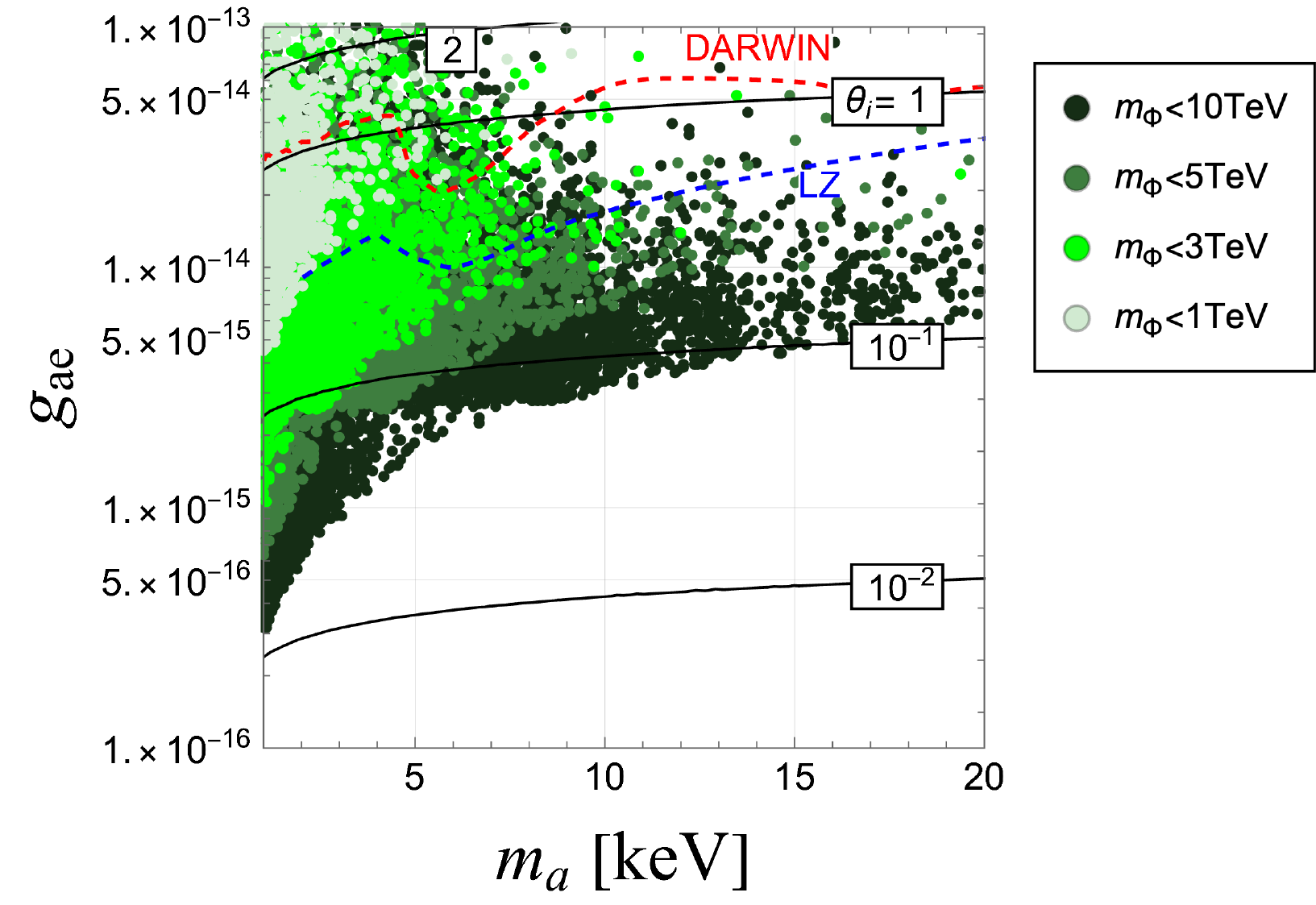}
  \caption{ 
  \red{The points satisfying all the theoretical and experimental constraints including the current X-ray bounds are shown.}
  The masses of the extra Higgs bosons are shown by different colors of the points. 
The contours of the initial misalignment angle for the axion to explain the observed DM abundance are shown by the thin solid black and dashed black lines,
\red{where we set $t_{\beta_1}=1$.}
\red{The projected limits of DARWIN~\cite{DARWIN:2016hyl} and LZ~\cite{LZ:2021xov} are shown by dashed lines. }
   }
  \label{FIG:ALPrelc}
 \end{figure}

The numerical results for $\theta_i$ to explain the observed DM abundance are shown in Fig.~\ref{FIG:ALPrelc} in the plane of the axion mass and the axion-electron coupling $g_{ae}$. 
Here the decay constant $f_a$ is converted to $g_{ae}$ through Eq.~\eqref{eq:gae}, and 
\red{$\tb{1}$ is chosen as $\tb{1}=1$ for the thin solid black line.}
Other parameters are set as $g_{\ast}= g_{s}(T_{\rm osc})=61.75,$ the Hubble parameter $H_0=(9.778{\rm Gyr})^{-1}h$ and the plank mass $m_{pl}=1.22\times 10^{19}$ GeV.
We use $g_{s}(T_0)=3.91$, $T_0\simeq 2.35\times 10^{-4}~{\rm eV}$ for the present photon temperature and effective number of entropic degree of freedom. 
One can see that the initial misalignment angle $\theta_i$ satisfying the observed  density of DM is ${\cal O}(1)$ when $g_{ae} = {\cal O}(10^{-14})$. 
For smaller $g_{ae}$, $\theta_i$ must be smaller than unity, which requires a mild fine-tuning. The  abundance scales as $\Omega_{a} \propto m_a^{1/2}g_{ae}^2$, so that it is not sensitive to $m_a$  compared with $g_{ae}$.


\red{\subsubsection*{Correlation between axion-electron coupling and extra Higgs boson masses}}
\begin{figure}[t!]
  \centering
  \includegraphics[scale=0.5]{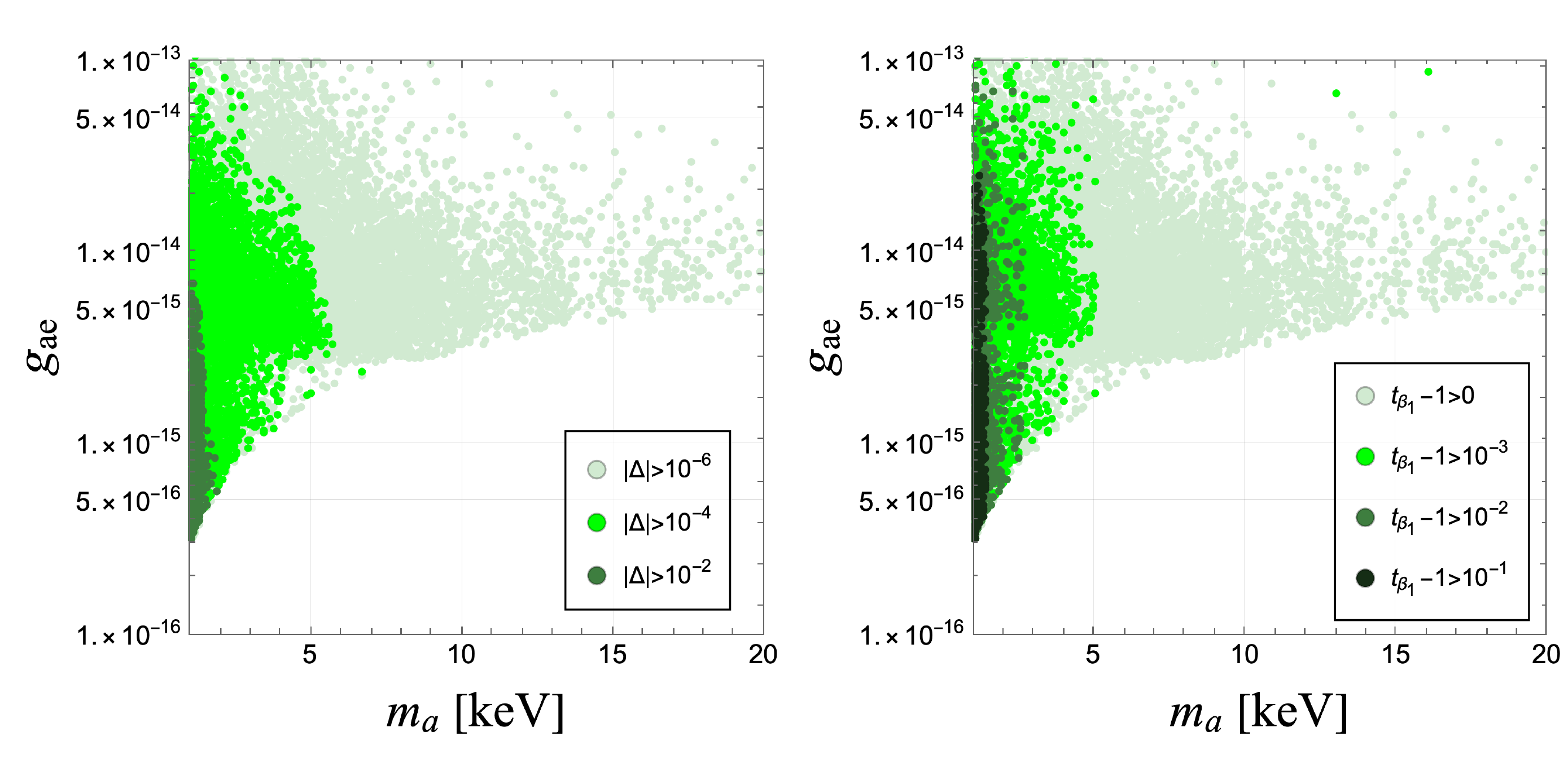}
  \caption{ 
  \red{
    We show parameter points that satisfy all the restrictions and could be further probed by future X-ray observations (see also Fig.~\ref{FIG:gagamXray}). 
  The magnitudes of $\Delta$ and $t_{\beta_1}$ are indicated by the different colors of the points in the left and right panels, respectively. 
}
   }
  \label{FIG:delgagam}
 \end{figure}
\red{For the evaluation of the mass of the extra Higgs bosons, we scan the parameters in the range given in Eq.\eqref{eq:scangae}, setting all the mass parameters equal, i.e., Eq.~\eqref{eq:degenemass}, and taking the alignment limit. }
In Fig.~\ref{FIG:ALPrelc}, the allowed parameter points satisfying all the constraints are also shown  by the green points \red{in the plane $(m_a,g_{ae})$}. 
The intensity of the color corresponds to the range of $m_{\Phi}$, i.e., $m_{\Phi}<1~{\rm TeV}$ (lightest green), $m_{\Phi}<3~{\rm TeV}$ (lighter green), $m_{\Phi}<5~{\rm TeV}$ (moss green), and $m_{\Phi}<10~{\rm TeV}$ (dark green). 
\red{We also show the future sensitivity of the direct DM search experiments, DARWIN~\cite{DARWIN:2016hyl} and LZ~\cite{LZ:2021xov}, by the dashed red and blue lines, respectively.}
\red{As can be seen from the figure, in the parameter region of $10{\rm keV}\lesssim m_{a} \lesssim 20{\rm keV}$ and $2\times 10^{-14}\lesssim g_{ae} \lesssim 1\times 10^{-13}$, there are few allowed parameter points.
This is because many parameter points in this region are excluded by the current limits of the X-ray observations. 
}
 The maximum value of $m_a$ is determined for a fixed $g_{ae}$ and $m_{\Phi}$, e.g. we obtain \green{$m_a\lesssim 15~{\rm keV}$} for $g_{ae}=5\times 10^{-15}$ and $m_{\Phi}=10~{\rm TeV}$. 
Since $m_a$ is inversely proportional to the  decay constant, the \red{maximum value of} $m_a$ decreases  as $g_{ae}$ becomes smaller. 

Interestingly, there is a correlation between 
the mass of the extra Higgs boson $m_{\Phi}$ and the axion-electron coupling $g_{ae}$ through the axion mass. 
Thus, one can obtain the information of the mass spectrum for the heavy Higgs bosons from the axion searches if it is detected or some anomalies are indicated \red{in the direct searches and/or future X-ray observations}. 
\red{For example}, unidentified X-ray line at around 3.5 keV was reported from observations of the galaxy clusters~\cite{Bulbul:2014sua,Boyarsky:2014jta} and galaxies~\cite{Boyarsky:2014jta,Boyarsky:2014ska} (also see the recent review in Ref.~\cite{Boyarsky:2018tvu}), which may be originated from the decay of axion  into photons~\cite{Higaki:2014zua,Jaeckel:2014qea}.
As one can see from  Fig.~\ref{FIG:ALPrelc}, if such a hint of axion is confirmed around \red{e.g., $(m_a,g_{ae})\sim (7~{\rm keV},7\times 10^{-14})~\mbox{or}~(2.5~{\rm keV},1\times 10^{-14})$}, the favored parameter region should be \red{$m_{\Phi}\gtrsim 3 ~\mbox{or}~1~{\rm TeV}$}. 
In other words, the lower bound on the mass of additional Higgs bosons can be derived from \red{the direct searches of the axion} and/or the X-ray observatories \red{in future}. 

\begin{figure}[t]
  \centering
  \includegraphics[scale=0.7]{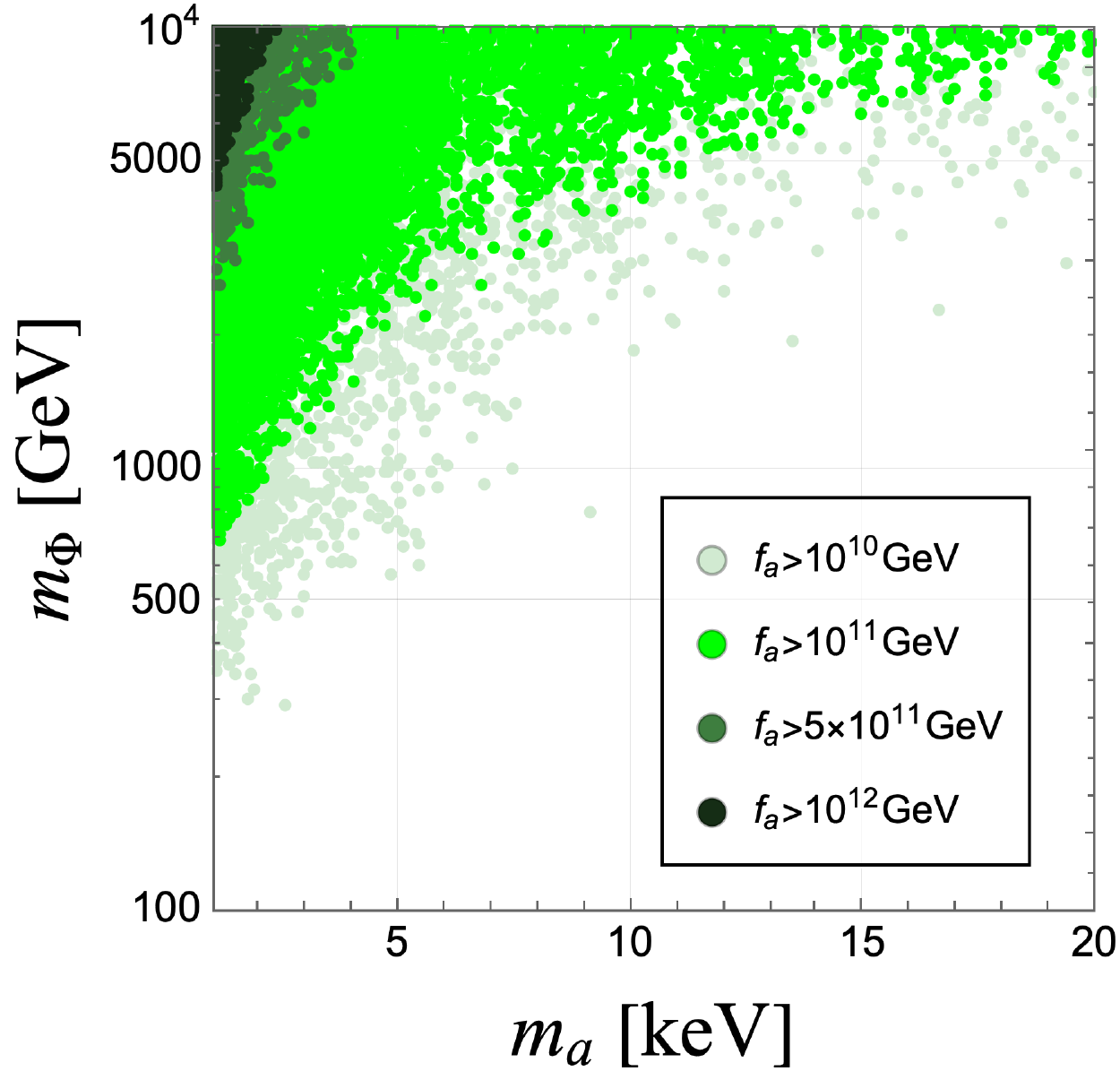}
    \caption{ Minimal value of the mass of heavy Higgs bosons that satisfy all constraint discussed in this paper is shown as a function of $m_a$, where the axion-electron coupling is fixed to $\green{g_{ae}\simeq 4\times 10^{-14}}$. 
  \green{Sensitivities of X-ray observations by Athena, Theseus XGIS-X, Theseus SXI, and eROSITA are denoted by magenta, blue, red and brown dashed lines.}
   }
  \label{FIG:ma_mphimin}
 \end{figure}

\red{
In order to reveal which parameter points are probed by the future X-ray observations, 
we show in~ Fig.\ref{FIG:delgagam} the parameter points that satisfy the
 current X-ray bounds and other theoretical and experimental limits and that are within the sensitivity reach of future X-ray observations
(this corresponds to the parameter points located below the blue solid lines and above dashed lines in Fig.~\ref{FIG:gagamXray}.). 
Intensity of color denotes the magnitudes  of $|\Delta|$  and $t_{\beta_1}$  in the left and right panels, respectively.
As can be seen from the left panel, if $m_a\gtrsim 5{\rm keV}$, one needs somewhat large tuning of $|\Delta|$, i.e., $10^{-6}\lesssim|\Delta|\lesssim 10^{-4}$. 
On the other hand, if $m_a\lesssim 5{\rm keV}$,  larger  $\Delta$ is possible, i.e., $|\Delta|\gtrsim 10^{-4} \mathchar`-\ 10^{-2}$.
Comparing the left and right panels, one can see that the dependence on $\tb{1}$ is similar to that of $|\Delta|$. 
This is because the quantity $\Delta$ is basically controlled by $\tb{1}$ and $M_{{12}}$ and $\tb{1}\sim 1$ means that the effect of mixing among the axion and the CP-odd Higgs bosons are small. 
Comparing Figs.~\ref{FIG:ALPrelc} and \ref{FIG:delgagam}, one can see that the distribution of points is almost the same, with only fewer points satisfying condition $|\Delta|>10^{-6}$ in Fig.~\ref{FIG:ALPrelc}.
Thus, one can conclude that most of the parameter points are surveyed by the future X-ray observations. This can be understood by noting that small $g_{a \gamma}$ is realized only when the two contributions are nearly canceled with each other.

Before closing this section, we show in Fig.~\ref{FIG:ma_mphimin} another example of the correlation between the mass of the additional Higgs bosons $m_{\Phi}$ and the axion mass. 
All the parameter points correspond to the ones within the reach of the future X-ray observations. 
Here, the intensity of the color represents different values of the decay constant $f_{a}$. 
Thus, if we fix $f_{a}$ and $m_{a}$, the minimal value of $m_{\Phi}$ is determined. 
The minimum value of $m_{\Phi}$ increases for heavier axion mass since it requires larger $M_{12}$ for a fixed $f_a$.
}

\subsection{Mixing angles}
\begin{figure}[t]
  \centering
  \includegraphics[scale=0.7]{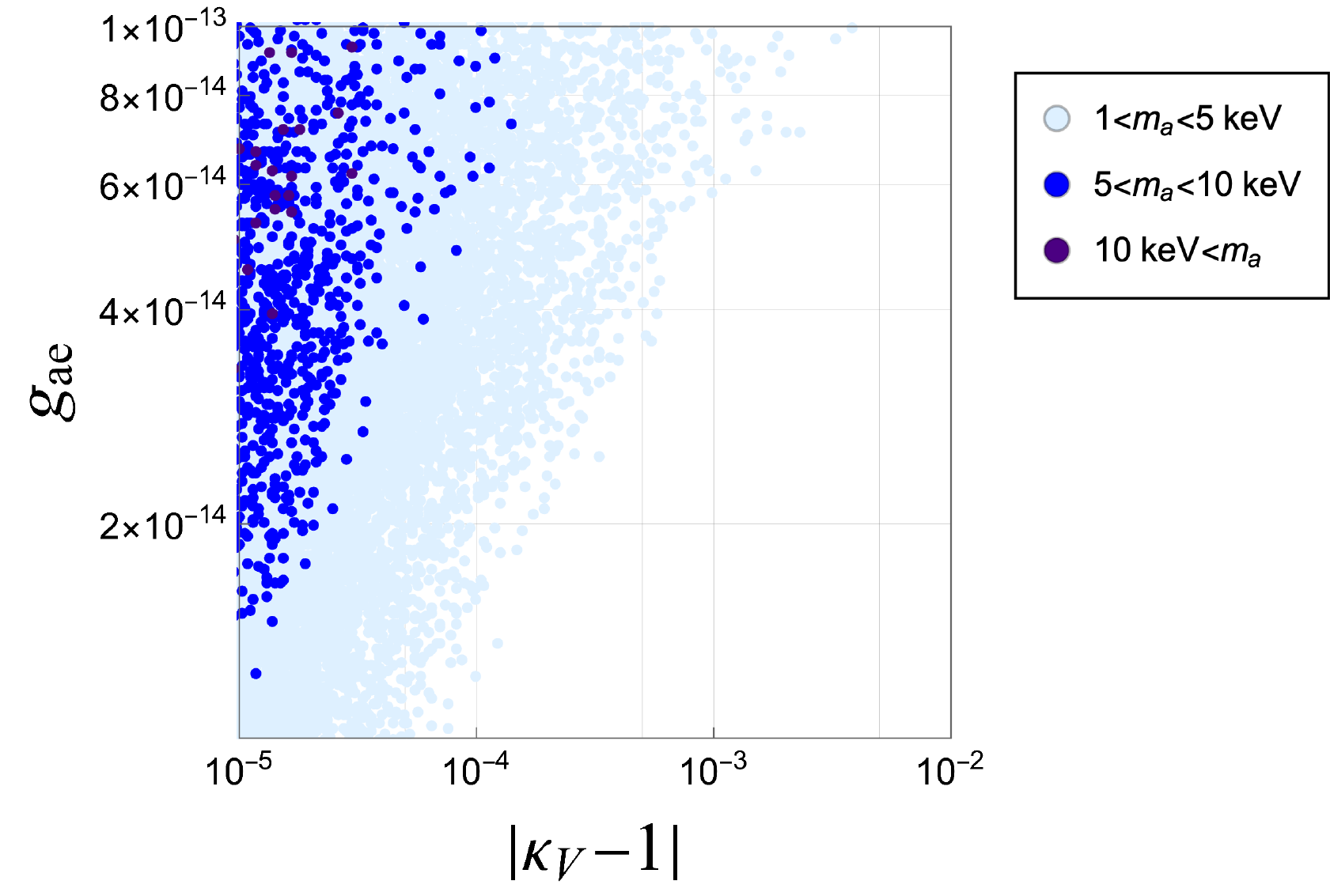}
  \caption{ Correlation between the axion electron coupling $g_{ae}$ and the deviation in the coupling constant of the SM-like Higgs boson with the weak gauge bosons $|\kappa_V-1|$, \red{where the all the constraint discussed in Sec.~\ref{sec:con} and and the current X-ray bounds are imposed}. 
  The intensity of the color denotes the range of axion mass; (dark blue): $10{\rm \,keV}<m_a$, (blue): $5<m_a<10 {\rm \, keV}$, and (light blue): $1<m_a<5 {\rm\, keV}$. 
   }
  \label{FIG:gaekappa}
 \end{figure}
 
  We also investigate the impact on the alignment parameters for the CP-even Higgs bosons, $\alpha_1$ and  $\alpha_2$ in case that the axion is detected (or indicated) in the future X-ray observatories. 
 To demonstrate this, we numerically evaluate the scaling factor for the weak gauge boson coupling of the SM-like Higgs boson $k^{H_1}_V$, which is defined in Eq.~\eqref{eq:kv}.
 The mixing angle are scanned in the range 
 \begin{align}
\alpha_{1}-\beta_1=\pm[10^{-5},1]\:,\quad
\alpha_{2}-\beta_2=\pm[10^{-5},1]\;.
 \end{align}
 The scan range for the other parameters is taken from Eq.~\eqref{eq:scangae}, assuming that all dimensionful parameters are degenerate. 

The numerical results for the correlation between the axion-electron coupling and the scaling factor $\kappa_V$ are shown in Fig.~\ref{FIG:gaekappa}, where the darkness of color corresponds to the range of axion mass;  
(dark blue): $10~{\rm keV}<m_a$, (blue): $5<m_a<10~{\rm keV}$, and (light blue): $1<m_a<5~{\rm keV}$. 
\red{We note that many parameter points with $m_{a}>10{\rm keV}$ are excluded by the constraint from the current X-ray observations. }
One sees that requiring axion mass to be larger than 10~keV makes almost alignment limit, i.e., 
$|\kappa_V-1|\lesssim \red{4}\times 10^{-5}$.
This consequence comes from the fact that the maximum size of the soft breaking parameters $M_{ij}^{(\prime)2}$ is limited by the constraint from the RG evolution of the scalar couplings as $\alpha_1$ and $\alpha_2$ deviate from the alignment limit. 
Also, smaller $g_{ae}$ makes the constraint tighter, so that the possible deviation of $|\kappa_V-1|$ becomes small.
\green{
We also numerically checked that the deviation over 1\% is difficult even if the axion mass is within the range of $ m_a\lesssim1$ keV.
This is mainly due to the lower bound of $m_\Phi > 200~{\rm GeV}$ in the parameter scan range Eq.~\eqref{eq:scangae}. 
Under the assumption the perturbativity for the running scalar couplings is satisfied, larger $|k_V-1|$ requires lighter extra Higgs bosons.
}

The axion with $m_a\simeq5~{\rm keV}$ can be probed by the future-X ray observatories such as eROSITA, Athena, and Theseus XGIS-X \red{and direct searches such as LZ, and DARWIN.} 
If there are some indications in the future observatories, 
one can set the upper bound for $\kappa_V-1$; e.g., 
\green{$\kappa_V-1\simeq \red{1} \times 10^{-2}\%$ $(2 \times 10^{-3}\%)$}
for $g_{ae}\simeq1\times 10^{-13}$ $(2\times 10^{-14})$.
The predicted deviations is not so large compare with the usual extended Higgs models~(see, e.g.,~Refs.\cite{Kanemura:2014bqa,Barklow:2017suo}). 
Hence, if one finds the deviations in $hVV$ coupling over 1\% in the future collider experiments, such as HL-LHC, ILC, CPEPC, and FCC, 
3HDM with B-L Higgs bosons can be ruled out. 

In this way, axion searches by the future X-ray observatories \red{and the direct detection} potentially have the impact on probing the extra Higgs bosons and the mixing parameters for the CP-even Higgs bosons, and eventually can narrow down the structure of the Higgs sector. 

\section{Conclusions}\label{sec:conclusion}
We have investigated phenomenological implications of the axion DM  based on the model with anomaly-free global flavor symmetry, which was originally proposed in Ref.~\cite{Nakayama:2014cza}. 
To build a concrete renormalizable model that includes the anomaly-free axion, we have considered the three Higgs doublet model with three $B-L$ Higgs fields, in which a global $U(1)_F$ flavor symmetry is imposed. 
In particular, we have focused on the axion with the mass of order ${\cal O}(1)$\,keV.
Such an axion DM scenario is promising because it can explain the reported excess in the electron recoil events of the XENON1T experiment, and because it can also be probed by the future X-ray observatories such as eROSITA, Athena, Theseus, and XRISM. 

\red{We have revealed that in this concrete model the axion-photon coupling involves the breaking of $U(1)_{F}$ and the mixing between the axion and the CP-odd Higgs bosons as seen in Eq.\eqref{eq:gagamfull}.
Thus, even for the anomaly-free axion, the anomalous coupling (the first term of \eqref{eq:gagamfull}) is not completely canceled out unless the effects of $U(1)_{F}$ breaking and the mixing are absent. 
This leads to the anomaly-free axion being more severely constrained by the X-ray observations depending on the model parameters as seen in Fig.~\ref{FIG:gagamXray}. To put it another way, future X-ray observation experiments will be able to detect anomaly-free axion more easily.  }

In order to investigate the possibility of narrowing the range of model parameters based on the nature of the axion (mass and coupling),
we have surveyed the mass spectrum of the axion and the extra Higgs bosons in the allowed parameter space satisfying the theoretical constraints and the experimental constraints given in Sec.~\ref{sec:con}. 
We have found that there are correlations among the ratio of the VEVs  $t_{\beta_2}$ and the mass of the extra Higgs bosons $m_{\Phi}$, provided that the axion has properties indicated by the XENON1T excess.  
As a result, the upper bound on $t_{\beta_2}$ is given depending on the scale of $m_{\Phi}$. 

We have also discussed the impact of the axion searches in the future X-ray observatories \red{and direct detection} on the extended Higgs sector. In particular, we have clarified
the correlation among the axion coupling with electron $g_{ae}$, the axion mass, and the extra Higgs boson mass.
We have revealed that the lower bound on the mass of the extra Higgs bosons can be obtained if the axion with a mass of order keV is indeed detected (or indicated) in the future X-ray observatories \red{and direct detection}. 
In addition, we have demonstrated that the axion searches by the X-ray observatories \red{and direct detection} can restrict the deviations of the SM-like Higgs boson couplings with the weak gauge bosons from the SM predictions.
Thus, the 3HDM with the $B-L$ Higgs fields can be probed by the synergy of the axion and extra Higgs boson searches.

\acknowledgments
The present work is supported by  JSPS Core-to-Core Program (grant number: JPJSCCA20200002) (F.T.), JSPS KAKENHI Grant Numbers 17H02878 (F.T.), 20H01894 (F.T. and K.S.), 20H05851 (F.T.), and 21K20363 (K.S.).

\appendix

\section{Relations for the parameters in the Higgs potential} \label{ap:lambda}

Scalar quartic couplings $\lambda_{i}$ $(i=1-9)$ and mass parameters $m^2_{ii}$ $(i=11,22,33)$ in the Higgs potential can be written in terms of the masses of Higgs bosons and the mixing angles.

Using the stationary conditions for the CP -even component fields of $\Phi_i$ $(i=1,2,3)$ and $S_m$ $(m=0,1,\bar{2})$, one can write mass parameters $m^2_{ii}$ in terms of other potential parameters as 
\begin{align}
\label{eq:m11}
    m_{11}^2&=
    \frac{M^{2}_{12} v_{2}^2}{v^2}+\frac{M^{\prime 2}_{13} v_{3}^2}{v^2}
    -\frac{1}{2}\left\{2\lambda_1v_1^2+v_2^2(\lambda_4+\lambda_7)+(\lambda_5+\lambda_8)v_{3}^2+\lambda_{10} \frac{v_{2} v_{3}^2}{v_1}\right\}
    \\ \notag
    &-\frac{1}{2} (\kappa_{0\phi 1} v_{S_0}^2+\kappa_{1\phi 1} v_{S_1}^2+\kappa_{\bar{2}\phi 1} v_{S_{\bar{2}}}^2)\;, \notag \\
    m_{22}^2&=
    \frac{M^{2}_{12} v_{1}^2}{v^2}+\frac{M^{\prime 2}_{23} v_{3}^2}{v^2}
    -\frac{1}{2}\left\{v_{1}^2 (\lambda_4+\lambda_7)+2 \lambda_{2} v_{2}^2+v_{3}^2 (\lambda_{6}+\lambda_9)+\lambda_{10} \frac{v_{1} v_{3}^2}{v_2} \right\}\;,
    \notag \\
    &-\frac{1}{2} (\kappa_{1\phi 2} v_{S_1}^2+\kappa_{\bar{2}\phi 2} v_{S_{\bar{2}}}^2)\;,  \\
    \label{eq:m33}
    m_{33}^2&= 
    \frac{ M^{\prime 2}_{13} v_{1}^2}{v^2}+\frac {M^{\prime 2}_{23} v_{2}^2}{v^2}
    -\frac{1}{2}\left\{2 \lambda_3 v_{3}^2+v_1^2(\lambda_5+\lambda_8)+v_{2}^2 (\lambda_{6}+\lambda_9) +2 \lambda_{10} v_{1} v_{2} \right\}
    \notag \\
    &-\frac{1}{2}( \kappa_{1\phi 3} v_{S_1}^2+\kappa_{\bar{2}\phi 3} v_{S_{\bar{2}}}^2)\;.
    \end{align}
Form the CP-even Higgs boson sector with Eq.~\eqref{e:msdiag}, $\lambda_{1}$ - $\lambda_{6}$ can be expressed as,
\begin{subequations}
	\label{e:lam1to6}
	\begin{eqnarray}
	\lambda_1  &=& \frac{m_{H_{1}}^2}{2 v^2} \frac{c^2_{\alpha_1} c^2_{\alpha_2} }{c^2_{\beta_1} c^2_{\beta_2}}
	+ \frac{m_{H_2}^2}{2 v^2 c^2_{\beta_1} c^2_{\beta_2} }  \left( c_{\alpha_1} s_{\alpha_2}  s_{\alpha_3} + s_{\alpha_1} c_{\alpha_3}  \right)^2
	+ \frac{m_{H_3}^2}{2 v^2 c^2_{\beta_1} c^2_{\beta_2}} \left(c_{\alpha_1}   s_{\alpha_2}  c_{\alpha_3} - s_{\alpha_1} s_{\alpha_3}  \right)^2 \nonumber \\
	&&+\frac{\tan\beta_1  \tan^{2} \beta_2}{4c^2_{\beta_1}}   \lambda_{10}   - \frac{M_{12}^2}{2 v^2} {t^2_{\beta_1}}
	-  \frac{1}{2 v^2}  \frac{t^2_{\beta_2}}{ c^2_{\beta_1}}M_{13}^{\prime 2}
\,, \\
	\lambda_2  &=& \frac{m_{H_{1}}^2}{2 v^2}\frac{s^2_{\alpha_1} c^2_{\alpha_2}  } {s^2_{\beta_1}c^2_{\beta_2}  }
	+\frac{m_{H_2}^2}{2 v^2 s^2_{\beta_1} c^2_{\beta_2}} \left(c_{\alpha_1} c_{\alpha_3} - s_{\alpha_1}s_{\alpha_2} s_{\alpha_3} \right)^2
	+\frac{m_{H_3}^2}{2 v^2 s^2_{\beta_1} c^2_{\beta_2}} \left(c_{\alpha_1} s_{\alpha_3} + s_{\alpha_1}s_{\alpha_2} c_{\alpha_3} \right)^2\nonumber \\
	&&+\frac{ \tan^{2}\beta_2}{4s^2_{\beta_1}\tan \beta_1 } \lambda_{10}  - \frac{M_{12}^2}{2 v^2} {\cot^2\beta_1}
	-  \frac{1}{2 v^2}  \frac{t^2_{\beta_2}}{ s^2_{\beta_1} }M_{23}^{\prime 2}
	\,, \\
	\lambda_3  &=& \frac{m_{H_{1}}^2}{2v^2} \frac{s^2_{\alpha_2}}{s^2_{\beta_2}}
	+\frac{m_{H_2}^2 c^2_{\alpha_2} s^2_{\alpha_3}}{2v^2 s^2_{\beta_2}}
	+\frac{m_{H_3}^2 c^2_{\alpha_2} c^2_{\alpha_3}}{2v^2 s^2_{\beta_2}}
	 - \frac{1}{2 v^2}  \frac{c^2_{\beta_1}}{ t^2_ {\beta_2}  }M_{13}^{\prime 2} 
	-  \frac{1}{2 v^2}  \frac{s^2_{\beta_1}}{t^2_{\beta_2}}M_{23}^{\prime 2}
	 \,,  \\
	\lambda_4  &=&
	\frac{1}{4v^2 s_{2\beta_1} c^2_{\beta_2}}\left[\left(m_{H_2}^2-m_{H_3}^2\right) \left\{(-3 + c_{2\alpha_2})s_{2\alpha_1}c_{2\alpha_3} - 4 c_{2\alpha_1} s_{\alpha_2} s_{2\alpha_3}\right\}
	-2\left(m_{H_2}^2+m_{H_3}^2\right)s_{2\alpha_1} c^2_{\alpha_2}  \right] \nonumber \\
	&&+\frac{m_{H_{1}}^2}{v^2}\frac{  s_{2\alpha_1}c^2_{\alpha_2}}{s_{2\beta_1} c^2_{\beta_2}}
	- \frac{\tan^{2}\beta_2}{ s_{2\beta_1}} \lambda_{10} -\lambda_{7}  + \frac{M_{12}^2}{v^2}   \,, \\
	\lambda_5 &=& \frac{m_{H_{1}}^2}{ v^2 } \frac{c_{\alpha_1} s_{2\alpha_2}}{c_{\beta_1} s_{2\beta_2}}
	-\frac{m_{H_2}^2}{v^2c_{\beta_1} s_{2\beta_2}} \left(c_{\alpha_1} s_{2\alpha_2} s^2_{\alpha_3}  + s_{\alpha_1}  c_{\alpha_2}s_{2\alpha_3} \right)
    +\frac{m_{H_3}^2}{v^2c_{\beta_1} s_{2\beta_2}} \left( s_{\alpha_1} c_{\alpha_2}  s_{2\alpha_3}-c_{\alpha_1} s_{2\alpha_2} c^2_{\alpha_3}\right)  \nonumber \\
	&& -\lambda_{10} \tan\beta_1- \lambda_8
	+ \frac{1}{v^2}  M_{13}^{\prime 2}
 \,, \\
	\lambda_6  &=&  \frac{ m_{H_{1}}^2 }{ v^2}\frac{s_{\alpha_1}s_{2\alpha_2} } {s_{\beta_1} s_{2\beta_2}}
	+ \frac{m_{H_2}^2 }{v^2}\frac{c_{\alpha_2}}{s_{\beta_1} s_{2\beta_2}} \left(-2s_{\alpha_1} s_{\alpha_2} s^2_{\alpha_3}  + c_{\alpha_1} s_{2\alpha_3} \right)
	- \frac{m_{H_3}^2}{v^2} \frac{c_{\alpha_2}} {s_{\beta_1}s_{2\beta_2} } \left(2s_{\alpha_1} s_{\alpha_2} c^2_{\alpha_3}  + c_{\alpha_1} s_{2\alpha_3} \right)\nonumber \\
	&&  -\lambda_{10} \cot\beta_1-\lambda_{9} + \frac{1}{ v^2} M_{23}^{\prime 2} 
	\,,
	\end{eqnarray}
\end{subequations}
On the other hand, one can get the following equations for $\lambda_{7}$ - $\lambda_{9}$ from the charged Higgs sector with Eqs.~\eqref{eq:mch1}-\eqref{eq:gammap},
\begin{subequations} \label{e:lam789}
	\begin{eqnarray}
 2\lambda_7v^2  &=&
 \frac{1}{c_{\beta_2}} (m_{H^\pm_1}^2-m_{H^\pm_2}^2) \left\{4 \cot{2 \beta_1} t_{\beta_2} s_{2 {\gamma_+}}+(c_{2 \beta_2}-3) \frac{1}{c_{\beta_2}} c_{2 {\gamma_+}}\right\} \notag \\
&+&4 M^{2}_{12}-2 (m_{H^\pm_1}^2+m_{H^\pm_2}^2)
  \,, \\
\lambda_8v^2  &=&
2 M^{\prime 2}_{13}-2 m_{H^\pm_1}^2 s_{\gamma_+}^2-2 m_{H^\pm_2}^2 c_{\gamma_+}^2+t_{\beta_1} \frac{s_{2 {\gamma_+}}}{s_{\beta_2}}  (m^2_{H^\pm_1}-m^2_{H^\pm_2}) -\lambda_{10} v^2 t_{\beta_1}\;,
  \\
\lambda_9v^2  &=&
2 M^{\prime 2}_{23}-2 m_{H^\pm_1}^2 s_{\gamma_+}^2-2 m_{H^\pm_2}^2 c_{\gamma_+}^2+\frac{1}{t_{\beta_1}} \frac{s_{2 {\gamma_+}}}{s_{\beta_2}} (m^2_{H^\pm_2}-m^2_{H^\pm_1}) -\lambda_{10} v^2\frac{1}{t_{\beta_1}} \,.
	\end{eqnarray}
\end{subequations}

We also get the following expressions for $\lambda_{1\mathchar`-6}$ in the alignment limit, $\alpha_{1}=\beta_{1},\ \alpha_{2}=\beta_{2}$,
\begin{align}
4\lambda_{1}v^{2}
&=2 t_{\beta_1}^2 \left\{\frac{1}{c_{\beta_2}^2} (m_{H_2}^2 c_{\alpha_3}^2+m_{H_3}^2 s_{\alpha_3}^2)-M^{2}_{12}\right\}\notag \\
&+t_{\beta_2}^2 \left\{\frac{1}{c^2_{\beta_1}} (\lambda_{10} v^2 t_{\beta_1}-2 M^{\prime 2}_{13})+2 m_{H_2}^2 s_{\alpha_3}^2+2 m_{H_3}^2 c_{\alpha_3}^2\right\} \notag \\
&+2 m_{H_1}^2+2 s_{2 \alpha_3} t_{\beta_1} t_{\beta_2} \frac{1}{c_{\beta_2}} (m_{H_2}^2-m_{H_3}^2) 
\;,\\
4\lambda_{2}v^{2}
&=2 \frac{1}{t^2_{\beta_1}} \left\{\frac{1}{c^2_{\beta_2}} (m_{H_2}^2 c_{\alpha_3}^2+m_{H_3}^2 s_{\alpha_3}^2)-M^{2}_{12}\right\} \notag \\
&+t_{\beta_2}^2 \left\{\frac{1}{s^2_{\beta_1}} (\lambda_{10} v^2 \cot{\beta_1}-2 M^{\prime 2}_{23})+2 m_{H_2}^2 s_{\alpha_3}^2+2 m_{H_3}^2 c_{\alpha_3}^2\right\} \notag \\
&+2 m_{H_1}^2-2 s_{2 \alpha_3} \frac{t_{\beta_2}}{t_{\beta_1}c_{\beta_2}} (m_{H_2}^2-m_{H_3}^2) 
\;, \\
2\lambda_{3}v^{2}
&=\frac{1}{t^2_{\beta_2}} (-M^{\prime 2}_{13} c_{\beta_1}^2-M^{\prime 2}_{23} s_{\beta_1}^2+m_{H_2}^2 s_{\alpha_3}^2+m_{H_3}^2 c_{\alpha_3}^2)+m_{H_1}^2
 \\
4\lambda_{4}v^{2}
&=-2 \left\{2 M^{2}_{12}-2 m_{H_1}^2+m_{H_2}^2+m_{H_3}^2-2 (m_{H^\pm_1}^2+m_{H^\pm_2}^2)\right\} \notag \\
&+4 \cot{2 \beta_1} t_{\beta_2} \frac{1}{c_{\beta_2}} \left\{s_{2 \alpha_3} (m_{H_3}^2-m_{H_2}^2)+2 s_{2 {\gamma_+}} (m_{H^\pm_2}^2-m_{H^\pm_1}^2)\right\}\notag \\
&+(c_{2 \beta_2}-3) \frac{1}{c_{\beta_2}^2}\left\{c_{2 \alpha_3} (m^2_{H_2}-m^2_{H_3}) +2 c_{2 {\gamma_+}} (m_{H^\pm_2}^2-m_{H^\pm_1}^2)\right\}-2 \lambda_{10} v^2  \frac{t_{\beta_2}^2}{s_{\beta_1}c_{\beta_1}} 
\;, \\
\lambda_{5}v^2
&=-M^{\prime 2}_{13}+m_{H_1}^2-m_{H_2}^2 s_{\alpha_3}^2-m_{H_3}^2 c_{\alpha_3}^2+2 m_{H^\pm_1}^2 s_{\gamma_+}^2+2 m_{H^\pm_2}^2 c_{\gamma_+}^2\notag \\
&+ \frac{t_{\beta_1}}{s_{\beta_2}} \left\{s_{\alpha_3} c_{\alpha_3} (m_{H_3}^2-m_{H_2}^2)+s_{2 {\gamma_+}} (m_{H^\pm_2}^2-m_{H^\pm_1}^2)\right\}
\;,\\
\lambda_{6}v^{2}
&=-M^{\prime 2}_{23}+m_{H_1}^2-m_{H_2}^2 s_{\alpha_3}^2-m_{H_3}^2 c_{\alpha_3}^2+2 m_{H^\pm_1}^2 s_{\gamma_+}^2+2 m_{H^\pm_2}^2 c_{\gamma_+}^2 \notag \\
&+\frac{1}{t_{\beta_1}s_{\beta_2}}
\left\{s_{\alpha_3} c_{\alpha_3}   (m_{H_2}^2-m^2_{H_3})+ s_{2 {\gamma_+}} (m^2_{H^\pm_1}-m^2_{H^\pm_2})\right\} 
\;.
\end{align}

\section{RGE $\beta$ functions for 3HDM}\label{ap:RGE}
We here give the $\beta$ functions at one-loop in the 3HDM. All of them are evaluated by using {\tt SARAH}~\cite{Staub:2009bi,Staub:2010jh,Staub:2012pb,Staub:2013tta}. 
Depending on the Type of Yukawa interactions, i.e., Type-A or Type-B, the contributions from the lepton Yukawa couplings are changed \green{(see Eq.~\eqref{eq:Yukawatype})}. 

The beta function for the gauge boson couplings is given by

\begin{align}
\beta_{g_1}^{(1)} & =
\frac{43}{10} g_{1}^{3}\;, \\
\beta_{g_2}^{(1)} & =
-\frac{17}{6} g_{2}^{3}\;, \\
\beta_{g_3}^{(1)} & =
-7 g_{3}^{3} \;,
\end{align}
  where $g_{3}=g_{s}$, $g_{2}=g$ and $g_{1}=C g^{\prime}$ with the Clebsch-Gordan coefficient $C^{2}=5/3 $.
\begin{align}
\beta_{\lambda_1}^{(1)} & =
+\frac{27}{200} g_{1}^{4} +\frac{9}{20} g_{1}^{2} g_{2}^{2} +\frac{9}{8} g_{2}^{4} -\frac{9}{5} g_{1}^{2} \lambda_1 -9 g_{2}^{2} \lambda_1 +24 \lambda_{1}^{2} +2 \lambda_{4}^{2} +2 \lambda_{5}^{2} +2 \lambda_4 \lambda_7 +\lambda_{7}^{2}+2 \lambda_5 \lambda_8 +\lambda_{8}^{2}\nonumber \\
 &+4 \lambda_1 y_{\ellp}^{2} -2 y_{\ellp}^{4}\;,  \\
 \beta_{\lambda_2}^{(1)} & =
+\frac{27}{200} g_{1}^{4} +\frac{9}{20} g_{1}^{2} g_{2}^{2} +\frac{9}{8} g_{2}^{4} -\frac{9}{5} g_{1}^{2} \lambda_2 -9 g_{2}^{2} \lambda_2 +24 \lambda_{2}^{2} +2 \lambda_{4}^{2} +2 \lambda_{6}^{2} +2 \lambda_4 \lambda_7 +\lambda_{7}^{2}+2 \lambda_6 \lambda_9 +\lambda_{9}^{2}\nonumber \\
 &+4 \lambda_2y_{e}^{2} -2y_{e}^{4}\;, \\
 \beta_{\lambda_3}^{(1)} & =
+\frac{27}{200} g_{1}^{4} +\frac{9}{20} g_{1}^{2} g_{2}^{2} +\frac{9}{8} g_{2}^{4} -\frac{9}{5} g_{1}^{2} \lambda_3 -9 g_{2}^{2} \lambda_3 +24 \lambda_{3}^{2} +2 \lambda_{5}^{2} +2 \lambda_{6}^{2} +2 \lambda_5 \lambda_8 +\lambda_{8}^{2}+2 \lambda_6 \lambda_9 +\lambda_{9}^{2}\nonumber \\
 &+2 |\lambda_{10}|^2 +12 \lambda_3 y_{b}^{2} +4 \lambda_3 y_{\ell}^{2}
 +12 \lambda_3y_{t}^{2}  -6 y_{b}^{4}
 -2y_{\ell}^{4}  -6 y_{t}^{4}\;, \\
 \beta_{\lambda_4}^{(1)} & =
+\frac{27}{100} g_{1}^{4} -\frac{9}{10} g_{1}^{2} g_{2}^{2} +\frac{9}{4} g_{2}^{4} -\frac{9}{5} g_{1}^{2} \lambda_4 -9 g_{2}^{2} \lambda_4 +12 \lambda_1 \lambda_4 +12 \lambda_2 \lambda_4 +4 \lambda_{4}^{2} +4 \lambda_5 \lambda_6 +4 \lambda_1 \lambda_7 +4 \lambda_2 \lambda_7 \nonumber \\
 &+2 \lambda_{7}^{2} +2 \lambda_6 \lambda_8 +2 \lambda_5 \lambda_9 +2 |\lambda_{10}|^2 +2 \lambda_4 y_{e}^{2} +2 \lambda_4 y_{\ellp}^{2}\;, \\
 \beta_{\lambda_5}^{(1)} & =
+\frac{27}{100} g_{1}^{4} -\frac{9}{10} g_{1}^{2} g_{2}^{2} +\frac{9}{4} g_{2}^{4} -\frac{9}{5} g_{1}^{2} \lambda_5 -9 g_{2}^{2} \lambda_5 +12 \lambda_1 \lambda_5 +12 \lambda_3 \lambda_5 +4 \lambda_{5}^{2} +4 \lambda_4 \lambda_6 +2 \lambda_6 \lambda_7 +4 \lambda_1 \lambda_8 \nonumber \\
 &+4 \lambda_3 \lambda_8 +2 \lambda_{8}^{2} +2 \lambda_4 \lambda_9 +2 |\lambda_{10}|^2 +6 \lambda_5 y_{b}^{2}
 +2 \lambda_5 y_{\ellp}^{2} +2 \lambda_5 y_{\ell}^{2} +6 \lambda_5 y_{t}^{2} \;, \\
 \beta_{\lambda_6}^{(1)} & =
+\frac{27}{100} g_{1}^{4} -\frac{9}{10} g_{1}^{2} g_{2}^{2} +\frac{9}{4} g_{2}^{4} +4 \lambda_4 \lambda_5 -\frac{9}{5} g_{1}^{2} \lambda_6 -9 g_{2}^{2} \lambda_6 +12 \lambda_2 \lambda_6 +12 \lambda_3 \lambda_6 +4 \lambda_{6}^{2} +2 \lambda_5 \lambda_7 +2 \lambda_4 \lambda_8 \nonumber \\
 &+4 \lambda_2 \lambda_9 +4 \lambda_3 \lambda_9 +2 \lambda_{9}^{2} +2 |\lambda_{10}|^2 +6 \lambda_6 y_b^2 +2 \lambda_6 y_e^2  +2 \lambda_6 y_{\ell}^2 +6 \lambda_6y_t^2 \;,
  \\
 \beta_{\lambda_7}^{(1)} & =
+\frac{9}{5} g_{1}^{2} g_{2}^{2} -\frac{9}{5} g_{1}^{2} \lambda_7 -9 g_{2}^{2} \lambda_7 +4 \lambda_1 \lambda_7 +4 \lambda_2 \lambda_7 +8 \lambda_4 \lambda_7 +4 \lambda_{7}^{2} +2 \lambda_8 \lambda_9 +2 |\lambda_{10}|^2 \nonumber \\
 &+2 \lambda_7 y_e^2 +2 \lambda_7 y_{\ellp}^2 \;, 
 \\ 
 \beta_{\lambda_8}^{(1)} & =
+\frac{9}{5} g_{1}^{2} g_{2}^{2} -\frac{9}{5} g_{1}^{2} \lambda_8 -9 g_{2}^{2} \lambda_8 +4 \lambda_1 \lambda_8 +4 \lambda_3 \lambda_8 +8 \lambda_5 \lambda_8 +4 \lambda_{8}^{2} +2 \lambda_7 \lambda_9 +6 \lambda_8 y_b^2 \nonumber \\
 & + 8 |\lambda_{10}|^2+2 \lambda_8 y_{\ellp}^2 +2 \lambda_8 y_{\ell}^2 +6 \lambda_8y_t^2 
  \;,   \\
 \beta_{\lambda_9}^{(1)} & =
+\frac{9}{5} g_{1}^{2} g_{2}^{2} +2 \lambda_7 \lambda_8 -\frac{9}{5} g_{1}^{2} \lambda_9 -9 g_{2}^{2} \lambda_9 +4 \lambda_2 \lambda_9 +4 \lambda_3 \lambda_9 +8 \lambda_6 \lambda_9 +4 \lambda_{9}^{2} +6 \lambda_9 y_b^2 \nonumber \\
 &  +2 \lambda_9 y_e^2 + 8 |\lambda_{10}|^2 
 +2 \lambda_9 y_{\ell}^2 +6 \lambda_9y_t^2 \;, \\
 \beta_{\lambda_{10}}^{(1)} & =
-\frac{9}{5} g_{1}^{2} \lambda_{10} -9 g_{2}^{2} \lambda_{10} +4 \lambda_{10} \lambda_3 +2 \lambda_{10} \lambda_4 +4 \lambda_{10} \lambda_5 +4 \lambda_{10} \lambda_6 +2 \lambda_{10} \lambda_7 +6 \lambda_{10} \lambda_8 +6 \lambda_{10} \lambda_9  \nonumber \\
 & +6 \lambda_{10} y_b^2+\lambda_{10} y_e^2 +\lambda_{10} y_{\ellp}^2  +2 \lambda_{10} y_{\ell}^2 +6 \lambda_{10}y_t^2\;,
\end{align}
\begin{align}
\beta_{y_t}^{(1)} & =
-\frac{3}{2} \Big(- {y_t ^{3}}  + {y_t   y_b^{2}}\Big)\nonumber \\
 &+y_t \Big(-\frac{17}{20} g_{1}^{2} -\frac{9}{4} g_{2}^{2} -8 g_{3}^{2} +3 y_b^2 +y_{\ell}^2+3y_t^2 \Big) \;,\\
 \beta_{y_b}^{(1)} & =
+\frac{3}{2} \Big(- {y_b  y_t^{2}}  + {y_b^{3}}\Big)\nonumber \\
 &+y_b \Big(-\frac{1}{4} g_{1}^{2} -\frac{9}{4} g_{2}^{2} -8 g_{3}^{2} +3 y_b^2 +y_{\ell}^2 +3y_t^2 \Big)\;, 
 \end{align}
\begin{align}
 \beta_{y_{\ell}}^{(1)} & =
\frac{3}{2} {y_{\ell}^{3} }  + y_{\ell} \Big(3 y_b^2  + 3y_t^2  -\frac{9}{4} g_{1}^{2}  -\frac{9}{4} g_{2}^{2}  +y_{\ell}^{2}\Big)\;,\\
\beta_{y_{\ellp}}^{(1)} & =
\frac{1}{4} \Big(6 {y_{\ellp} ^{3} }  + y_{\ellp} \Big(4y_{\ellp}^{2}  -9 \Big(g_{1}^{2} + g_{2}^{2}\Big)\Big)\Big)\;, \\
\beta_{y_e}^{(1)} & =
\frac{1}{4} \Big(6 {y_e ^{3} }  + y_e \Big(4y_{e}^{2}  -9 \Big(g_{1}^{2} + g_{2}^{2}\Big)\Big)\Big)\;,
\end{align}

The running coupling constant for an any coupling $X$ is derived from the RGE equation
\begin{align}
\frac{dX}{d\log\mu}=\frac{1}{16\pi^{2}}\beta_{X}^{(1)}.
\end{align}
For the gauge boson couplings, the RGE equation can be analytically solved
as
\begin{align}
g_{i}[\log\mu]=g_{i}\left(1-2g_{i}^{2}\frac{1}{(16\pi^{2})}c_{g_{i}}^{(1)} \log(\mu/\mu_{0}) \right)^{-\frac{1}{2}}
\quad (i=1,2,3)\;,
\end{align}
where $c_{g_{i}}^{(1)}=\beta_{g_{i}}^{(1)}/g_{i}^{3}$ and the initial scale as defined by $\mu_{0}$.
We take $\mu_{0}=m_{Z}$ in the numerical calculations.

\section{Electroweak oblique parameters}\label{ap:ST}
We derive here the formulae for the electroweak $S$ and $T$ parameters in the 3HDM with B-L Higgs bosons.
To this end, we first evaluate scalar boson loop contributions to the transverse part of the gauge boson two-point functions.
For the weak gauge bosons, they are written in terms of the Passarino-Veltman functions~\cite{Passarino:1978jh} as
\begin{align}\label{eq:piWWS}
(16\pi^{2})\Pi_{WW}^{S,\; 3HDM}&=\frac{g^{2} } {4}\
\Bigg[
\sum_{i=1}^3\sum_{j=1}^3(R_{S}R_{+}^{\dagger})^{2}_{ij}B_{5}(q^{2};m^{}_{H^{}_{i}},m^{}_{{\cal H}^{\pm}_{j}})
+\sum_{i=1}^4\sum_{j=1}^3(R^{z}_{P}R_{+}^{\dagger})^{2}_{ij}B_{5}(q^{2};m^{}_{{\cal A}_{i}},m^{}_{H_{j}})
 \notag \\
&+4m_{W}^{2}\sum_{i=1}^3\Big(\sum_{j=1}^3\frac{v_{j}}{v}(R_{S}^{\dagger})_{ji}\Big)^{2}B_{0}(q^{2};m^{}_{H_{i}},m^{}_{W}) \notag \\
&+\sum_{i=2}^3\left\{1-(R_{P}^{z}R_{P}^{z \dagger} )_{ii}\right\}A(m^{}_{{\cal A}^{}_{i}})
+(R_{P}^{z}R_{P}^{z \dagger} )_{44}A(m^{}_{{ a}_{}})\Bigg], \\ \label{eq:piZZS}
(16\pi^{2})\Pi_{ZZ}^{S,\; 3HDM}=&\frac{g^{2}_{Z} } {4}\
\Bigg[
c_{2W}^{2}\sum^3_{i=1}B_{5}(q^{2};m^{}_{{\cal H}^{\pm}_{i}},m^{}_{{\cal H}^{\pm}_{i}})
+\sum_{i=1}^3\sum_{j=1}^4(R_{S}R_{p}^{z\dagger})^{2}_{ij}B_{5}(q^{2};m^{}_{H_{i}},m^{}_{{\cal A}_{j}}) \notag \\
&+4m_{Z}^{2}\sum_{i=1}^3\Big(\sum_{j=1}^3\frac{v_{j}}{v}(R_{S}^{\dagger})_{ji}\Big)^{2}B_{0}(q^{2};m^{}_{H_{i}},m_{Z})\notag \\
&+\sum_{i=2}^3\left\{1-(R_{P}^{z}R_{P}^{z \dagger} )_{ii}\right\}A(m^{}_{{\cal A}^{}_{i}})
+(R_{P}^{z}R_{P}^{z \dagger} )_{44}A(m^{}_{{ a}_{}})
\Bigg]
\end{align}
where $R_{P}^{z}$ is 4 $\times$ 3 matrix for the CP -odd scalars ${\cal A}_{i}$,i.e.,
\begin{equation}
\begin{pmatrix}
G^{0}\\
A_{1}\\
A_{2}\\
a
\end{pmatrix}=
R_{p}^{z}
\begin{pmatrix}
z_{1}\\
z_{2}\\
z_{3}\\
\end{pmatrix}.
\end{equation}
It can be extracted from the original $4\times 4$ orthogonal matrix $R_{P}$, removing the 4th column vector, and satisfies  $(R_{p}^{z})^{T}R_{p}^{z}=I_{3\times 3}$ while $R_{p}^{z}(R_{p}^{z})^{T}\neq I_{4\times 4}$.
The parameter $m_{{\cal A}_j}$ denotes the mass for the CP-odd scalar bosons, i.e.,$(m_{{\cal A}_1},m_{{\cal A}_2},m_{{\cal A}_3},m_{{\cal A}_4})=(m_{G^0},m_{A_1},m_{A_2},m_{a})$.
\green{$m_{{\cal H}^\pm_j}$ is the mass for the charged scalar bosons, $(m_{{\cal H}^\pm_1},m_{{\cal H}^\pm_2},m_{{\cal H}^\pm_3})=(m_{G^\pm},m_{H_1^\pm},m_{H_2^\pm})$.}
\green{We note that the last two terms in Eqs.\eqref{eq:piWWS} and \eqref{eq:piZZS} vanish in the limit of $\gamma_{2,3}\to 0$. Thus, hereafter, we just drop them. }

To extract new physics contributions to the oblique parameters, we need to subtract the SM contributions from $\Pi_{WW,ZZ}^{S}$.
It is written by
\begin{align}
(16\pi^{2})\Pi_{WW}^{S,\;SM}=\frac{g^{2} }{4}\Big[B_{5}(q^{2};m_{h},m_{G^{\pm}})
+B_{5}(q^{2};m_{G^{0}},m_{G^{\pm}})
+4m_{W}^{2}B_{0}(q^{2};m_{h},m_{W})\Big]\;, \\
(16\pi^{2})\Pi_{ZZ}^{S,\;SM}=\frac{g^{2}_{Z} }{4}\Big[c_{2W}^{2}B_{5}(q^{2};m_{G^{\pm}},m_{G^{\pm}})
+B_{5}(q^{2};m_{h},m_{G^{0}})
+4m_{Z}^{2}B_{0}(q^{2};m_{h},m_{Z})\Big]\;,
\end{align}
with $g_Z=(2m_Z)/v$, 
\green{where $h$ denotes the Higgs boson with the mass of 125 GeV in the SM.}
Thus, new physics contributions are expressed by
\begin{equation}
\Pi_{WW}^{S}= \Pi_{WW}^{S,\;3HDM}-\Pi_{WW}^{S,\;SM}\;, \quad
\Pi_{ZZ}^{S}= \Pi_{ZZ}^{S,\;3HDM}-\Pi_{ZZ}^{S,\;SM}\;.
\end{equation}

For the photon-photon two two-point functions and the photon and Z boson mixing two-point function,
the new physics contributions just stem from the  charged Higgs boon loops as
\begin{align}
(16\pi^{2})\Pi_{Z\gamma}^{S}&=\frac{e g_{Z}}{2}
\Big[
B_{5}(q^2;m_{H^{\pm}_{1}},m_{H^{\pm}_{1}})+B_{5}(q^2;m_{H^{\pm}_{2}},m_{H^{\pm}_{2}})
\Big]-(16\pi^{2})\frac{s_{W}}{c_{W}}\Pi_{\gamma\gamma}^{S}(q^{2}),\ \\
(16\pi^{2})\Pi_{\gamma\gamma}^{S}&=e^{2}
\Big[
B_{5}(q^2;m_{H^{\pm}_{1}},m_{H^{\pm}_{1}})+B_{5}(q^2;m_{H^{\pm}_{2}},m_{H^{\pm}_{2}})
\Big].
\end{align}

We use the following definition of the electroweak oblique parameters \cite{Hagiwara:1994pw},
\begin{align}
  \label{eq:Spara}
\frac{1}{16\pi}S&=
\frac{1}{m_Z^2}\left\{\Pi_T^{3Q}(m_Z^2)-\Pi_T^{3Q}(0)\right\}
-\frac{1}{m_Z^2}\left\{\Pi_T^{33}(m_Z^2)-\Pi_T^{33}(0)
\right\}\;,\\
\frac{\alpha_{\rm em}}{4\sqrt{2}G_F}T&=
\Pi_T^{33}(0)-\Pi_T^{11}(0)\;, \\
\label{eq:Upara}
\frac{1}{16\pi}U&=
\frac{1}{m_Z^2}\left\{\Pi_T^{33}(m_Z^2)-\Pi_T^{33}(0)\right\}
-\frac{1}{m_W^2}\left\{\Pi_T^{11}(m_W^2)-\Pi_T^{11}(0)
\right\}\;,
\end{align}
where the functions $\Pi^{AB}_T$ is defined by
\begin{align}
  \Pi_{WW}&=g^2\Pi^{11}_T(q^2)\;,\\
  \Pi_{\gamma\gamma}&=e^2\Pi^{QQ}_T(q^2)\;, \\
  \Pi_{Z\gamma}&=e g_Z\left\{\Pi^{3Q}_T(q^2)-s^2_W\Pi^{QQ}_T(q^2) \right\}\;, \\
  \Pi_{ZZ}&
  =g_Z^2\left\{\Pi^{33}_T(q^2)-2s_W^2\Pi^{3Q}_T(q^2)+s^4_W\Pi^{QQ}_T(q^2) \right\}\;.
\end{align}
In the expressions of Eqs.~\eqref{eq:Spara}-\eqref{eq:Upara}, real part for the $\Pi^{AB}_T(q^2)$ are taken and the relevant pinch terms are added. 
These yield the concrete expressions for the oblique parameters to the new physics contributions in the 3HDM, 
\begin{align}\label{eq:DeltaS}
  \Delta S&=\frac{1}{4\pi}\Bigg[
-\Delta B_5(m_Z^2;m_{H^\pm_1},m_{H^\pm_1})  
-\Delta B_5(m_Z^2;m_{H^\pm_2},m_{H^\pm_2})  \notag\\
&+\sum_{i=1}^3\sum_{j=1}^4(R_S({R}^{z}_P)^\dagger)^2_{ij}\Delta B_5(m_Z^2;m_{H_i},m_{{\cal A}_j})\notag \\
&+4{m_Z^2}\sum_{i=1}^3\left(\sum_{k=1}^3 {\frac{v_k}{v}(R_S^\dagger)_{ki}}\right)^2\Delta B_0(m_Z^2;m_{H_i},m_Z) \notag \\
&-\Delta B_5(m_Z^2;m_{h},m_{G^0})
-4m_Z^2\Delta B_0(m_Z^2;m_{h},m_Z)
\Bigg]\;,
\end{align}
\begin{align}\label{eq:DeltaT}
\Delta T&=\frac{\sqrt{2}G_F}{\alpha_{\rm em}}\frac{1}{16\pi^2}\Bigg[
  \sum_{i=1}^3\sum_{j=1}^4(R_S(R^z_P)^\dagger)^2_{ij}B_5(0;m_{H_i},m_{{\cal A}_j}) \notag \\
  &+4{m_Z^2}\sum_{i=1}^3\left(\sum_{k=1}^3 {\frac{v_k}{v}(R_S^\dagger)_{ki}}\right)^2B_0(0;m_{H_i},m_Z) \notag \\
  &-B_5(0;m_{h},m_{G^0})-4m_Z^2B_0(0;m_{h},m_Z) \notag \\
  &-\sum_{i=1}^3\sum_{j=1}^3(R_SR^\dagger_+)^2_{ij}B_5(0;m_{{ H}_i},m_{{\cal H}^\pm_j})-\sum_{i=1}^4\sum_{j=1}^3(R^z_PR^\dagger_+)^2_{ij}B_5(0;m_{{\cal A}_i},m_{{\cal H}^\pm_j}) \notag \\
  &-4{m_W^2}\sum_{i=1}^3\left(\sum_{k=1}^3 {\frac{v_k}{v}(R_S^\dagger)_{ki}}\right)^2B_0(0;m_{H_i},m_W) \notag \\
  &+B_5(0,m_{h},m_{G^\pm})+B_5(0,m_{G^0},m_{G^\pm})
  +4m_W^2B_0(0,m_{h},m_{W})
  \Bigg]\; .
\end{align}

\section{The analytic expressions for the ${B}_s$-$\bar{B}_s$ mixing }\label{ap:DMs}
We here give the analytical expressions which are relevant to the mass difference $\Delta M_s$ for the mass eigenstates in the ${B}_s$-$\bar{B}_s$ system, i.e., Eq.~\eqref{eq:delMs} in the main text. 

The effective Lagrangian is written by 
\begin{align}
\mathcal{L}_{\rm eff}=\frac{2G_F^2m_W^2}{16\pi^2}
C_{VLL}\mathcal{O}_{VLL}
+{\rm h.c.}
\end{align}
with the operator
\begin{equation}
{O}_{VLL}=\bar{s}^\alpha \gamma_\mu (1-\gamma_5)b^\alpha \bar{s}^\beta \gamma^\mu (1-\gamma_5) b^{\beta}.
\end{equation}
The Wilson coefficient is separated into three parts,
\begin{align}
C_{VLL}=C_{VLL}^{WW}+2\sum_{i=1,2}C_{VLL}^{WH^\pm_i}+\sum_{i=1,2}\sum_{j=1,2}C_{VLL}^{H^\pm_iH^\pm_j},
\end{align}
where $C_{VLL}^{WW}$ corresponds to the contribution from the box diagram with two virtual $W$ bosons, which is the same with SM contributions~\cite{Inami:1980fz,Buras:1990fn}. 
The second (third) term comes from the box diagrams with virtual $W$ boson and charged Higgs bosons (two virtual charged Higgs bosons). 
The analytical expressions for each coefficient are given by
\begin{align}
C_{VLL}^{WW}&=
\frac{x_t}{4} \left[1+\frac{9}{1-{x_t}}-\frac{6}{(1-{x_t})^2}-\frac{6 x_t^2 \log
({x_t})}{(1-{x_t})^3}\right]\;, \\
C_{VLL}^{WH^\pm_i}&= (\xi^{q}_{H^\pm_i})^2
\frac{ x_t^2}{4}\Bigg[
-\frac{{x_t}-4}{({x_t}-1) ({x_{H_i^\pm}}-{x_t})}
+\frac{({x_{H_i^\pm}}-4) {x_{H_i^\pm}} \log ({x_{H_i^\pm}})}{({x_{H_i^\pm}}-1) ({x_{H_i^\pm}}-{x_t})^2}\;,
\notag \\
&+\frac{-3x_t^2+x_{H^\pm_i}(x_t^2-2x_t+4) }{(x_{t}-1)^2 ({x_{H_i^\pm}}-{x_t})^2}\log ({x_t})
\Bigg]\;, \\
C_{VLL}^{H^\pm_iH^\pm_j}&= (\xi^{q}_{H^\pm_i}\xi^{q}_{H^\pm_j})^2
\frac{x_t^2}{4}\Bigg[
\frac{{x_t}}{({x_{H_i^\pm}}-{x_t}) ({x_{H_j^\pm}}-{x_t})}
+\frac{x_{H_i^\pm}^2 \log ({x_{H_i^\pm}})}{({x_{H_i^\pm}}-{x_{H_j^\pm}}) ({x_{H_i^\pm}}-{x_t})^2} \notag \\
&-\frac{x_{H_j^\pm}^2 \log ({x_{H_j^\pm}})}{({x_{H_i^\pm}}-{x_{H_j^\pm}}) ({x_{H_j^\pm}}-{x_t})^2}
-\frac{{x_t} \log ({x_t}) ({x_{H_i^\pm}} ({x_t}-2 {x_{H_j^\pm}})+{x_{H_j^\pm}} {x_t})}{({x_{H_i^\pm}}-{x_t})^2 ({x_{H_j^\pm}}-{x_t})^2}
\Bigg]\;.
\end{align}
where $x_{H_i^\pm}$ and $x_t$ denote the mass fraction $x_{H_i^\pm}=m^2_{H_i^\pm}/m_W^2$ (i=1,2) and $x_{t}=\bar{m}^2_{t}(\mu_0)/m_W^2$ with $\bar{m}_{t}(\mu_0)$ being the running mass at the scale $\mu_0$.
The terms with $x_b=m_b^2/m_W^2$ are neglected in these expressions. 
We note that the limit $x_{H_j^\pm}\to x_{H_i^\pm} $ in $C_{VLL}^{H^\pm_iH^\pm_j}$ with the replacement $\xi_{H^\pm_j}= \xi_{H^\pm_i} $ reproduces the result of 2HDMs~\cite{Barger:1989fj,Buras:1989ui,Chang:2015rva,Enomoto:2015wbn}.

\section{Lepton couplings of the anomaly-free axion}\label{ap:axioncoup}
In this Appendix, we derive the lepton couplings for the axion, following Ref.~\cite{Saikawa:2019lng}.

The scalar fields are defined by
\begin{align}
\Phi_{i}=\frac{v_{i}}{\sqrt{2}}\exp\left[{i\frac{a_{i}}{v_{i}}}\right]
\begin{pmatrix}
1\\
0
\end{pmatrix}
\;,\quad
S_{m}=\frac{v_{S_{m}}}{\sqrt{2}}\exp\left[{i\frac{a_{S_{m}}}{v_{S_{m}}}}\right]\;,\quad
\end{align}
where the indices run $i=1\mathchar`-3$, $ m=1,\bar{2}$ . 
The $U(1)_{F}$ transformation for these scalar fields and fermion fields forms
\begin{align}
\Phi_{i}&\to e ^{iQ_{\Phi_{i}}\xi } \Phi_{i}\;,\quad
S_{m}\to e ^{iQ_{S_{m}}\xi } S_{m}\;, \\ 
\psi_{L/R}&\to e ^{iQ_{\psi_{L/R}}\xi } \psi_{L/R}\;, \\ 
\end{align}
with $\xi =[0,2\pi]$. 
The Nether current for the $U(1)_{F}$ symmetry is given by
\begin{align}
\left.j^{'}_{\mu}\right|_{\Phi_{i}}&=i Q_{\Phi_{i}}
\left\{(\partial_{\mu} \Phi_{i}^{\dagger})\Phi_{i}-\Phi_{i}^{\dagger}(\partial_{\mu}\Phi_{i})\right\}
\;,\quad
\left.j^{'}_{\mu}\right|_{S_{m}}=i Q_{S_{m}}
\left\{(\partial_{\mu} S_{m}^{\ast})S_{m}-S_{m}^{\ast}(\partial_{\mu}S_{m})\right\}\;. \\
\left.j^{'}_{\mu}\right|_{\psi_{L/R}}&=- Q_{\psi_{L/R}}
\bar{\psi}_{L/R}\gamma_{\mu}\psi_{L/R}.
\end{align}
From these equations, we can write the  $U(1)_{F}$ current 
\begin{align}
\label{eq:current}
j^{'}_{\mu}=v_{a}\partial_{\mu}a'- \sum_{\psi}\left\{Q_{\psi_{L}}
\bar{\psi}_{L}\gamma_{\mu}\psi_{L}+Q_{\psi_{R}}
\bar{\psi}_{R}\gamma_{\mu}\psi_{R}\right\}
\end{align}
Here, we have defined the axion field
\begin{align}
a'=\frac{1}{v_{a}}\left(\sum_{i}Q_{\Phi_i}v_{i}a_{i}+\sum_{i}Q_{S_m}v_{S_{m}}a_{S_{m}}\right)
\end{align}
with $v_{a}$ being $v_{a}^{2}=\sum_{i}Q_{i}v^{2}_{i}+\sum_{m}Q_{S}v^{2}_{S_{m}}$.

The physical axion field should be defined in such a way that it does not mix with the NG boson $G^{0}$, which is absorbed by the longitudinal component of $Z^{0}$~\cite{Bardeen:1977bd,Srednicki:1985xd}.
The NG boson $G^{0}$ is defined by
\begin{align}
G^{0}=v^{-1}\sum_{i}2Y_{i}v_{i}a_{i}
\end{align}
with $v^{2}=\sum_{i}(2Y_{i}v_{i})^{2}=(246{\rm GeV})^{2}$ and $Y_{i}$ being the hypercharge for $\Phi_{i}$ .
Hence, the physical axion field $a$ is defined by
\begin{align}
a=a'-(v_{a}v)^{-1}\sum_{i}Q_{\Phi_{i}}2Y_{i}v_{i}^{2}G^{0}\;,
\end{align}
where we have used that hypercharge for the singlet fields is zero. 
This modifies the $U(1)_{F}$ Eq.~\eqref{eq:current} current  as ~\cite{Bardeen:1977bd,Srednicki:1985xd,Saikawa:2019lng}
\begin{align}
j_{\mu}=v_{a}\partial_{\mu}a- \sum_{\psi}\left\{Q'_{\psi_{L}}
\bar{\psi}_{L}\gamma_{\mu}\psi_{L}+Q'_{\psi_{R}}
\bar{\psi}_{R}\gamma_{\mu}\psi_{R}\right\}
\end{align}
where 
\begin{align}
Q'_{\psi_{L/R}}=Q_{\psi_{L/R}}-\frac{1}{v^{2}}\left(\sum_{i}Q_{i}2Y_{i}v_{i}^{2} \right)2Y_{\psi_{L/R}}.
\end{align}

Using the obtained $U(1)_{F}$ current, the axion-lepton interaction is written by
\begin{align}
{\cal L}&=\frac{\partial_{\mu}a}{v_{a}}j^{\mu}\\
&\ni \frac{\partial_{\mu}a}{v_{a}}\left[\bar{L_{\ell}}Q^{L}_{\ell \ell'}\gamma_{\mu}L_{\ell'}
+\bar{{\ell}_{R}}Q^{R}_{\ell \ell'}\gamma_{\mu}{\ell'_{R}}\right]
\end{align}
where
\begin{align}
Q^{L}_{}=
\begin{pmatrix}
Q'_{L_{\tau}}&0&0 \\
0&Q'_{L_{\mu}}&0 \\
0&0&Q'_{L_{e}} \\
\end{pmatrix}\; ,\quad
Q^{R}_{}=
\begin{pmatrix}
Q'_{{\tau_{R}}}&0&0 \\
0&Q'_{{\mu_{R}}}&0 \\
0&0&Q'_{{e_{R}}} \\
\end{pmatrix}\;. 
\end{align}
This can be rewritten by the following form
\begin{align}
{\cal L}=
-\frac{\partial_{\mu} a}{2f_{a}}\sum_{\ell}\sum_{\ell^{'}}
\bar{\ell}\gamma_{\mu}\big[
(C_{a l}^{V})_{\ell\ell'}-(C_{a l}^{A})_{\ell\ell'}\gamma_{5}
\big]{\ell'}\;.
\end{align}
where we define $f_{a}=v_{a}$. 
The coefficients $C_{a l}^{V}$ and $C_{a l}^{A}$ are defined by
\begin{align}
C_{a l}^{V}&=U^{\dagger}_{l_{L}}Q_{{L}}U^{}_{l_{L}}+U^{\dagger}_{l_{R}}Q_{{R}}U^{}_{l_{R}} \\
C_{a l}^{A}&=U^{\dagger}_{l_{L}}Q_{{L}}U^{}_{l_{L}}-U^{\dagger}_{l_{R}}Q_{{R}}U^{}_{l_{R}} 
\end{align}
with the unitarity matrices $U^{}_{l_{L/R}}$ for the fermion fields $l_{L/R}$.

We calculate the coefficients $C_{a l}^{V}$ and $C_{a l}^{A}$ in the case of Type-B: 
\begin{align}
U^{\dagger}_{l_{L}}Q_{{L}}U^{}_{l_{L}}&=
U^{\dagger}_{l_{L}}
\begin{pmatrix}
1-3\frac{v_{1}^{2}-v_{2}^{2}}{v^{2}}&0&0 \\
0&-1-3\frac{v_{1}^{2}-v_{2}^{2}}{v^{2}}&0 \\
0&0&-3\frac{v_{1}^{2}-v_{2}^{2}}{v^{2}} \\
\end{pmatrix}
U^{\dagger}_{l_{L}}  \notag \\
&\equiv
(V_{l_{L}})_{ij}-3\frac{v_{1}^{2}-v_{2}^{2}}{v^{2}}\delta_{ij} \\
U^{\dagger}_{l_{R}}Q_{{R}}U^{}_{l_{R}}&=
U^{\dagger}_{l_{R}}
\begin{pmatrix}
-2-6\frac{v_{1}^{2}-v_{2}^{2}}{v^{2}}&0&0 \\
0&2-6\frac{v_{1}^{2}-v_{2}^{2}}{v^{2}}&0 \\
0&0&-6\frac{v_{1}^{2}-v_{2}^{2}}{v^{2}} \\
\end{pmatrix}
U^{\dagger}_{l_{R}}  \notag \\
&\equiv
(V_{l_{R}})_{ij}-6\frac{v_{1}^{2}-v_{2}^{2}}{v^{2}}\delta_{ij}. 
\end{align}
In the end, we get
\begin{align}
C_{al}^{V}&=(V_{l_{L}})_{ij}+(V_{l_{R}})_{ij}\;, \\
C_{al}^{A}&=3(c_{\beta_{1} }^{2}-s_{\beta_{1} }^{2} )c_{\beta_{2}}^{2}\delta_{ij}+ (V_{l_{L}})_{ij}-(V_{l_{R}})_{ij}\;.
\end{align}
We have defined the unitary matrices $V_{l_{L}}$, $V_{l_{R}}$ as
\begin{align}
V_{l_{L}}&=
U^{\dagger}_{l_{L}}
\begin{pmatrix}
1&0&0 \\
0&-1&0 \\
0&0&0 \\
\end{pmatrix}
U^{\dagger}_{l_{L}}\;, \quad
V_{l_{R}}=
U^{\dagger}_{l_{R}}
\begin{pmatrix}
-2&0&0 \\
0&2&0 \\
0&0&0 \\
\end{pmatrix}
U^{\dagger}_{l_{R}}\;.  
\end{align}
The diagonal component of $C_{al}^{V}$ vanish by the equation of motion for fermion fields. 
For the model presented in sec.~2, the unitary matrices $U^{}_{\psi_{L/R}}$ are identity matrix since the off-diagonal components of the lepton Yukawa matrix are zero. 
One then obtains the \red{axion-lepton couplings},
\red{
\begin{align}\label{eq:gae}
g_{ae}&=\frac{m_{e}}{f_{a}}\left[3(c_{\beta_{1} }^{2}-s_{\beta_{1} }^{2} )c_{\beta_{2}}^{2}+3 \right]\;, \\ \label{eq:gamu}
g_{a\mu}&=\frac{m_{\mu}}{f_{a}}\left[3(c_{\beta_{1} }^{2}-s_{\beta_{1} }^{2} )c_{\beta_{2}}^{2}-3 \right]\;, \\ \label{eq:gatau}
g_{a\tau}&=\frac{m_{\tau}}{f_{a}}\left[3(c_{\beta_{1} }^{2}-s_{\beta_{1} }^{2} )c_{\beta_{2}}^{2} \right]\;.
\end{align}
}
In the limit of $m_{12} \to 0$, these agree with the numerical results of (\ref{eq:gaeex1}) which is obtained by diagonalizing the mass matrix ${\cal M}_{P}^2$. When $m_{12} \ne 0$, the actual axion-electron coupling becomes slightly smaller than the above estimate.
The approximate expression for the 
axion-\red{lepton} coupling (\ref{gae}) is obtained in the limit of
$c_{\beta_2} \to 0$ \red{or $t_{\beta_1}\to1$}  since the flavor charge of the electron
is $q_e = 1-(-2) = 3$.  


\bibliography{references}

\providecommand{\href}[2]{#2}\begingroup\raggedright\begin{thebibliography}{100}

\bibitem{Peccei:1977hh}
R.~D. Peccei and H.~R. Quinn, {\it {CP Conservation in the Presence of
  Instantons}},  {\em Phys. Rev. Lett.} {\bf 38} (1977) 1440--1443.

\bibitem{Peccei:1977ur}
R.~D. Peccei and H.~R. Quinn, {\it {Constraints Imposed by CP Conservation in
  the Presence of Instantons}},  {\em Phys. Rev. D} {\bf 16} (1977) 1791--1797.

\bibitem{Weinberg:1977ma}
S.~Weinberg, {\it {A New Light Boson?}},  {\em Phys. Rev. Lett.} {\bf 40}
  (1978) 223--226.

\bibitem{Wilczek:1977pj}
F.~Wilczek, {\it {Problem of Strong $P$ and $T$ Invariance in the Presence of
  Instantons}},  {\em Phys. Rev. Lett.} {\bf 40} (1978) 279--282.

\bibitem{Preskill:1982cy}
J.~Preskill, M.~B. Wise, and F.~Wilczek, {\it {Cosmology of the Invisible
  Axion}},  {\em Phys. Lett. B} {\bf 120} (1983) 127--132.

\bibitem{Abbott:1982af}
L.~F. Abbott and P.~Sikivie, {\it {A Cosmological Bound on the Invisible
  Axion}},  {\em Phys. Lett. B} {\bf 120} (1983) 133--136.

\bibitem{Dine:1982ah}
M.~Dine and W.~Fischler, {\it {The Not So Harmless Axion}},  {\em Phys. Lett.
  B} {\bf 120} (1983) 137--141.

\bibitem{Jaeckel:2010ni}
J.~Jaeckel and A.~Ringwald, {\it {The Low-Energy Frontier of Particle
  Physics}},  {\em Ann. Rev. Nucl. Part. Sci.} {\bf 60} (2010) 405--437,
  [\href{http://arxiv.org/abs/1002.0329}{{\tt arXiv:1002.0329}}].

\bibitem{Ringwald:2012hr}
A.~Ringwald, {\it {Exploring the Role of Axions and Other WISPs in the Dark
  Universe}},  {\em Phys. Dark Univ.} {\bf 1} (2012) 116--135,
  [\href{http://arxiv.org/abs/1210.5081}{{\tt arXiv:1210.5081}}].

\bibitem{Arias:2012az}
P.~Arias, D.~Cadamuro, M.~Goodsell, J.~Jaeckel, J.~Redondo, and A.~Ringwald,
  {\it {WISPy Cold Dark Matter}},  {\em JCAP} {\bf 06} (2012) 013,
  [\href{http://arxiv.org/abs/1201.5902}{{\tt arXiv:1201.5902}}].

\bibitem{Graham:2015ouw}
P.~W. Graham, I.~G. Irastorza, S.~K. Lamoreaux, A.~Lindner, and K.~A. van
  Bibber, {\it {Experimental Searches for the Axion and Axion-Like Particles}},
   {\em Ann. Rev. Nucl. Part. Sci.} {\bf 65} (2015) 485--514,
  [\href{http://arxiv.org/abs/1602.00039}{{\tt arXiv:1602.00039}}].

\bibitem{Marsh:2015xka}
D.~J.~E. Marsh, {\it {Axion Cosmology}},  {\em Phys. Rept.} {\bf 643} (2016)
  1--79, [\href{http://arxiv.org/abs/1510.07633}{{\tt arXiv:1510.07633}}].

\bibitem{Irastorza:2018dyq}
I.~G. Irastorza and J.~Redondo, {\it {New experimental approaches in the search
  for axion-like particles}},  {\em Prog. Part. Nucl. Phys.} {\bf 102} (2018)
  89--159, [\href{http://arxiv.org/abs/1801.08127}{{\tt arXiv:1801.08127}}].

\bibitem{DiLuzio:2020wdo}
L.~Di~Luzio, M.~Giannotti, E.~Nardi, and L.~Visinelli, {\it {The landscape of
  QCD axion models}},  {\em Phys. Rept.} {\bf 870} (2020) 1--117,
  [\href{http://arxiv.org/abs/2003.01100}{{\tt arXiv:2003.01100}}].

\bibitem{Ayala:2014pea}
A.~Ayala, I.~Dom\'\i{}nguez, M.~Giannotti, A.~Mirizzi, and O.~Straniero, {\it
  {Revisiting the bound on axion-photon coupling from Globular Clusters}},
  {\em Phys. Rev. Lett.} {\bf 113} (2014), no.~19 191302,
  [\href{http://arxiv.org/abs/1406.6053}{{\tt arXiv:1406.6053}}].

\bibitem{Capozzi:2020cbu}
F.~Capozzi and G.~Raffelt, {\it {Axion and neutrino bounds improved with new
  calibrations of the tip of the red-giant branch using geometric distance
  determinations}},  {\em Phys. Rev. D} {\bf 102} (2020), no.~8 083007,
  [\href{http://arxiv.org/abs/2007.03694}{{\tt arXiv:2007.03694}}].

\bibitem{Straniero:2020iyi}
O.~Straniero, C.~Pallanca, E.~Dalessandro, I.~Dominguez, F.~R. Ferraro,
  M.~Giannotti, A.~Mirizzi, and L.~Piersanti, {\it {The RGB tip of galactic
  globular clusters and the revision of the axion-electron coupling bound}},
  {\em Astron. Astrophys.} {\bf 644} (2020) A166,
  [\href{http://arxiv.org/abs/2010.03833}{{\tt arXiv:2010.03833}}].

\bibitem{Battich:2016htm}
T.~Battich, A.~H. C\'orsico, L.~G. Althaus, M.~M. Miller~Bertolami, and
  M.~M.~M. Bertolami, {\it {First axion bounds from a pulsating helium-rich
  white dwarf star}},  {\em JCAP} {\bf 08} (2016) 062,
  [\href{http://arxiv.org/abs/1605.07668}{{\tt arXiv:1605.07668}}].

\bibitem{Corsico:2016okh}
A.~H. C\'orsico, A.~D. Romero, L.~G. Althaus, E.~Garc\'\i{}a-Berro, J.~Isern,
  S.~O. Kepler, M.~M. Miller~Bertolami, D.~J. Sullivan, and P.~Chote, {\it {An
  asteroseismic constraint on the mass of the axion from the period drift of
  the pulsating DA white dwarf star L19-2}},  {\em JCAP} {\bf 07} (2016) 036,
  [\href{http://arxiv.org/abs/1605.06458}{{\tt arXiv:1605.06458}}].

\bibitem{MillerBertolami:2014rka}
M.~M. Miller~Bertolami, B.~E. Melendez, L.~G. Althaus, and J.~Isern, {\it
  {Revisiting the axion bounds from the Galactic white dwarf luminosity
  function}},  {\em JCAP} {\bf 10} (2014) 069,
  [\href{http://arxiv.org/abs/1406.7712}{{\tt arXiv:1406.7712}}].

\bibitem{Higaki:2016yqk}
T.~Higaki, K.~S. Jeong, N.~Kitajima, and F.~Takahashi, {\it {Quality of the
  Peccei-Quinn symmetry in the Aligned QCD Axion and Cosmological
  Implications}},  {\em JHEP} {\bf 06} (2016) 150,
  [\href{http://arxiv.org/abs/1603.02090}{{\tt arXiv:1603.02090}}].

\bibitem{Farina:2016tgd}
M.~Farina, D.~Pappadopulo, F.~Rompineve, and A.~Tesi, {\it {The photo-philic
  QCD axion}},  {\em JHEP} {\bf 01} (2017) 095,
  [\href{http://arxiv.org/abs/1611.09855}{{\tt arXiv:1611.09855}}].

\bibitem{Higaki:2015jag}
T.~Higaki, K.~S. Jeong, N.~Kitajima, and F.~Takahashi, {\it {The QCD Axion from
  Aligned Axions and Diphoton Excess}},  {\em Phys. Lett. B} {\bf 755} (2016)
  13--16, [\href{http://arxiv.org/abs/1512.05295}{{\tt arXiv:1512.05295}}].

\bibitem{Nakayama:2014cza}
K.~Nakayama, F.~Takahashi, and T.~T. Yanagida, {\it {Anomaly-free flavor models
  for Nambu\textendash{}Goldstone bosons and the 3.5keV X-ray line signal}},
  {\em Phys. Lett. B} {\bf 734} (2014) 178--182,
  [\href{http://arxiv.org/abs/1403.7390}{{\tt arXiv:1403.7390}}].

\bibitem{XENON:2020rca}
{\bf XENON} Collaboration, E.~Aprile et~al., {\it {Excess electronic recoil
  events in XENON1T}},  {\em Phys. Rev. D} {\bf 102} (2020), no.~7 072004,
  [\href{http://arxiv.org/abs/2006.09721}{{\tt arXiv:2006.09721}}].

\bibitem{Takahashi:2020bpq}
F.~Takahashi, M.~Yamada, and W.~Yin, {\it {XENON1T Excess from Anomaly-Free
  Axionlike Dark Matter and Its Implications for Stellar Cooling Anomaly}},
  {\em Phys. Rev. Lett.} {\bf 125} (2020), no.~16 161801,
  [\href{http://arxiv.org/abs/2006.10035}{{\tt arXiv:2006.10035}}].

\bibitem{Weinberg:1976hu}
S.~Weinberg, {\it {Gauge Theory of CP Violation}},  {\em Phys. Rev. Lett.} {\bf
  37} (1976) 657.

\bibitem{Branco:1980sz}
G.~C. Branco, {\it {Spontaneous {CP} Nonconservation and Natural Flavor
  Conservation: A Minimal Model}},  {\em Phys. Rev. D} {\bf 22} (1980) 2901.

\bibitem{Glashow:1976nt}
S.~L. Glashow and S.~Weinberg, {\it {Natural Conservation Laws for Neutral
  Currents}},  {\em Phys. Rev. D} {\bf 15} (1977) 1958.

\bibitem{Ivanov:2011ae}
I.~P. Ivanov, V.~Keus, and E.~Vdovin, {\it {Abelian symmetries in
  multi-Higgs-doublet models}},  {\em J. Phys. A} {\bf 45} (2012) 215201,
  [\href{http://arxiv.org/abs/1112.1660}{{\tt arXiv:1112.1660}}].

\bibitem{Ivanov:2012fp}
I.~P. Ivanov and E.~Vdovin, {\it {Classification of finite reparametrization
  symmetry groups in the three-Higgs-doublet model}},  {\em Eur. Phys. J. C}
  {\bf 73} (2013), no.~2 2309, [\href{http://arxiv.org/abs/1210.6553}{{\tt
  arXiv:1210.6553}}].

\bibitem{Ivanov:2012ry}
I.~P. Ivanov and E.~Vdovin, {\it {Discrete symmetries in the
  three-Higgs-doublet model}},  {\em Phys. Rev. D} {\bf 86} (2012) 095030,
  [\href{http://arxiv.org/abs/1206.7108}{{\tt arXiv:1206.7108}}].

\bibitem{Keus:2013hya}
V.~Keus, S.~F. King, and S.~Moretti, {\it {Three-Higgs-doublet models:
  symmetries, potentials and Higgs boson masses}},  {\em JHEP} {\bf 01} (2014)
  052, [\href{http://arxiv.org/abs/1310.8253}{{\tt arXiv:1310.8253}}].

\bibitem{Ivanov:2015mwl}
I.~P. Ivanov and J.~P. Silva, {\it {$CP$-conserving multi-Higgs model with
  irremovable complex coefficients}},  {\em Phys. Rev. D} {\bf 93} (2016),
  no.~9 095014, [\href{http://arxiv.org/abs/1512.09276}{{\tt
  arXiv:1512.09276}}].

\bibitem{deMedeirosVarzielas:2019rrp}
I.~de~Medeiros~Varzielas and I.~P. Ivanov, {\it {Recognizing symmetries in a
  3HDM in a basis-independent way}},  {\em Phys. Rev. D} {\bf 100} (2019),
  no.~1 015008, [\href{http://arxiv.org/abs/1903.11110}{{\tt
  arXiv:1903.11110}}].

\bibitem{Darvishi:2019dbh}
N.~Darvishi and A.~Pilaftsis, {\it {Classifying Accidental Symmetries in
  Multi-Higgs Doublet Models}},  {\em Phys. Rev. D} {\bf 101} (2020), no.~9
  095008, [\href{http://arxiv.org/abs/1912.00887}{{\tt arXiv:1912.00887}}].

\bibitem{deMedeirosVarzielas:2021zqs}
I.~de~Medeiros~Varzielas, I.~P. Ivanov, and M.~Levy, {\it {Exploring
  multi-Higgs models with softly broken large discrete symmetry groups}},  {\em
  Eur. Phys. J. C} {\bf 81} (2021), no.~10 918,
  [\href{http://arxiv.org/abs/2107.08227}{{\tt arXiv:2107.08227}}].

\bibitem{Varzielas:2022lye}
I.~d.~M. Varzielas and D.~Ivo, {\it {Softly-broken $A_4$ or $S_4$ 3HDMs with
  stable states}},  \href{http://arxiv.org/abs/2202.00681}{{\tt
  arXiv:2202.00681}}.

\bibitem{Ferreira:2017tvy}
P.~M. Ferreira, I.~P. Ivanov, E.~Jim\'enez, R.~Pasechnik, and H.~Ser\^odio,
  {\it {CP4 miracle: shaping Yukawa sector with CP symmetry of order four}},
  {\em JHEP} {\bf 01} (2018) 065, [\href{http://arxiv.org/abs/1711.02042}{{\tt
  arXiv:1711.02042}}].

\bibitem{Ivanov:2021pnr}
I.~P. Ivanov and S.~A. Obodenko, {\it {Constraining CP4 3HDM with Top Quark
  Decays}},  {\em Universe} {\bf 7} (2021), no.~6 197,
  [\href{http://arxiv.org/abs/2104.11440}{{\tt arXiv:2104.11440}}].

\bibitem{Yagyu:2016whx}
K.~Yagyu, {\it {Higgs boson couplings in multi-doublet models with natural
  flavour conservation}},  {\em Phys. Lett. B} {\bf 763} (2016) 102--107,
  [\href{http://arxiv.org/abs/1609.04590}{{\tt arXiv:1609.04590}}].

\bibitem{Das:2019yad}
D.~Das and I.~Saha, {\it {Alignment limit in three Higgs-doublet models}},
  {\em Phys. Rev. D} {\bf 100} (2019), no.~3 035021,
  [\href{http://arxiv.org/abs/1904.03970}{{\tt arXiv:1904.03970}}].

\bibitem{Boto:2021qgu}
R.~Boto, J.~C. Rom\~ao, and J.~a.~P. Silva, {\it {Current bounds on the type-Z
  Z3 three-Higgs-doublet model}},  {\em Phys. Rev. D} {\bf 104} (2021), no.~9
  095006, [\href{http://arxiv.org/abs/2106.11977}{{\tt arXiv:2106.11977}}].

\bibitem{Chakraborti:2021bpy}
M.~Chakraborti, D.~Das, M.~Levy, S.~Mukherjee, and I.~Saha, {\it {Prospects for
  light charged scalars in a three-Higgs-doublet model with Z3 symmetry}},
  {\em Phys. Rev. D} {\bf 104} (2021), no.~7 075033,
  [\href{http://arxiv.org/abs/2104.08146}{{\tt arXiv:2104.08146}}].

\bibitem{Cree:2011uy}
G.~Cree and H.~E. Logan, {\it {Yukawa alignment from natural flavor
  conservation}},  {\em Phys. Rev. D} {\bf 84} (2011) 055021,
  [\href{http://arxiv.org/abs/1106.4039}{{\tt arXiv:1106.4039}}].

\bibitem{Akeroyd:2016ssd}
A.~G. Akeroyd, S.~Moretti, K.~Yagyu, and E.~Yildirim, {\it {Light charged Higgs
  boson scenario in 3-Higgs doublet models}},  {\em Int. J. Mod. Phys. A} {\bf
  32} (2017), no.~23n24 1750145, [\href{http://arxiv.org/abs/1605.05881}{{\tt
  arXiv:1605.05881}}].

\bibitem{Akeroyd:2020nfj}
A.~G. Akeroyd, S.~Moretti, T.~Shindou, and M.~Song, {\it {CP asymmetries of
  ${\overline B}\to X_s/X_d\gamma$ in models with three Higgs doublets}},  {\em
  Phys. Rev. D} {\bf 103} (2021), no.~1 015035,
  [\href{http://arxiv.org/abs/2009.05779}{{\tt arXiv:2009.05779}}].

\bibitem{Logan:2020mdz}
H.~E. Logan, S.~Moretti, D.~Rojas-Ciofalo, and M.~Song, {\it {CP violation from
  charged Higgs bosons in the three Higgs doublet model}},  {\em JHEP} {\bf 07}
  (2021) 158, [\href{http://arxiv.org/abs/2012.08846}{{\tt arXiv:2012.08846}}].

\bibitem{Akeroyd:2021fpf}
A.~G. Akeroyd, H.~E. Logan, S.~Moretti, D.~Rojas-Ciofalo, T.~Shindou, and
  M.~Song, {\it {CP-Violation in the 3-Higgs Doublet Model: CP-Asymmetries from
  Charged Higgs Bosons and Electric Dipole Moments}},
  \href{http://arxiv.org/abs/2111.11931}{{\tt arXiv:2111.11931}}.

\bibitem{Davoudiasl:2019lcg}
H.~Davoudiasl, I.~M. Lewis, and M.~Sullivan, {\it {Higgs Troika for Baryon
  Asymmetry}},  {\em Phys. Rev. D} {\bf 101} (2020), no.~5 055010,
  [\href{http://arxiv.org/abs/1909.02044}{{\tt arXiv:1909.02044}}].

\bibitem{Davoudiasl:2021syn}
H.~Davoudiasl, I.~M. Lewis, and M.~Sullivan, {\it {Multi-TeV signals of
  baryogenesis in a Higgs troika model}},  {\em Phys. Rev. D} {\bf 104} (2021),
  no.~1 015024, [\href{http://arxiv.org/abs/2103.12089}{{\tt
  arXiv:2103.12089}}].

\bibitem{Darvishi:2021txa}
N.~Darvishi, M.~R. Masouminia, and A.~Pilaftsis, {\it {Maximally symmetric
  three-Higgs-doublet model}},  {\em Phys. Rev. D} {\bf 104} (2021), no.~11
  115017, [\href{http://arxiv.org/abs/2106.03159}{{\tt arXiv:2106.03159}}].

\bibitem{Das:2021oik}
D.~Das, P.~M. Ferreira, A.~P. Morais, I.~Padilla-Gay, R.~Pasechnik, and J.~P.
  Rodrigues, {\it {A three Higgs doublet model with symmetry-suppressed flavour
  changing neutral currents}},  {\em JHEP} {\bf 11} (2021) 079,
  [\href{http://arxiv.org/abs/2106.06425}{{\tt arXiv:2106.06425}}].

\bibitem{Das:2014fea}
D.~Das and U.~K. Dey, {\it {Analysis of an extended scalar sector with $S_3$
  symmetry}},  {\em Phys. Rev. D} {\bf 89} (2014), no.~9 095025,
  [\href{http://arxiv.org/abs/1404.2491}{{\tt arXiv:1404.2491}}]. [Erratum:
  Phys.Rev.D 91, 039905 (2015)].

\bibitem{Calibbi:2020jvd}
L.~Calibbi, D.~Redigolo, R.~Ziegler, and J.~Zupan, {\it {Looking forward to
  lepton-flavor-violating ALPs}},  {\em JHEP} {\bf 09} (2021) 173,
  [\href{http://arxiv.org/abs/2006.04795}{{\tt arXiv:2006.04795}}].

\bibitem{Han:2020dwo}
C.~Han, M.~L. L\'opez-Ib\'a\~nez, A.~Melis, O.~Vives, and J.~M. Yang, {\it
  {Anomaly-free leptophilic axionlike particle and its flavor violating
  tests}},  {\em Phys. Rev. D} {\bf 103} (2021), no.~3 035028,
  [\href{http://arxiv.org/abs/2007.08834}{{\tt arXiv:2007.08834}}].

\bibitem{Han:2022iig}
C.~Han, M.~L. L\'opez-Ib\'a\~nez, A.~Melis, O.~Vives, and J.~M. Yang, {\it
  {Anomaly-free ALP from non-Abelian flavor symmetry}},
  \href{http://arxiv.org/abs/2203.16376}{{\tt arXiv:2203.16376}}.

\bibitem{Takahashi:2020uio}
F.~Takahashi, M.~Yamada, and W.~Yin, {\it {What if ALP dark matter for the
  XENON1T excess is the inflaton}},  {\em JHEP} {\bf 01} (2021) 152,
  [\href{http://arxiv.org/abs/2007.10311}{{\tt arXiv:2007.10311}}].

\bibitem{Minkowski:1977sc}
P.~Minkowski, {\it {$\mu \to e\gamma$ at a Rate of One Out of $10^{9}$ Muon
  Decays?}},  {\em Phys. Lett. B} {\bf 67} (1977) 421--428.

\bibitem{Yanagida:1979as}
T.~Yanagida, {\it {Horizontal gauge symmetry and masses of neutrinos}},  {\em
  Conf. Proc. C} {\bf 7902131} (1979) 95--99.

\bibitem{Ramond:1979py}
P.~Ramond, {\it {The Family Group in Grand Unified Theories}},  in {\em
  {International Symposium on Fundamentals of Quantum Theory and Quantum Field
  Theory}}, 2, 1979.
\newblock \href{http://arxiv.org/abs/hep-ph/9809459}{{\tt hep-ph/9809459}}.

\bibitem{Glashow:1979nm}
S.~L. Glashow, {\it {The Future of Elementary Particle Physics}},  {\em NATO
  Sci. Ser. B} {\bf 61} (1980) 687.

\bibitem{tHooft:1979rat}
G.~'t~Hooft, {\it {Naturalness, chiral symmetry, and spontaneous chiral
  symmetry breaking}},  {\em NATO Sci. Ser. B} {\bf 59} (1980) 135--157.

\bibitem{Volkas:1988cm}
R.~R. Volkas, A.~J. Davies, and G.~C. Joshi, {\it {NATURALNESS OF THE INVISIBLE
  AXION MODEL}},  {\em Phys. Lett. B} {\bf 215} (1988) 133--138.

\bibitem{Georgi:1978ri}
H.~Georgi and D.~V. Nanopoulos, {\it {Suppression of Flavor Changing Effects
  From Neutral Spinless Meson Exchange in Gauge Theories}},  {\em Phys. Lett.
  B} {\bf 82} (1979) 95--96.

\bibitem{Lavoura:1994fv}
L.~Lavoura and J.~P. Silva, {\it {Fundamental CP violating quantities in a
  SU(2) x U(1) model with many Higgs doublets}},  {\em Phys. Rev. D} {\bf 50}
  (1994) 4619--4624, [\href{http://arxiv.org/abs/hep-ph/9404276}{{\tt
  hep-ph/9404276}}].

\bibitem{Lavoura:1994yu}
L.~Lavoura, {\it {Signatures of discrete symmetries in the scalar sector}},
  {\em Phys. Rev. D} {\bf 50} (1994) 7089--7092,
  [\href{http://arxiv.org/abs/hep-ph/9405307}{{\tt hep-ph/9405307}}].

\bibitem{Davidson:2005cw}
S.~Davidson and H.~E. Haber, {\it {Basis-independent methods for the
  two-Higgs-doublet model}},  {\em Phys. Rev. D} {\bf 72} (2005) 035004,
  [\href{http://arxiv.org/abs/hep-ph/0504050}{{\tt hep-ph/0504050}}]. [Erratum:
  Phys.Rev.D 72, 099902 (2005)].

\bibitem{ATLAS:2021vrm}
{\bf ATLAS} Collaboration, {\it {Combined measurements of Higgs boson
  production and decay using up to $139$ fb$^{-1}$ of proton-proton collision
  data at $\sqrt{s}= 13$ TeV collected with the ATLAS experiment}}, .

\bibitem{CMS:2020gsy}
{\bf CMS} Collaboration, {\it {Combined Higgs boson production and decay
  measurements with up to 137 fb$^{-1}$ of proton-proton collision data at
  $\sqrt s$ = 13 TeV}}, .

\bibitem{Gunion:2002zf}
J.~F. Gunion and H.~E. Haber, {\it {The CP conserving two Higgs doublet model:
  The Approach to the decoupling limit}},  {\em Phys. Rev. D} {\bf 67} (2003)
  075019, [\href{http://arxiv.org/abs/hep-ph/0207010}{{\tt hep-ph/0207010}}].

\bibitem{Carena:2013ooa}
M.~Carena, I.~Low, N.~R. Shah, and C.~E.~M. Wagner, {\it {Impersonating the
  Standard Model Higgs Boson: Alignment without Decoupling}},  {\em JHEP} {\bf
  04} (2014) 015, [\href{http://arxiv.org/abs/1310.2248}{{\tt
  arXiv:1310.2248}}].

\bibitem{Darvishi:2020teg}
N.~Darvishi and A.~Pilaftsis, {\it {Natural Alignment in Multi-Higgs Doublet
  Models}},  {\em PoS} {\bf CORFU2019} (2020) 064,
  [\href{http://arxiv.org/abs/2004.04505}{{\tt arXiv:2004.04505}}].

\bibitem{Darvishi:2022zag}
N.~Darvishi, M.~R. Masouminia, and A.~Pilaftsis, {\it {Higgs-Sector Predictions
  from Maximally Symmetric multi-Higgs Doublet Models}},  in {\em {7th
  Symposium on Prospects in the Physics of Discrete Symmetries}}, 1, 2022.
\newblock \href{http://arxiv.org/abs/2201.00600}{{\tt arXiv:2201.00600}}.

\bibitem{Passarino:1978jh}
G.~Passarino and M.~J.~G. Veltman, {\it {One Loop Corrections for e+ e-
  Annihilation Into mu+ mu- in the Weinberg Model}},  {\em Nucl. Phys. B} {\bf
  160} (1979) 151--207.

\bibitem{Pospelov:2008jk}
M.~Pospelov, A.~Ritz, and M.~B. Voloshin, {\it {Bosonic super-WIMPs as
  keV-scale dark matter}},  {\em Phys. Rev. D} {\bf 78} (2008) 115012,
  [\href{http://arxiv.org/abs/0807.3279}{{\tt arXiv:0807.3279}}].

\bibitem{THESEUS:2017qvx}
{\bf THESEUS} Collaboration, L.~Amati et~al., {\it {The THESEUS space mission
  concept: science case, design and expected performances}},  {\em Adv. Space
  Res.} {\bf 62} (2018) 191--244, [\href{http://arxiv.org/abs/1710.04638}{{\tt
  arXiv:1710.04638}}].

\bibitem{THESEUS:2017wvz}
{\bf THESEUS} Collaboration, G.~Stratta et~al., {\it {THESEUS: a key space
  mission concept for Multi-Messenger Astrophysics}},  {\em Adv. Space Res.}
  {\bf 62} (2018) 662--682, [\href{http://arxiv.org/abs/1712.08153}{{\tt
  arXiv:1712.08153}}].

\bibitem{Barret:2018qft}
D.~Barret et~al., {\it {The Athena X-ray Integral Field Unit}},  {\em Proc.
  SPIE Int. Soc. Opt. Eng.} {\bf 10699} (2018) 106991G,
  [\href{http://arxiv.org/abs/1807.06092}{{\tt arXiv:1807.06092}}].

\bibitem{eROSITA:2012lfj}
{\bf eROSITA} Collaboration, A.~Merloni et~al., {\it {eROSITA Science Book:
  Mapping the Structure of the Energetic Universe}},
  \href{http://arxiv.org/abs/1209.3114}{{\tt arXiv:1209.3114}}.

\bibitem{XRISMScienceTeam:2020rvx}
{\bf XRISM Science Team} Collaboration, {\it {Science with the X-ray Imaging
  and Spectroscopy Mission (XRISM)}},
  \href{http://arxiv.org/abs/2003.04962}{{\tt arXiv:2003.04962}}.

\bibitem{LZ:2021xov}
{\bf LZ} Collaboration, D.~S. Akerib et~al., {\it {Projected sensitivities of
  the LUX-ZEPLIN experiment to new physics via low-energy electron recoils}},
  {\em Phys. Rev. D} {\bf 104} (2021), no.~9 092009,
  [\href{http://arxiv.org/abs/2102.11740}{{\tt arXiv:2102.11740}}].

\bibitem{DARWIN:2016hyl}
{\bf DARWIN} Collaboration, J.~Aalbers et~al., {\it {DARWIN: towards the
  ultimate dark matter detector}},  {\em JCAP} {\bf 11} (2016) 017,
  [\href{http://arxiv.org/abs/1606.07001}{{\tt arXiv:1606.07001}}].

\bibitem{Klimenko:1984qx}
K.~G. Klimenko, {\it {On Necessary and Sufficient Conditions for Some Higgs
  Potentials to Be Bounded From Below}},  {\em Theor. Math. Phys.} {\bf 62}
  (1985) 58--65.

\bibitem{Lee:1977eg}
B.~W. Lee, C.~Quigg, and H.~B. Thacker, {\it {Weak Interactions at Very
  High-Energies: The Role of the Higgs Boson Mass}},  {\em Phys. Rev. D} {\bf
  16} (1977) 1519.

\bibitem{Bento:2017eti}
M.~P. Bento, H.~E. Haber, J.~C. Rom\~ao, and J.~a.~P. Silva, {\it {Multi-Higgs
  doublet models: physical parametrization, sum rules and unitarity bounds}},
  {\em JHEP} {\bf 11} (2017) 095, [\href{http://arxiv.org/abs/1708.09408}{{\tt
  arXiv:1708.09408}}].

\bibitem{Peskin:1991sw}
M.~E. Peskin and T.~Takeuchi, {\it {Estimation of oblique electroweak
  corrections}},  {\em Phys. Rev. D} {\bf 46} (1992) 381--409.

\bibitem{Grimus:2007if}
W.~Grimus, L.~Lavoura, O.~M. Ogreid, and P.~Osland, {\it {A Precision
  constraint on multi-Higgs-doublet models}},  {\em J. Phys. G} {\bf 35} (2008)
  075001, [\href{http://arxiv.org/abs/0711.4022}{{\tt arXiv:0711.4022}}].

\bibitem{Grimus:2008nb}
W.~Grimus, L.~Lavoura, O.~M. Ogreid, and P.~Osland, {\it {The Oblique
  parameters in multi-Higgs-doublet models}},  {\em Nucl. Phys. B} {\bf 801}
  (2008) 81--96, [\href{http://arxiv.org/abs/0802.4353}{{\tt
  arXiv:0802.4353}}].

\bibitem{Hagiwara:1994pw}
K.~Hagiwara, S.~Matsumoto, D.~Haidt, and C.~S. Kim, {\it {A Novel approach to
  confront electroweak data and theory}},  {\em Z. Phys. C} {\bf 64} (1994)
  559--620, [\href{http://arxiv.org/abs/hep-ph/9409380}{{\tt hep-ph/9409380}}].
  [Erratum: Z.Phys.C 68, 352 (1995)].

\bibitem{Hahn:1998yk}
T.~Hahn and M.~Perez-Victoria, {\it {Automatized one loop calculations in
  four-dimensions and D-dimensions}},  {\em Comput. Phys. Commun.} {\bf 118}
  (1999) 153--165, [\href{http://arxiv.org/abs/hep-ph/9807565}{{\tt
  hep-ph/9807565}}].

\bibitem{Haller:2018nnx}
J.~Haller, A.~Hoecker, R.~Kogler, K.~M\"onig, T.~Peiffer, and J.~Stelzer, {\it
  {Update of the global electroweak fit and constraints on two-Higgs-doublet
  models}},  {\em Eur. Phys. J. C} {\bf 78} (2018), no.~8 675,
  [\href{http://arxiv.org/abs/1803.01853}{{\tt arXiv:1803.01853}}].

\bibitem{HFLAV:2019otj}
{\bf HFLAV} Collaboration, Y.~S. Amhis et~al., {\it {Averages of b-hadron,
  c-hadron, and $\tau $-lepton properties as of 2018}},  {\em Eur. Phys. J. C}
  {\bf 81} (2021), no.~3 226, [\href{http://arxiv.org/abs/1909.12524}{{\tt
  arXiv:1909.12524}}].

\bibitem{Adel:1993ah}
K.~Adel and Y.-P. Yao, {\it {$O(\alpha_s)$ calculation of the decays $b \to s +
  \gamma$ and $b \to s + g$}},  {\em Phys. Rev. D} {\bf 49} (1994) 4945--4948,
  [\href{http://arxiv.org/abs/hep-ph/9308349}{{\tt hep-ph/9308349}}].

\bibitem{Misiak:1994zw}
M.~Misiak and M.~Munz, {\it {Two loop mixing of dimension five flavor changing
  operators}},  {\em Phys. Lett. B} {\bf 344} (1995) 308--318,
  [\href{http://arxiv.org/abs/hep-ph/9409454}{{\tt hep-ph/9409454}}].

\bibitem{Ali:1995bi}
A.~Ali and C.~Greub, {\it {Photon energy spectrum in B ---\ensuremath{>} X(s) +
  gamma and comparison with data}},  {\em Phys. Lett. B} {\bf 361} (1995)
  146--154, [\href{http://arxiv.org/abs/hep-ph/9506374}{{\tt hep-ph/9506374}}].

\bibitem{Pott:1995if}
N.~Pott, {\it {Bremsstrahlung corrections to the decay b ---\ensuremath{>} s
  gamma}},  {\em Phys. Rev. D} {\bf 54} (1996) 938--948,
  [\href{http://arxiv.org/abs/hep-ph/9512252}{{\tt hep-ph/9512252}}].

\bibitem{Greub:1996tg}
C.~Greub, T.~Hurth, and D.~Wyler, {\it {Virtual O (alpha-s) corrections to the
  inclusive decay b ---\ensuremath{>} s gamma}},  {\em Phys. Rev. D} {\bf 54}
  (1996) 3350--3364, [\href{http://arxiv.org/abs/hep-ph/9603404}{{\tt
  hep-ph/9603404}}].

\bibitem{Chetyrkin:1996vx}
K.~G. Chetyrkin, M.~Misiak, and M.~Munz, {\it {Weak radiative B meson decay
  beyond leading logarithms}},  {\em Phys. Lett. B} {\bf 400} (1997) 206--219,
  [\href{http://arxiv.org/abs/hep-ph/9612313}{{\tt hep-ph/9612313}}]. [Erratum:
  Phys.Lett.B 425, 414 (1998)].

\bibitem{Greub:1997hf}
C.~Greub and T.~Hurth, {\it {Two loop matching of the dipole operators for b
  ---\ensuremath{>} s gamma and b ---\ensuremath{>} s g}},  {\em Phys. Rev. D}
  {\bf 56} (1997) 2934--2949, [\href{http://arxiv.org/abs/hep-ph/9703349}{{\tt
  hep-ph/9703349}}].

\bibitem{Buras:1997xf}
A.~J. Buras, A.~Kwiatkowski, and N.~Pott, {\it {Next-to-leading order matching
  for the magnetic photon penguin operator in the $B \to X_s \gamma$ decay}},
  {\em Nucl. Phys. B} {\bf 517} (1998) 353--373,
  [\href{http://arxiv.org/abs/hep-ph/9710336}{{\tt hep-ph/9710336}}].

\bibitem{Buras:2001mq}
A.~J. Buras, A.~Czarnecki, M.~Misiak, and J.~Urban, {\it {Two loop matrix
  element of the current current operator in the decay B ---\ensuremath{>} X(s)
  gamma}},  {\em Nucl. Phys. B} {\bf 611} (2001) 488--502,
  [\href{http://arxiv.org/abs/hep-ph/0105160}{{\tt hep-ph/0105160}}].

\bibitem{Misiak:2006zs}
M.~Misiak et~al., {\it {Estimate of $\mathcal{B} (\bar B \to X_s \gamma)$ at
  $O(\alpha_s^2)$}},  {\em Phys. Rev. Lett.} {\bf 98} (2007) 022002,
  [\href{http://arxiv.org/abs/hep-ph/0609232}{{\tt hep-ph/0609232}}].

\bibitem{Misiak:2020vlo}
M.~Misiak, A.~Rehman, and M.~Steinhauser, {\it {Towards $ \overline{B}\to
  {X}_s\gamma $ at the NNLO in QCD without interpolation in m$_{c}$}},  {\em
  JHEP} {\bf 06} (2020) 175, [\href{http://arxiv.org/abs/2002.01548}{{\tt
  arXiv:2002.01548}}].

\bibitem{Ciuchini:1997xe}
M.~Ciuchini, G.~Degrassi, P.~Gambino, and G.~F. Giudice, {\it {Next-to-leading
  QCD corrections to $B \to X_s \gamma$: Standard model and two Higgs doublet
  model}},  {\em Nucl. Phys. B} {\bf 527} (1998) 21--43,
  [\href{http://arxiv.org/abs/hep-ph/9710335}{{\tt hep-ph/9710335}}].

\bibitem{Borzumati:1998tg}
F.~Borzumati and C.~Greub, {\it {2HDMs predictions for anti-B ---\ensuremath{>}
  X(s) gamma in NLO QCD}},  {\em Phys. Rev. D} {\bf 58} (1998) 074004,
  [\href{http://arxiv.org/abs/hep-ph/9802391}{{\tt hep-ph/9802391}}].

\bibitem{Borzumati:1998nx}
F.~Borzumati and C.~Greub, {\it {Two Higgs doublet model predictions for anti-B
  ---\ensuremath{>} X(s) gamma in NLO QCD: Addendum}},  {\em Phys. Rev. D} {\bf
  59} (1999) 057501, [\href{http://arxiv.org/abs/hep-ph/9809438}{{\tt
  hep-ph/9809438}}].

\bibitem{Ciafaloni:1997un}
P.~Ciafaloni, A.~Romanino, and A.~Strumia, {\it {Two loop QCD corrections to
  charged Higgs mediated b ---\ensuremath{>} s gamma decay}},  {\em Nucl. Phys.
  B} {\bf 524} (1998) 361--376,
  [\href{http://arxiv.org/abs/hep-ph/9710312}{{\tt hep-ph/9710312}}].

\bibitem{Bobeth:1999ww}
C.~Bobeth, M.~Misiak, and J.~Urban, {\it {Matching conditions for $b \to s
  \gamma$ and $b \to s gluon$ in extensions of the standard model}},  {\em
  Nucl. Phys. B} {\bf 567} (2000) 153--185,
  [\href{http://arxiv.org/abs/hep-ph/9904413}{{\tt hep-ph/9904413}}].

\bibitem{Hermann:2012fc}
T.~Hermann, M.~Misiak, and M.~Steinhauser, {\it {$\bar{B}\to X_s \gamma$ in the
  Two Higgs Doublet Model up to Next-to-Next-to-Leading Order in QCD}},  {\em
  JHEP} {\bf 11} (2012) 036, [\href{http://arxiv.org/abs/1208.2788}{{\tt
  arXiv:1208.2788}}].

\bibitem{Misiak:2015xwa}
M.~Misiak et~al., {\it {Updated NNLO QCD predictions for the weak radiative
  B-meson decays}},  {\em Phys. Rev. Lett.} {\bf 114} (2015), no.~22 221801,
  [\href{http://arxiv.org/abs/1503.01789}{{\tt arXiv:1503.01789}}].

\bibitem{Barger:1989fj}
V.~D. Barger, J.~L. Hewett, and R.~J.~N. Phillips, {\it {New Constraints on the
  Charged Higgs Sector in Two Higgs Doublet Models}},  {\em Phys. Rev. D} {\bf
  41} (1990) 3421--3441.

\bibitem{Enomoto:2015wbn}
T.~Enomoto and R.~Watanabe, {\it {Flavor constraints on the Two Higgs Doublet
  Models of Z$_{2}$ symmetric and aligned types}},  {\em JHEP} {\bf 05} (2016)
  002, [\href{http://arxiv.org/abs/1511.05066}{{\tt arXiv:1511.05066}}].

\bibitem{CKMfit2021}
\url{http://ckmfitter.in2p3.fr/www/results/plots_spring21/num/ckmEval_results_spring21.html}.

\bibitem{Bloch:2020uzh}
I.~M. Bloch, A.~Caputo, R.~Essig, D.~Redigolo, M.~Sholapurkar, and T.~Volansky,
  {\it {Exploring new physics with O(keV) electron recoils in direct detection
  experiments}},  {\em JHEP} {\bf 01} (2021) 178,
  [\href{http://arxiv.org/abs/2006.14521}{{\tt arXiv:2006.14521}}].

\bibitem{Charles:2020dfl}
J.~Charles, S.~Descotes-Genon, Z.~Ligeti, S.~Monteil, M.~Papucci, K.~Trabelsi,
  and L.~Vale~Silva, {\it {New physics in $B$ meson mixing: future sensitivity
  and limitations}},  {\em Phys. Rev. D} {\bf 102} (2020), no.~5 056023,
  [\href{http://arxiv.org/abs/2006.04824}{{\tt arXiv:2006.04824}}].

\bibitem{LHCb:2018roe}
{\bf LHCb} Collaboration, R.~Aaij et~al., {\it {Physics case for an LHCb
  Upgrade II - Opportunities in flavour physics, and beyond, in the HL-LHC
  era}},  \href{http://arxiv.org/abs/1808.08865}{{\tt arXiv:1808.08865}}.

\bibitem{BelleIIVXD}
Belle II VXD open workshop on possible future technologies for a collider with
  $4 \times 1036~{\rm cm^{-2}s^{-1}}$ luminosity,
  \url{https://indico.cern.ch/event/810687/}.

\bibitem{deBlas:2019rxi}
J.~de~Blas et~al., {\it {Higgs Boson Studies at Future Particle Colliders}},
  {\em JHEP} {\bf 01} (2020) 139, [\href{http://arxiv.org/abs/1905.03764}{{\tt
  arXiv:1905.03764}}].

\bibitem{Djouadi:2005gi}
A.~Djouadi, {\it {The Anatomy of electro-weak symmetry breaking. I: The Higgs
  boson in the standard model}},  {\em Phys. Rept.} {\bf 457} (2008) 1--216,
  [\href{http://arxiv.org/abs/hep-ph/0503172}{{\tt hep-ph/0503172}}].

\bibitem{Djouadi:2005gj}
A.~Djouadi, {\it {The Anatomy of electro-weak symmetry breaking. II. The Higgs
  bosons in the minimal supersymmetric model}},  {\em Phys. Rept.} {\bf 459}
  (2008) 1--241, [\href{http://arxiv.org/abs/hep-ph/0503173}{{\tt
  hep-ph/0503173}}].

\bibitem{Fujii:2017vwa}
K.~Fujii et~al., {\it {Physics Case for the 250 GeV Stage of the International
  Linear Collider}},  \href{http://arxiv.org/abs/1710.07621}{{\tt
  arXiv:1710.07621}}.

\bibitem{Cepeda:2019klc}
M.~Cepeda et~al., {\it {Report from Working Group 2}: {Higgs Physics at the
  HL-LHC and HE-LHC}},  {\em CERN Yellow Rep. Monogr.} {\bf 7} (2019) 221--584,
  [\href{http://arxiv.org/abs/1902.00134}{{\tt arXiv:1902.00134}}].

\bibitem{Goncalves:2018qas}
D.~Gon\c{c}alves, T.~Han, F.~Kling, T.~Plehn, and M.~Takeuchi, {\it {Higgs
  boson pair production at future hadron colliders: From kinematics to
  dynamics}},  {\em Phys. Rev. D} {\bf 97} (2018), no.~11 113004,
  [\href{http://arxiv.org/abs/1802.04319}{{\tt arXiv:1802.04319}}].

\bibitem{Boyarsky:2018tvu}
A.~Boyarsky, M.~Drewes, T.~Lasserre, S.~Mertens, and O.~Ruchayskiy, {\it
  {Sterile neutrino Dark Matter}},  {\em Prog. Part. Nucl. Phys.} {\bf 104}
  (2019) 1--45, [\href{http://arxiv.org/abs/1807.07938}{{\tt
  arXiv:1807.07938}}].

\bibitem{Tremaine:1979we}
S.~Tremaine and J.~E. Gunn, {\it {Dynamical Role of Light Neutral Leptons in
  Cosmology}},  {\em Phys. Rev. Lett.} {\bf 42} (1979) 407--410.

\bibitem{Boyarsky:2008ju}
A.~Boyarsky, O.~Ruchayskiy, and D.~Iakubovskyi, {\it {A Lower bound on the mass
  of Dark Matter particles}},  {\em JCAP} {\bf 03} (2009) 005,
  [\href{http://arxiv.org/abs/0808.3902}{{\tt arXiv:0808.3902}}].

\bibitem{Gorbunov:2008ka}
D.~Gorbunov, A.~Khmelnitsky, and V.~Rubakov, {\it {Constraining sterile
  neutrino dark matter by phase-space density observations}},  {\em JCAP} {\bf
  10} (2008) 041, [\href{http://arxiv.org/abs/0808.3910}{{\tt
  arXiv:0808.3910}}].

\bibitem{Savchenko:2019qnn}
D.~Savchenko and A.~Rudakovskyi, {\it {New mass bound on fermionic dark matter
  from a combined analysis of classical dSphs}},  {\em Mon. Not. Roy. Astron.
  Soc.} {\bf 487} (2019), no.~4 5711--5720,
  [\href{http://arxiv.org/abs/1903.01862}{{\tt arXiv:1903.01862}}].

\bibitem{Shi:1998km}
X.-D. Shi and G.~M. Fuller, {\it {A New dark matter candidate: Nonthermal
  sterile neutrinos}},  {\em Phys. Rev. Lett.} {\bf 82} (1999) 2832--2835,
  [\href{http://arxiv.org/abs/astro-ph/9810076}{{\tt astro-ph/9810076}}].

\bibitem{Serpico:2005bc}
P.~D. Serpico and G.~G. Raffelt, {\it {Lepton asymmetry and primordial
  nucleosynthesis in the era of precision cosmology}},  {\em Phys. Rev. D} {\bf
  71} (2005) 127301, [\href{http://arxiv.org/abs/astro-ph/0506162}{{\tt
  astro-ph/0506162}}].

\bibitem{Shaposhnikov:2008pf}
M.~Shaposhnikov, {\it {The nuMSM, leptonic asymmetries, and properties of
  singlet fermions}},  {\em JHEP} {\bf 08} (2008) 008,
  [\href{http://arxiv.org/abs/0804.4542}{{\tt arXiv:0804.4542}}].

\bibitem{Laine:2008pg}
M.~Laine and M.~Shaposhnikov, {\it {Sterile neutrino dark matter as a
  consequence of nuMSM-induced lepton asymmetry}},  {\em JCAP} {\bf 06} (2008)
  031, [\href{http://arxiv.org/abs/0804.4543}{{\tt arXiv:0804.4543}}].

\bibitem{Canetti:2012kh}
L.~Canetti, M.~Drewes, T.~Frossard, and M.~Shaposhnikov, {\it {Dark Matter,
  Baryogenesis and Neutrino Oscillations from Right Handed Neutrinos}},  {\em
  Phys. Rev. D} {\bf 87} (2013) 093006,
  [\href{http://arxiv.org/abs/1208.4607}{{\tt arXiv:1208.4607}}].

\bibitem{Graham:2013gfa}
P.~W. Graham and S.~Rajendran, {\it {New Observables for Direct Detection of
  Axion Dark Matter}},  {\em Phys. Rev. D} {\bf 88} (2013) 035023,
  [\href{http://arxiv.org/abs/1306.6088}{{\tt arXiv:1306.6088}}].

\bibitem{Redondo:2013lna}
J.~Redondo and G.~Raffelt, {\it {Solar constraints on hidden photons
  re-visited}},  {\em JCAP} {\bf 08} (2013) 034,
  [\href{http://arxiv.org/abs/1305.2920}{{\tt arXiv:1305.2920}}].

\bibitem{Fabbrichesi:2020wbt}
M.~Fabbrichesi, E.~Gabrielli, and G.~Lanfranchi, {\it {The Dark Photon}},
  \href{http://arxiv.org/abs/2005.01515}{{\tt arXiv:2005.01515}}.

\bibitem{Boyarsky:2007ge}
A.~Boyarsky, D.~Malyshev, A.~Neronov, and O.~Ruchayskiy, {\it {Constraining DM
  properties with SPI}},  {\em Mon. Not. Roy. Astron. Soc.} {\bf 387} (2008)
  1345, [\href{http://arxiv.org/abs/0710.4922}{{\tt arXiv:0710.4922}}].

\bibitem{Foster:2021ngm}
J.~W. Foster, M.~Kongsore, C.~Dessert, Y.~Park, N.~L. Rodd, K.~Cranmer, and
  B.~R. Safdi, {\it {Deep Search for Decaying Dark Matter with XMM-Newton
  Blank-Sky Observations}},  {\em Phys. Rev. Lett.} {\bf 127} (2021), no.~5
  051101, [\href{http://arxiv.org/abs/2102.02207}{{\tt arXiv:2102.02207}}].

\bibitem{Morgan:2020rwc}
C.~Thorpe-Morgan, D.~Malyshev, A.~Santangelo, J.~Jochum, B.~J\"ager, M.~Sasaki,
  and S.~Saeedi, {\it {THESEUS insights into axionlike particles, dark photon,
  and sterile neutrino dark matter}},  {\em Phys. Rev. D} {\bf 102} (2020),
  no.~12 123003, [\href{http://arxiv.org/abs/2008.08306}{{\tt
  arXiv:2008.08306}}].

\bibitem{Dekker:2021bos}
A.~Dekker, E.~Peerbooms, F.~Zimmer, K.~C.~Y. Ng, and S.~Ando, {\it {Searches
  for sterile neutrinos and axionlike particles from the Galactic halo with
  eROSITA}},  {\em Phys. Rev. D} {\bf 104} (2021), no.~2 023021,
  [\href{http://arxiv.org/abs/2103.13241}{{\tt arXiv:2103.13241}}].

\bibitem{Viel:2005qj}
M.~Viel, J.~Lesgourgues, M.~G. Haehnelt, S.~Matarrese, and A.~Riotto, {\it
  {Constraining warm dark matter candidates including sterile neutrinos and
  light gravitinos with WMAP and the Lyman-alpha forest}},  {\em Phys. Rev. D}
  {\bf 71} (2005) 063534, [\href{http://arxiv.org/abs/astro-ph/0501562}{{\tt
  astro-ph/0501562}}].

\bibitem{Irsic:2017ixq}
V.~Ir\v{s}i\v{c} et~al., {\it {New Constraints on the free-streaming of warm
  dark matter from intermediate and small scale Lyman-$\alpha$ forest data}},
  {\em Phys. Rev. D} {\bf 96} (2017), no.~2 023522,
  [\href{http://arxiv.org/abs/1702.01764}{{\tt arXiv:1702.01764}}].

\bibitem{Kamada:2019kpe}
A.~Kamada and K.~Yanagi, {\it {Constraining FIMP from the structure formation
  of the Universe: analytic mapping from $m_{WDM}$}},  {\em JCAP} {\bf 11}
  (2019) 029, [\href{http://arxiv.org/abs/1907.04558}{{\tt arXiv:1907.04558}}].

\bibitem{Lyth:1991ub}
D.~H. Lyth, {\it {Axions and inflation: Sitting in the vacuum}},  {\em Phys.
  Rev. D} {\bf 45} (1992) 3394--3404.

\bibitem{Visinelli:2009zm}
L.~Visinelli and P.~Gondolo, {\it {Dark Matter Axions Revisited}},  {\em Phys.
  Rev. D} {\bf 80} (2009) 035024, [\href{http://arxiv.org/abs/0903.4377}{{\tt
  arXiv:0903.4377}}].

\bibitem{Bulbul:2014sua}
E.~Bulbul, M.~Markevitch, A.~Foster, R.~K. Smith, M.~Loewenstein, and S.~W.
  Randall, {\it {Detection of An Unidentified Emission Line in the Stacked
  X-ray spectrum of Galaxy Clusters}},  {\em Astrophys. J.} {\bf 789} (2014)
  13, [\href{http://arxiv.org/abs/1402.2301}{{\tt arXiv:1402.2301}}].

\bibitem{Boyarsky:2014jta}
A.~Boyarsky, O.~Ruchayskiy, D.~Iakubovskyi, and J.~Franse, {\it {Unidentified
  Line in X-Ray Spectra of the Andromeda Galaxy and Perseus Galaxy Cluster}},
  {\em Phys. Rev. Lett.} {\bf 113} (2014) 251301,
  [\href{http://arxiv.org/abs/1402.4119}{{\tt arXiv:1402.4119}}].

\bibitem{Boyarsky:2014ska}
A.~Boyarsky, J.~Franse, D.~Iakubovskyi, and O.~Ruchayskiy, {\it {Checking the
  Dark Matter Origin of a 3.53 keV Line with the Milky Way Center}},  {\em
  Phys. Rev. Lett.} {\bf 115} (2015) 161301,
  [\href{http://arxiv.org/abs/1408.2503}{{\tt arXiv:1408.2503}}].

\bibitem{Higaki:2014zua}
T.~Higaki, K.~S. Jeong, and F.~Takahashi, {\it {The 7 keV axion dark matter and
  the X-ray line signal}},  {\em Phys. Lett. B} {\bf 733} (2014) 25--31,
  [\href{http://arxiv.org/abs/1402.6965}{{\tt arXiv:1402.6965}}].

\bibitem{Jaeckel:2014qea}
J.~Jaeckel, J.~Redondo, and A.~Ringwald, {\it {3.55 keV hint for decaying
  axionlike particle dark matter}},  {\em Phys. Rev. D} {\bf 89} (2014) 103511,
  [\href{http://arxiv.org/abs/1402.7335}{{\tt arXiv:1402.7335}}].

\bibitem{Kanemura:2014bqa}
S.~Kanemura, K.~Tsumura, K.~Yagyu, and H.~Yokoya, {\it {Fingerprinting
  nonminimal Higgs sectors}},  {\em Phys. Rev. D} {\bf 90} (2014) 075001,
  [\href{http://arxiv.org/abs/1406.3294}{{\tt arXiv:1406.3294}}].

\bibitem{Barklow:2017suo}
T.~Barklow, K.~Fujii, S.~Jung, R.~Karl, J.~List, T.~Ogawa, M.~E. Peskin, and
  J.~Tian, {\it {Improved Formalism for Precision Higgs Coupling Fits}},  {\em
  Phys. Rev. D} {\bf 97} (2018), no.~5 053003,
  [\href{http://arxiv.org/abs/1708.08912}{{\tt arXiv:1708.08912}}].

\bibitem{Staub:2009bi}
F.~Staub, {\it {From Superpotential to Model Files for FeynArts and
  CalcHep/CompHep}},  {\em Comput. Phys. Commun.} {\bf 181} (2010) 1077--1086,
  [\href{http://arxiv.org/abs/0909.2863}{{\tt arXiv:0909.2863}}].

\bibitem{Staub:2010jh}
F.~Staub, {\it {Automatic Calculation of supersymmetric Renormalization Group
  Equations and Self Energies}},  {\em Comput. Phys. Commun.} {\bf 182} (2011)
  808--833, [\href{http://arxiv.org/abs/1002.0840}{{\tt arXiv:1002.0840}}].

\bibitem{Staub:2012pb}
F.~Staub, {\it {SARAH 3.2: Dirac Gauginos, UFO output, and more}},  {\em
  Comput. Phys. Commun.} {\bf 184} (2013) 1792--1809,
  [\href{http://arxiv.org/abs/1207.0906}{{\tt arXiv:1207.0906}}].

\bibitem{Staub:2013tta}
F.~Staub, {\it {SARAH 4 : A tool for (not only SUSY) model builders}},  {\em
  Comput. Phys. Commun.} {\bf 185} (2014) 1773--1790,
  [\href{http://arxiv.org/abs/1309.7223}{{\tt arXiv:1309.7223}}].

\bibitem{Inami:1980fz}
T.~Inami and C.~S. Lim, {\it {Effects of Superheavy Quarks and Leptons in
  Low-Energy Weak Processes k(L) ---\ensuremath{>} mu anti-mu, K+
  ---\ensuremath{>} pi+ Neutrino anti-neutrino and K0
  \ensuremath{<}---\ensuremath{>} anti-K0}},  {\em Prog. Theor. Phys.} {\bf 65}
  (1981) 297. [Erratum: Prog.Theor.Phys. 65, 1772 (1981)].

\bibitem{Buras:1990fn}
A.~J. Buras, M.~Jamin, and P.~H. Weisz, {\it {Leading and Next-to-leading {QCD}
  Corrections to $\epsilon$ Parameter and $B^0 - \bar{B}^0$ Mixing in the
  Presence of a Heavy Top Quark}},  {\em Nucl. Phys. B} {\bf 347} (1990)
  491--536.

\bibitem{Buras:1989ui}
A.~J. Buras, P.~Krawczyk, M.~E. Lautenbacher, and C.~Salazar, {\it {B0 -
  Anti-B0 Mixing, {CP} Violation, $K^+ \to \pi^+ \nu \bar{\nu}$ and $B \to K
  \gamma X$ in a Two Higgs Doublet Model}},  {\em Nucl. Phys. B} {\bf 337}
  (1990) 284--312.

\bibitem{Chang:2015rva}
Q.~Chang, P.-F. Li, and X.-Q. Li, {\it {${B_s^0}$ \textendash{} ${\bar{B}}_s^0$
  mixing within minimal flavor-violating two-Higgs-doublet models}},  {\em Eur.
  Phys. J. C} {\bf 75} (2015), no.~12 594,
  [\href{http://arxiv.org/abs/1505.03650}{{\tt arXiv:1505.03650}}].

\bibitem{Saikawa:2019lng}
K.~Saikawa and T.~T. Yanagida, {\it {Stellar cooling anomalies and variant
  axion models}},  {\em JCAP} {\bf 03} (2020) 007,
  [\href{http://arxiv.org/abs/1907.07662}{{\tt arXiv:1907.07662}}].

\bibitem{Bardeen:1977bd}
W.~A. Bardeen and S.~H.~H. Tye, {\it {Current Algebra Applied to Properties of
  the Light Higgs Boson}},  {\em Phys. Lett. B} {\bf 74} (1978) 229--232.

\bibitem{Srednicki:1985xd}
M.~Srednicki, {\it {Axion Couplings to Matter. 1. CP Conserving Parts}},  {\em
  Nucl. Phys. B} {\bf 260} (1985) 689--700.

\end{thebibliography}\endgroup


\providecommand{\href}[2]{#2}\begingroup\raggedright\endgroup
\bibliographystyle{JHEP}

\end{document}